\documentclass[aps,prl,twocolumn]{revtex4-2}

\usepackage{graphicx}  
\usepackage{dcolumn}   
\usepackage{bm}        
\usepackage{amssymb}   
\usepackage{amsmath}
\usepackage{version}
\usepackage{braket}

\usepackage[mathscr]{euscript}
\usepackage{bbold}
\usepackage{xcolor}

\newcommand{\be}{\begin{equation}}
\newcommand{\ee}{\end{equation}}
\newcommand{\bea}{\begin{eqnarray}}
\newcommand{\eea}{\end{eqnarray}}

\newcommand{\ad}{\ensuremath{a^{\dagger}}}
\newcommand{\dd}{\ensuremath{d^{\dagger}}}

\newcommand{\Oc}{\ensuremath{\Omega_C}}
\newcommand{\Ox}{\ensuremath{\Omega_X}}
\newcommand{\wc}{\ensuremath{\omega_C}}
\newcommand{\wx}{\ensuremath{\omega_X}}

\newcommand{\gc}{\ensuremath{\gamma_C}}
\newcommand{\gx}{\ensuremath{\gamma_X}}

\DeclareMathAlphabet\mathbfcal{OMS}{cmsy}{b}{n}

\newcommand{\Eq}[1]{Eq.\,(\ref{#1})}
\newcommand{\Eqs}[2]{Eqs.\,(\ref{#1}) and (\ref{#2})}

\newcommand{\Fig}[1]{Fig.\,\ref{#1}}

\newcommand\hL{\hat{L}}
\newcommand\hU{\hat{U}}
\newcommand\hV{\hat{V}}
\newcommand\hO{\hat{\Omega}}

\newcommand{\Ec}{\mathbfcal{E}}
\newcommand{\bmu}{\pmb{\mu}}

\newcommand{\DN}{\ensuremath{{\cal S}}}

\newcommand{\NWM}{${\cal N}$WM}

\usepackage{xr} 
\begin{document}
\title{Quantum Mollow quadruplet in non-linear cavity-QED}

\author{Thomas Allcock}
\author{Wolfgang Langbein}
\author{Egor A. Muljarov}
\affiliation{School\,of\,Physics\,and\,Astronomy,\,Cardiff\,University,\,The Parade,\,Cardiff\,CF24\,3AA,\,United\,Kingdom}

\date{\today}

\begin{abstract}
We develop an exact analytical approach to the optical response of a quantum dot-microcavity system for arbitrary excitation strengths. The response is determined in terms of the complex amplitudes of transitions between the rungs of the Jaynes-Cummings ladder, explicitly isolating nonlinearities of different orders. Increasing the pulse area of the excitation field, we demonstrate the formation of a quantum Mollow quadruplet (QMQ), quantizing the semi-classical Mollow triplet into a coherent superposition of a large number of transitions between rungs of the ladder, with inner and outer doublets of the QMQ formed by densely lying inner and outer quantum transitions between the split rungs. Remarkably, a closed-form analytic approximation for the QMQ of any order of nonlinearity is found in the high-field low-damping limit.
\end{abstract}

\maketitle

The strong coupling regime of cavity quantum electrodynamics (QED), in which light-matter interaction dominates over any dissipation processes, is of both fundamental and technological interest.
It gives rise to a formation of mixed states of light and matter, called polaritons~\cite{HopfieldPR58}, and to observation of characteristic vacuum Rabi splitting~\cite{WeisbuchPRL92,HoodPRL98}.
The latter, being recently observed also in semiconductor quantum dots (QDs) coupled to an optical cavity mode (CM)~\cite{ReithmaierN04,YoshieN04}, is often referred to as linear classical effect, as it can be described  by a coupled oscillator model and studied in linear optics.
In a widely used two-level model of a QD, localized excitons are treated as fermions coupled to a bosonic CM. This coupling introduces a quantum nonlinearity~\cite{FinkN08,BishopNP08}, which results in an effective photon-photon interaction~\cite{BirnbaumN05} that can be naturally observed in nonlinear optical spectroscopy~\cite{SchusterN08,KasprzakNMa10}.

The interaction of a two-level system with a single bosonic CM is described by the Jaynes-Cummings (JC) model~\cite{JaynesIEEE63}. The eigenstates of the system form a JC ladder, with a splitting of the polariton-like doublet within each rung proportional to the square root of the rung number. The increasing higher rung splitting is evidence for quantum nonlinearity and quantum strong coupling. The latter was observed \cite{FaraonNPhy08,KasprzakNMa10} for QD excitons and also in other systems, such as superconducting circuits \cite{FinkN08}. A culmination of the quantum strong coupling is a quantum Mollow triplet (MT) forming in optical spectra of QDs with increasing optical excitation. The classical MT~\cite{MollowPR69} has been recently demonstrated in the coherent emission of QDs~\cite{MullerPRL07,AtesPRL09,VamivakasNP09,FlaggNP09,AstafievS10}. Physically, light trapped within a cavity interacts with a QD exciton in the same way as a classical coherent wave, the salient differentiating feature being the quantization of the cavity photon number. A theoretical study of the QD emission spectrum under incoherent optical excitation demonstrated a quantum MT formed due to a superposition of higher-rung transitions~\cite{LaussyPRB06,VallePRB09,VallePRL10}. This incoherent quantum MT in a QD-cavity system was shown~\cite{VallePRL10} to be different from the classical MT, but is hard to observe due to the short cavity lifetime in presently available structures, preventing the excitation of large photon numbers by the QD. In order to study the quantum MT, the nonlinear optical response of a coherently excited QD-cavity system is thus the observable in experimental reach.

We focus here on the four-wave mixing (FWM) and higher-order optical nonlinearities, which can be measured by heterodyne spectral interferometry~\cite{LangbeinOL06}.
For excitation with average photon numbers much lower than one, only the first two rungs of the JC ladder are relevant in the FWM response, with six optical transitions fully describing the dynamics of the system, as demonstrated by a good agreement with experiment~\cite{KasprzakNMa10,KasprzakNJP13,AlbertNC13}.
At higher excitations, deviations between theoretical predictions and experimental data have been attributed to higher rung contributions to nonlinear spectra, where also some signatures of a MT have been reported~\cite{KasprzakNJP13}. This experiment has been recently extended~\cite{GrollPRB20} to larger photon numbers and simulated using a time domain master equation.

In this Letter, we present an exact analytical approach to the nonlinear optical response of a QD-cavity system excited by a sequence of ultrashort pulses introducing an arbitrary number of cavity photons. A coherent optical pulse brings a cavity from its ground state into a Glauber coherent state, and its subsequent dynamics is rigorously represented by a superposition of optical transitions between the states of the JC ladder. At higher excitations, the interference of a large number of these transitions gradually transforms the nonlinear spectrum into a coherent quantum Mollow quadruplet (QMQ). To demonstrate this, we calculate the quantum dynamics with multiple precision arithmetic and take into account up to 800 rungs of the JC ladder to achieve convergence (see Sec.\,S.VII of \cite{SI}).
We provide a visualization of QMQ formation in the coherent dynamics involving an increasing number of rungs with increasing excitation. We furthermore present an analytic approximation in the low-damping limit, providing a closed-form solution for the nonlinear optical response of any order. It proves in particular that the line splitting and the linewidth of the outer (inner) doublet of the QMQ observed in nonlinear spectra are given, respectively, by $4\sqrt{n}g$ ($g/\sqrt{n}$) and $4g$ ($g/n$), where $n$ is the average number of excited cavity photons and $g$ is the QD-cavity coupling strength.

Let us start with the QD-cavity dynamics which is described~\cite{WisemanBook09,KasprzakNMa10,AlbertNC13} by the master equation (taking $\hbar=1$),
\be\label{eqn:liouville}
i\dot{\rho}= \hL\rho+[ V(t),\rho]\,,\ee
where $\rho$ is the density matrix (DM), and the Lindblad super-operator $\hL$ is given by
\bea\label{eqn:lindblad}
\hL\rho=&[H,\rho]&-i\gx(\dd d \rho+\rho \dd d-2d\rho \dd)\nonumber\\
&&-i\gc(\ad a \rho+\rho \ad a-2a\rho \ad)\,.\eea
Here, $H$ is the JC Hamiltonian of the QD-cavity system,
\be H=\Ox\dd d+\Oc\ad a+g(\dd a+\ad d)\,,
\label{eqn:hamiltonian}\ee
$\dd$ and $d$ are the creation and annihilation operators of the QD exciton, while $\ad$ and $a$ are those for the CM, having complex eigen-frequencies of $\omega_{X}=\Omega_{X}-i\gamma_{X}$ and $\omega_{C}=\Omega_{C}-i\gamma_{C}$, respectively.
The dipole coupling of the CM to the external classical electric field $\Ec(t)$ is described in the rotating wave approximation (consistent with the JC model) by an operator
\be V(t)=-\bmu\cdot \Ec(t) \ad - \bmu^\ast\!\!\cdot \Ec^\ast(t) a\,,
\label{eqn:Vdef} \ee
in which $\bmu$  is the effective dipole moment of the CM.
For the QD-cavity system excited by a sequence of ultrashort pulses, this interaction is well described by a series of $\delta$ functions,
\be \bmu\cdot \Ec(t)=\sum_j E_j \delta(t-t_j)\,,
\label{eqn:V}\ee
where  $E_j$ is the complex amplitude, known as pulse area, of the pulse arriving at time $t_j$. Excitations by longer pulses and even finite wave packets can be approximated with \Eq{eqn:V}, as shown in Sec.\,S.I  of~\cite{SI}.

For excitations in the form of \Eq{eqn:V}, the evolution of the DM is given by a time-ordered product of operators acting on the initial DM, each such operator consisting of a pulse operator $\hat{X}(E_j)$ due to pulse $j$ and a subsequent 
Lindblad dynamics during time $\tau$ between pulses (i.e. $\tau\leqslant t_{j+1}-t_j$):
\be \rho(t_j+\tau)=e^{-i \hL\tau}\hat{X}(E_j) \rho(t_j-0_+)\,,
\label{eqn:mastereqnsol}\ee
where $0_+$ is a positive infinitesimal. The pulse operator has the following explicit form
\be \hat{X}(E) \rho= e^{i(E\ad +E^\ast a)} \rho  e^{-i(E\ad +E^\ast a)}\,,
\label{equ:X}\ee
in which  $e^{i(E\ad +E^\ast a)}= e^{-|E|^2/2}e^{iE\ad}e^{iE^\ast a}$ is an operator transforming the cavity ground state into a Glauber coherent state~\cite{GlauberPR63} with the eigenvalue $iE$, as shown in Sec.\,S.I of~\cite{SI}. Hence the average number of photons in such a coherent state, given by the expectation value of $\ad a$, is  $|E|^2$. Due to the presence of multiple pulses, the DM in general will not be in the ground state at pulse arrival. To solve this problem analytically, we introduce an extended basis of Fock states $|\nu,n\rangle$ with the occupation numbers $\nu=0,\,1$ for the QD exciton and $n=0,\,1,\,2\dots$ for the CM. Using this basis, the DM can be written as
\be \rho=\sum_{\nu\nu' n n'}\rho^{\nu\nu'}_{nn'}|\nu,n\rangle \langle\nu',n'|\,,
\label{equ:rho}\ee
so that the total optical polarization takes the form
\be P(t)={\rm Tr}\{\rho(t) a\}= \sum_{\nu n} \rho^{\nu\nu}_{n, n-1}(t)\sqrt{n}\,.
\label{eqn:pol}\ee
Furthermore, as we show in Sec.\,S.I of~\cite{SI}, the pulse operator $\hat{X}(E)$
with a complex pulse area $E=|E|e^{i\varphi}$ transforms the elements $\rho^{\nu\nu'}_{nn'}$ of the DM according to
\be \left[\hat{X}(E) \rho\right]^{\nu\nu'}_{nn'}=\sum_{kk'}e^{i\varphi(n-k-n'+k')} C_{nk} C^\ast_{n'k'} \rho^{\nu\nu'}_{kk'}
\label{equ:pulse}\ee
with the transformation matrix having the {\em analytic} form
\be C_{nk}=i^{n-k} |E|^{n-k} \sqrt{\frac{k!}{n!}}L^{n-k}_{k}(|E|^2) e^{-|E|^2/2}\,,
\label{equ:Cnm}\ee
where $L^p_k(x)$ are the associated Laguerre polynomials.

The phase factor in \Eq{equ:pulse} determines the phase $\Phi$ 
of the optical response, which in turn fixes the number of steps
\mbox{$
\DN=\nu+n-(\nu'+n')
$}
between the rungs involved in the coherent dynamics. Even starting from the ground state $\rho_0=|0,0\rangle \langle0,0|$, an optical pulse distributes the excitation across all rungs of the JC ladder. However, choosing a particular phase $\Phi=\varphi \DN$, the subsequent Lindblad evolution does not mix elements of the DM corresponding to different \DN. In fact, the JC Hamiltonian conserves the particle number, so that without dissipation ($\gamma_X=\gamma_C=0$) the evolution between pulses does not alter the rung number on either side of the DM, in this way conserving \DN. Including the dissipation introduces on both sides of the DM simultaneous relaxation between neighboring rungs, again conserving \DN.

Similarly, with a number $J$ of pulses exciting the system, all phase channels can be treated independently, so that  one can select a phase $\Phi=\sum_{j=1}^J \DN_j \varphi_j$ of the optical polarization, determining the transitions between rungs present in the coherent dynamics following the pulses. These rungs of the JC ladder are separated by a distance $\sum_{j=1}^J \DN_j$. In the standard FWM polarization, excited by a sequence of three pulses, the selected phase channel after all pulses is given by $\Phi=\varphi_2+\varphi_3-\varphi_1$, corresponding to $\DN_1=-1$ and  $\DN_2=\DN_3=1$, and thus involving transitions between neighboring rungs only. The same phase selection procedure is applicable to the evolution of the system between pulses, in this way determining the pulse delay dynamics.

The evolution of the system between pulses ($t<0$) and after pulses ($t>0$) can also be described by explicit analytic expressions. Introducing a vector $\vec{\rho}$ comprising all relevant elements of the DM, i.e. those involved in the coherent dynamics for the selected phase, the time evolution after pulses is given by
\be \vec{\rho}(t)= e^{-i\hL t} \vec{\rho}(0_+)=\hU e^{-i\hO t}\hV  \vec{\rho}(0_+)\,,
\label{eqn:rhot}\ee
where the matrices $\hU$ and $\hV$ diagonalizing the Lindblad matrix, $\hL=\hU \hO \hV$, take
an {\em analytic} form (see \cite{SI}, Sec.\,S.II) in terms of $2\times2$ matrices $Y_N$ diagonalizing the complex Hamiltonian $H_N$ of the $N$-th rung,
\be H_N = \begin{bmatrix}
\wx+(N-1)\wc & \sqrt{N} g\\
\sqrt{N} g &N\wc \end{bmatrix} =Y_N
\begin{bmatrix}
\lambda_N^- & 0\\
0 &\lambda_N^+ \end{bmatrix} Y_N^{\rm T}\,,
\label{equ:HN} \ee
where $\lambda_N^\pm$ are the complex eigenvalues of $H_N$. The diagonal matrix $\hO$ in \Eq{eqn:rhot} consists of the eigenvalues of $\hL$ which are given by $\omega_r=\lambda_{N+\DN}^s-(\lambda_N^{s'})^\ast$, with a fixed $\DN$ and all possible sign combinations of $s,s'=\pm$ and rung numbers $N$, including the case of the ground state with $\lambda_0=0$. Consequently, the optical polarization \Eq{eqn:pol} and its Fourier transform take the following analytic form
\be P(t>0)=\sum_r A_r e^{-i\omega_r t}\,,\ \ \ \tilde P(\omega)=\sum_r \frac{iA_r}{\omega-\omega_r}
\label{eqn:polt}\ee
with the amplitudes $A_r=\sum_{i,j}(\vec{a})_i(\hat{U})_{ir}(\hat{V})_{rj}(\vec{\rho})_j$, according to \Eqs{eqn:pol}{eqn:rhot}. Note that the interaction of the QD exciton with a phonon environment is not included in this formalism. However, as has been recently demonstrated with the help of an asymptotically exact solution~\cite{MorreauPRB19}, the acoustic-phonon interaction can be incorporated in the QD-cavity QED by simply renormalizing the QD-cavity coupling strength $g\to g e^{-S_{\rm HR}/2}$ and the exciton energy $\Omega_X\to \Omega_X+\Omega_p$, where $S_{\rm HR}$ is the Huang-Rhys factor and $\Omega_p$ is the polaron shift. This provides a valid approximation for the cavity polarization if $g$ is much smaller than the typical energy of phonons coupled to the QD.

\begin{figure}
	\centering
	\includegraphics[width=\columnwidth]{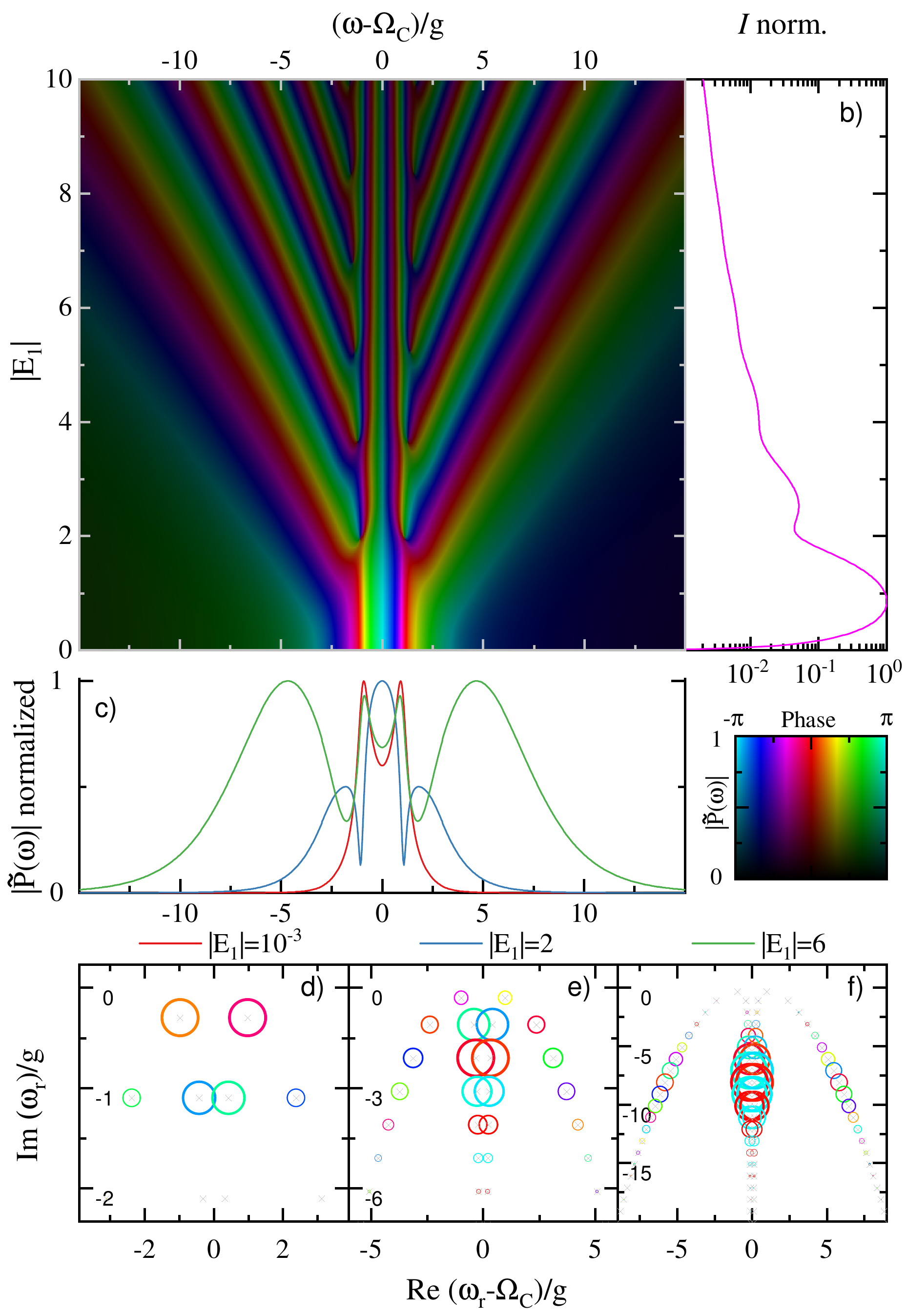}
	\caption{FWM response calculated for $|E_2|=0.001$ and varying $|E_1|$, with $\Omega_X=\Omega_C$, $\gamma_C=g/2$, and $\gamma_X=g/10$. (a) FWM spectrum $\tilde P(\omega)$ in a color plot with the hue giving the phase (see color scale) and the brightness giving the amplitude ${|\tilde{P}|}^{1/4}$. (b) spectrally integrated power $I=\int|\tilde{P}(\omega)|^2 d\omega$  versus $|E_1|$. (c) $|P(\omega)|$ normalized to 1, for selected $|E_1|$ as labelled. (d)--(f) optical transition frequencies $\omega_r$ and their complex amplitudes $A_r$ in $\tilde{P}(\omega)$, see \Eq{eqn:polt}, for different $|E_1|$ as given in (c). $\omega_r$ and $A_r$ are shown, respectively, by crosses in the complex $\omega$-plane and by circles centered at $\omega_r$  with an area proportional to $|A_r|$ and color given by the phase according to the scale in (a).}
	\label{fig:E1DgGg}
\end{figure}

\begin{figure}
	\centering
	\includegraphics[width=\columnwidth]{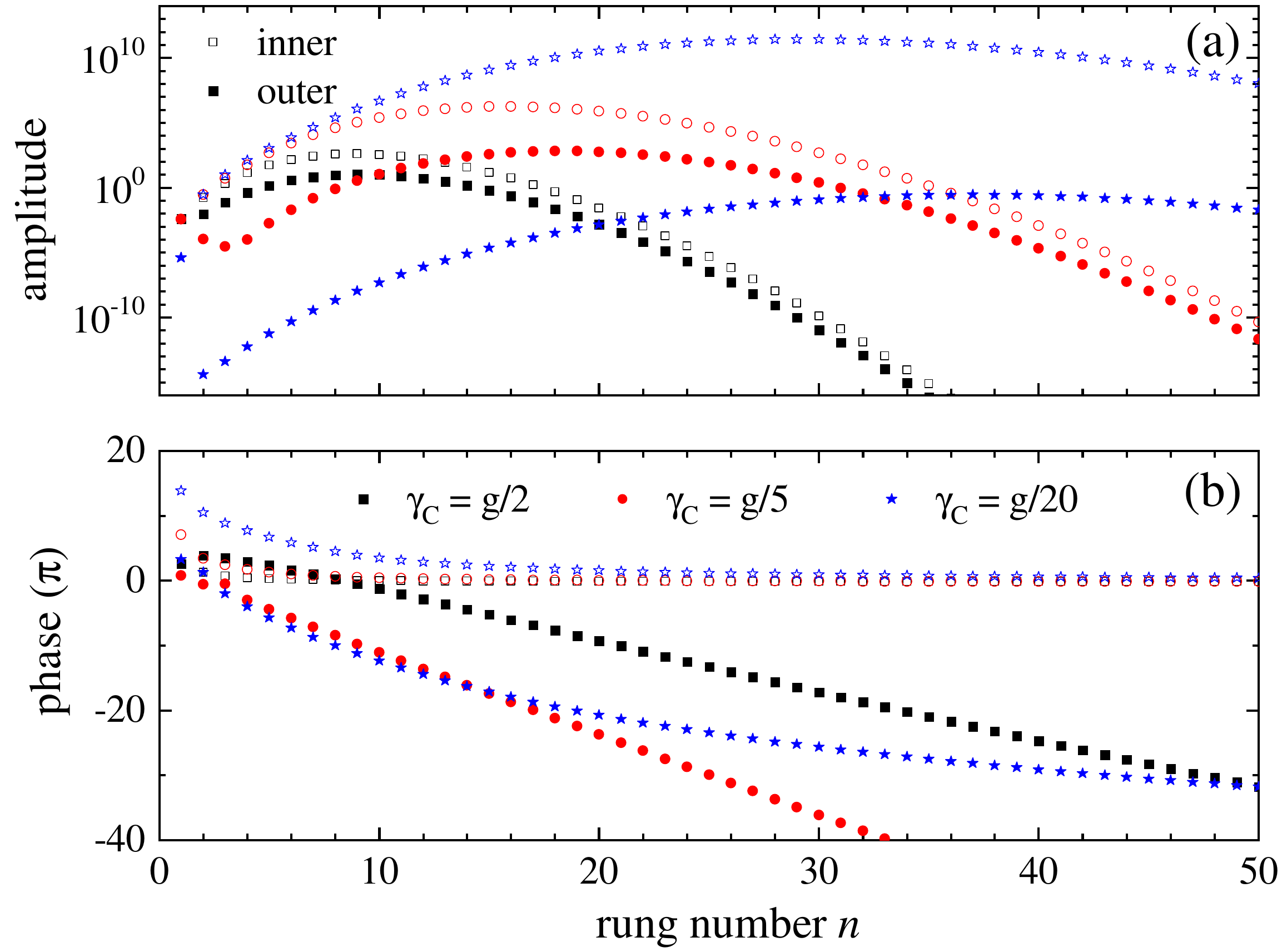}
	\caption{(a) Amplitude and (b) phase of $A_r^{(+)}(-1)^n$ for inner (empty symbols) and outer (full symbols) transitions with positive frequencies Re$(\omega_r)=\Delta_n^{ o,i}$, as functions of the rung number $n$, for $|E_1|=6$ and different values of $\gamma_C$ as given; other parameters as in \Fig{fig:E1DgGg}. Note that for negative frequencies $A_r^{(-)}= (A_r^{(+)})^\ast$.}
	\label{fig:AmpPhase}
\end{figure}

We consider below a general case of ${\cal N}$-wave mixing (\NWM) cavity polarization and  nonlinear spectrum given by \Eq{eqn:polt}. While the formalism described above is developed for any number of excitation pulses and arbitrary delay times between them, we focus here on degenerate \NWM,
generated by two optical pulses with complex pulse areas $E_1$ and $E_2$, so that ${\cal N}=|\DN_1|+|\DN_2|+1$ with $\DN=\DN_1+\DN_2=1$. The optical response of the system carries a phase $\Phi=\DN_1 \varphi_1+\DN_2 \varphi_2$ (FWM corresponds to $\Phi=2\varphi_2-\varphi_1$, with ${\cal N}=4$) and in the low-excitation regime is proportional
to a factor $ie^{i\Phi} |E_1|^{|\DN_1|}  |E_2|^{|\DN_2|}$ which we drop in all the results presented below unless otherwise stated. We assume for simplicity zero delay between the pulses and focus on the case of arbitrary $E_1$ and $|E_2|\ll1$,  and zero detuning $\Omega_X=\Omega_C$. The cases $|E_1|\ll1$ and arbitrary $E_2$, and $|E_1|=|E_2|$, as well as non-zero detuning, are considered in~\cite{SI} Sec.\,S.VI. In the below analytics we use $\Omega_C=0$ for brevity.

The FWM spectrum $\tilde P(\omega)$ calculated for  $\gamma_C=g/2$ is shown
in \Fig{fig:E1DgGg}a as a phase-amplitude color map for the pulse area $E_1$ from zero to $|E_1|=10$, exciting an average of up to 100 photons. $|\tilde P(\omega)|$ is displayed in \Fig{fig:E1DgGg}c for selected $|E_1|$, and the optical transitions between neighboring rungs which contribute to these spectra are shown in \Fig{fig:E1DgGg}d-f, in terms of the complex transition amplitudes $A_r$ (colored circles) and their frequencies $\omega_r$ (circle centers). In the low-excitation regime ($|E_1|=10^{-3}$), only the first two rungs contribute, and the spectrum shows a doublet due to the lowest-rung transitions, studied in detail in~\cite{KasprzakNMa10}.
For $|E_1|=2$, the spectrum is wider, having a central peak and sidebands, formed by a range of transitions strongest for rungs 1-5. For $|E_1|=6$, an outer doublet of increased separation and strength develops, the inner doublet reappears, and a large range of rungs are involved.

The transition frequencies are given by $ \omega_r=\pm\Delta_n^\sigma-i(2n\gamma_C+\gamma_X)$,
where $\sigma=o,\,i$, and $n$ is the rung number. For each rung $n>1$, there are two ``inner'' and  two ``outer'' transitions, corresponding to $\Delta_n^{i}=(\sqrt{n+1}-\sqrt{n})g$ and $\Delta_n^{o}=(\sqrt{n+1}+\sqrt{n})g$, respectively. Neglecting relaxation, the system excited with $|E_1|^2$ photons has a dominant contribution coming from rungs with $n\sim|E_1|^2$. This implies that as $E_1$ increases, the spectrum can consist of an inner doublet due to the inner transitions at $\omega\approx \pm \Delta_n^{i}$, close to zero, and an outer doublet at $\omega \approx \pm  \Delta_n^{o} \approx \pm 2g|E_1|$, thus forming the QMQ. Such a spectrum differs from the MT observed in a two-level system continuously driven by a classical light~\cite{MollowPR69}, by the formation of the inner doublet and the intrinsic linewidths, as we will see later. The separation of the outer doublet of $\tilde{P}(\omega)$ in \Fig{fig:E1DgGg}a grows almost linearly with $E_1$, similar to the sidebands in the MT. However, the observed splitting is somewhat smaller than $4g|E_1|$. A closer look into the complex transition amplitudes provided in \Fig{fig:AmpPhase}a reveals a Poisson-like distribution peaked at a lower rung number than excited ($n=|E_1|^2=36$), which is caused by relaxation. As $\gamma_C$ reduces, the maximum of the distribution gradually moves towards $n=|E_1|^2$. Importantly, the phase of $A_r$ across the rungs is also changing with $\gamma_C$, see \Fig{fig:AmpPhase}b. The phase determines whether the interference in the response \Eq{eqn:polt} is more constructive or destructive. This interference is so pronounced that calculating it numerically requires the use of multiprecision arithmetic and a large number of rungs -- for example, for $|E_1|=10$ the results shown use 1000 bits of precision and 500 rungs.

To better understand the observed QMQ, we develop an analytic approach to the \NWM\ response with $\DN_1=1-{\cal N}/2$ and $\DN_2={\cal N}/2$ in the limit of low damping ($\gamma_X,\gamma_C\ll g$), as  detailed in \cite{SI}, Sec.\,S.IV. Using \Eqs{equ:pulse}{equ:Cnm} we find all relevant elements of the DM after the pulses,
\be \rho^{00}_{n+1,n}(0_+)=\frac{e^{-\lambda}  \lambda^{n+1-m} L^{n+1-m}_m(\lambda)}{n!\sqrt{n+1}}\,,
\label{equ:rho00}\ee
where $\lambda=|E_1|^2$ and $m={\cal N}/2=\DN_2$.
In the limit of large pulse area, $\lambda\gg1$, the Poisson distribution in \Eq{equ:rho00} becomes Gaussian, with the mean rung number given by the mean photon number, $\langle n\rangle =\lambda$, and the mean square deviation $\langle n^2-\lambda^2\rangle =\lambda$\,. Around the maximum of this distribution,
the Laguerre polynomials are approximated as
\break
$
L^{n-m}_m(\lambda) \approx  \left(\lambda/2\right)^{{m}/{2}} H_m(z)/{m!}\,,
$
where $z=(n-\lambda)/\sqrt{2\lambda}$ and $ H_m(z)$ are Hermite polynomials, and the frequencies of the inner and outer transitions as
\be
\Delta_n^{ o}\approx 2\sqrt{\lambda}g+z\sqrt{2}g\,,
\ \ \ \
\Delta_n^{ i}\approx \frac{g}{2\sqrt{\lambda}}-z\frac{g}{2\sqrt{2}\lambda}\,.
\label{equ:Dio}\ee
This allows us to replace the transition frequencies in \Eq{eqn:polt} with $\omega_r=s \Delta^\sigma_n\approx \omega_{\sigma s}+z\gamma_{\sigma s}$, where the frequencies $\omega_{\sigma s}$ and the linewidths $\gamma_{\sigma s}$ are defined by \Eq{equ:Dio}, and $s=\pm$.
Note that the linewidth $\gamma_{\sigma s}$ is produced by a coherent superposition of many inner or outer transitions and is thus determined by their frequency dispersion with respect to the rung number $n$. The coherent dynamics after pulses can then be treated analytically, replacing  $\sum_n\to\sqrt{2\lambda}\int dz$ in \Eq{eqn:pol}, which results in the explicit form of the \NWM\ polarization:
\be P_{\sigma s}(t)=\sum_{\sigma=i,o} \sum_{s=\pm} \frac{1}{2} A^{(m)}_\sigma(\gamma_{\sigma s} t)^m e^{-i\omega_{\sigma s} t-(\gamma_{\sigma s} t)^2/4}\,,
\label{equ:Pt} \ee
where $A^{(m)}_o =(-i)^m/[4 m! (\sqrt{2\lambda})^m]$ and $A^{(m)}_i =4\lambda A^{(m)}_o$. 
Fourier transforming \Eq{equ:Pt} gives an {\it analytic} \NWM\ spectrum
\bea \tilde{P}(\omega)&=& \frac{(-i)^m}{4m!(\sqrt{2\lambda})^m}\left[ \bar{P}(\omega)+\bar{P}^\ast(-\omega)\right]
\label{equ:Pwtil}\,, \\  \nonumber
\bar{P}(\omega)&=&w_m\left(\frac{\omega-\sqrt{4\lambda}g}{\sqrt{2}g}\right)+
16\lambda^2 w_m\left(4\lambda\frac{\omega+g/\sqrt{4\lambda}}{\sqrt{2}g}\right)\,,\eea
where $w_m(z)= \frac{1}{2}\int_0^\infty t^m e^{izt}e^{-t^2/4} dt$ is a generalized Faddeeva function. \Eq{equ:Pwtil} also holds for the case of small $E_1$ and large $E_2$, by using instead $\lambda=|E_2|^2$, $m={\cal N}/2-1$, and dividing $\bar{P}(\omega)$ by $\lambda$.

\begin{figure}
	\centering
	\includegraphics[width=\columnwidth]{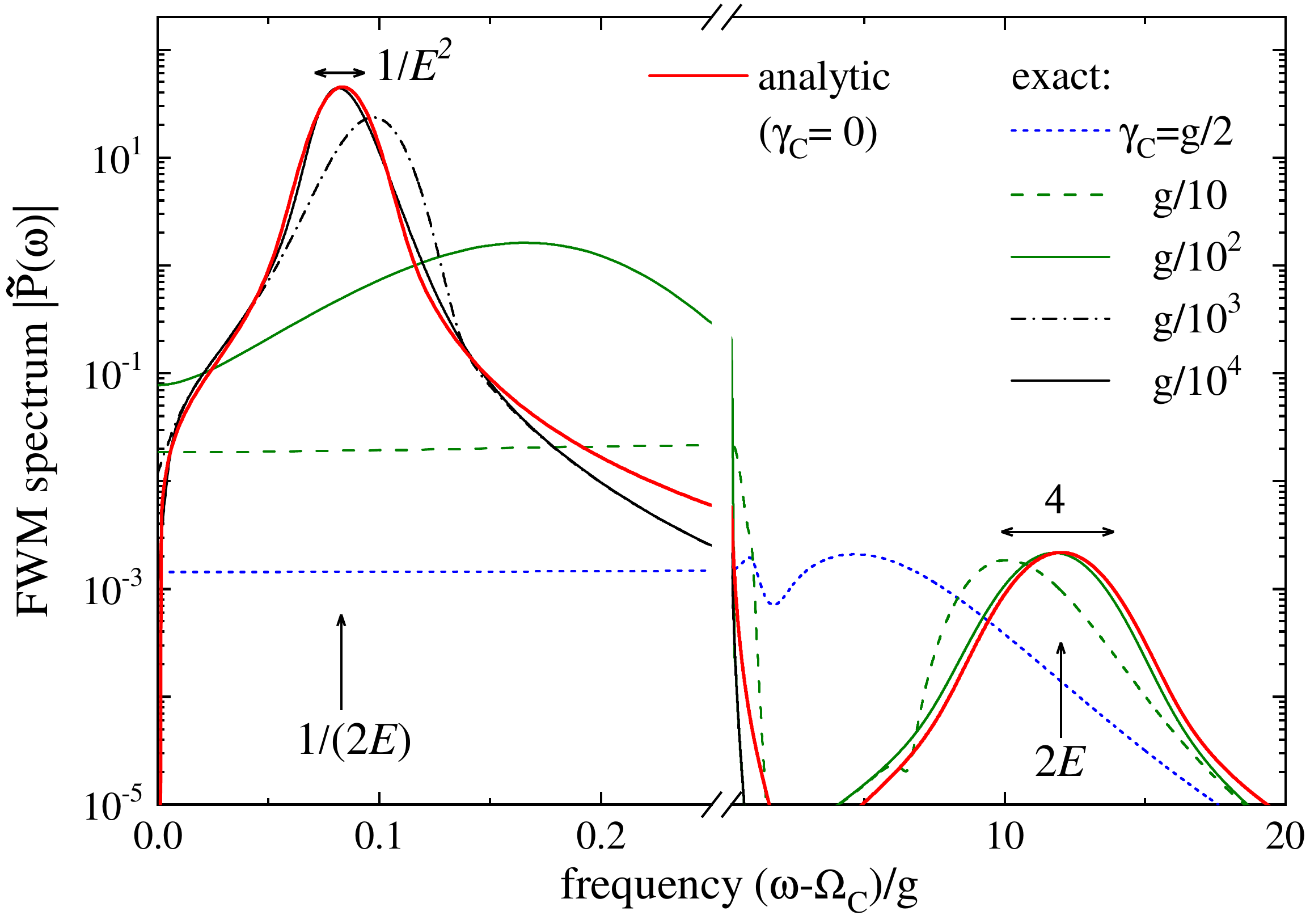}
	\caption{
	Analytic approximation \Eq{equ:Pwtil} (red curve)  and exact FWM spectrum for different $\gamma_C$ as given, for $E=\sqrt{\lambda}=|E_1|=6$. Vertical (horizontal) arrows show the position and FWHM of the spectral lines produced by the inner and outer transitions.
	}
	\label{fig:Analyt}
\end{figure}

\Fig{fig:Analyt} illustrates the analytic approximation \Eq{equ:Pwtil} for $|E_1|=6$, in comparison with the full calculation at  different values of $\gamma_C$, demonstrating good agreement in the limit of small damping. The first term in $\bar{P}$, produced by the outer transitions, describes the outer doublet of the QMQ, with a maximum at $\omega=\sqrt{4\lambda}g=2g|E_1|$, i.e. growing linearly with the pulse area, as discussed above, and a linewidth of $\sqrt{2}g$, corresponding to a full width at half maximum (FWHM) of about $4g$. The linewidth is independent of the pulse area, which is also seen in the full calculation  even for a rather large damping. This can be understood from the dispersion of the outer transitions, with the peak frequency of $\sim 2g\sqrt{\lambda}$, and the Gaussian distribution of transition amplitudes, with the root-mean-square width of $\sqrt{\lambda}$, leading to spectral width of $\sim g$,  independent of $\lambda$.
The inner transitions are in turn responsible for the inner doublet of the QMQ, which is replacing the central line of the MT. Its peak position $g/\sqrt{4\lambda}$ and FWHM $g/\lambda$ both decrease with pulse area. Notably, the relative amplitude of inner to outer doublet scales as $\lambda^2$, so that the inner doublet dominates at high pulse areas.
\begin{figure}
	\centering
	\includegraphics[width=\columnwidth]{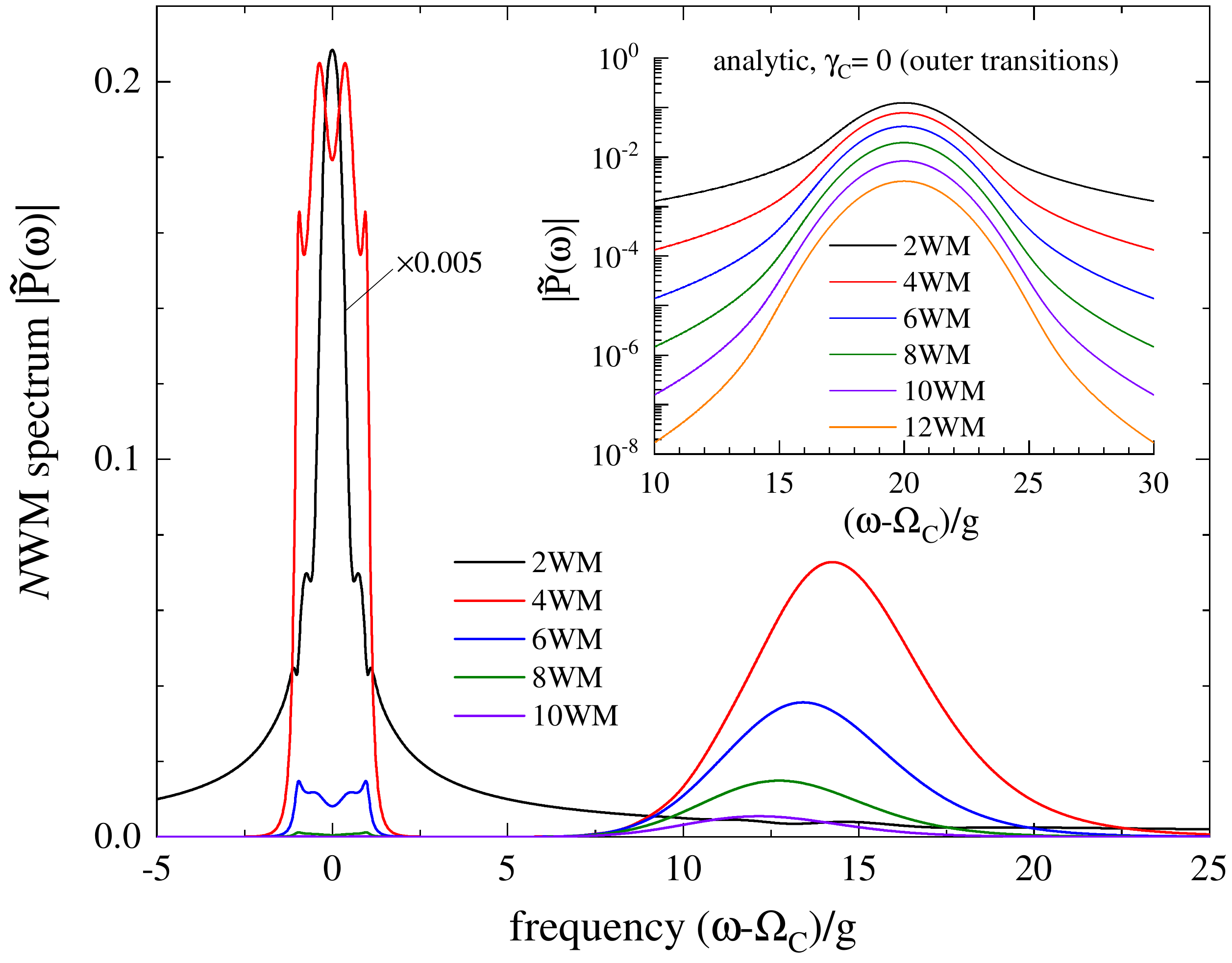}
	\caption{ \NWM\ spectra $|\tilde P(\omega)|$ of the response detected at \mbox{$\Phi=\varphi_1$} (${\cal N}=2$, black), $2\varphi_2-\varphi_1$ (${\cal N}=4$, red), $3\varphi_2-2\varphi_1$ (${\cal N}=6$, green), and  $4\varphi_2-3\varphi_1$ (${\cal N}=8$, blue), for $|E_1|=10$ and $\gamma_C=g/5$.
	Inset: Spectral line of the outer doublet of the QMQ for ${\cal N}$ up to 12, calculated using the analytic approximation \Eq{equ:Pwtil}. All spectra are multiplied with $|E_1|^{{\cal N}/2}$. }
	\label{fig:NWM}
\end{figure}

Let us now consider higher-order \NWM. The inset in \Fig{fig:NWM} shows the analytic spectrum of the outer transitions for ${\cal N}=2,\,4,\,6,\,8,\,10,$ and 12. While the FWHM almost does not change with ${\cal N}$, the spectral tails are getting more suppressed, which can be seen in the time domain as $t^{{\cal N}/2}$ rise of the polarization at short times, see \Eq{equ:Pt}.  The increase with ${\cal N}$ of the rise time is due to the fact that the optical non-linearity requires the excitation to be transferred from the cavity mode to the QD exciton and then back to the cavity, with the complexity of this  process increasing with ${\cal N}$. The $2$WM spectrum (with ${\cal N}=2$) also contains a linear-optical response which results in a long tail of the inner doublet; therefore the outer doublet is not well seen in the full spectrum in \Fig{fig:NWM}. The outer doublet is prominent in the FWM spectrum, dominates increasingly over the central band in the $6$WM (for the chosen parameters), and is getting weaker for the $8$WM and higher non-linearities, in accordance with the analytic results.

In conclusion, we find that interference of the spectrally dense transitions of a Jaynes-Cummings ladder sculpts the nonlinear spectra in a complex fashion:  A quantum Mollow quadruplet is formed by the dispersion of the transitions between the split rungs, with the inner and outer transitions giving rise to, respectively, inner and outer doublets. Remarkably, we have found in the high field and low damping limit a closed-form analytic approximation for the optical nonlinearity of any order, which explains, both qualitatively and quantitatively the observed quantum Mollow quadruplet.


T.A. acknowledges support by a PhD studentship funded by Cardiff University. This work was supported by the EPSRC under grant EP/M020479/1.


\begin{thebibliography}{0}%
\makeatletter
\providecommand \@ifxundefined [1]{%
 \@ifx{#1\undefined}
}%
\providecommand \@ifnum [1]{%
 \ifnum #1\expandafter \@firstoftwo
 \else \expandafter \@secondoftwo
 \fi
}%
\providecommand \@ifx [1]{%
 \ifx #1\expandafter \@firstoftwo
 \else \expandafter \@secondoftwo
 \fi
}%
\providecommand \natexlab [1]{#1}%
\providecommand \enquote  [1]{``#1''}%
\providecommand \bibnamefont  [1]{#1}%
\providecommand \bibfnamefont [1]{#1}%
\providecommand \citenamefont [1]{#1}%
\providecommand \href@noop [0]{\@secondoftwo}%
\providecommand \href [0]{\begingroup \@sanitize@url \@href}%
\providecommand \@href[1]{\@@startlink{#1}\@@href}%
\providecommand \@@href[1]{\endgroup#1\@@endlink}%
\providecommand \@sanitize@url [0]{\catcode `\\12\catcode `\$12\catcode
  `\&12\catcode `\#12\catcode `\^12\catcode `\_12\catcode `\%12\relax}%
\providecommand \@@startlink[1]{}%
\providecommand \@@endlink[0]{}%
\providecommand \url  [0]{\begingroup\@sanitize@url \@url }%
\providecommand \@url [1]{\endgroup\@href {#1}{\urlprefix }}%
\providecommand \urlprefix  [0]{URL }%
\providecommand \Eprint [0]{\href }%
\providecommand \doibase [0]{https://doi.org/}%
\providecommand \selectlanguage [0]{\@gobble}%
\providecommand \bibinfo  [0]{\@secondoftwo}%
\providecommand \bibfield  [0]{\@secondoftwo}%
\providecommand \translation [1]{[#1]}%
\providecommand \BibitemOpen [0]{}%
\providecommand \bibitemStop [0]{}%
\providecommand \bibitemNoStop [0]{.\EOS\space}%
\providecommand \EOS [0]{\spacefactor3000\relax}%
\providecommand \BibitemShut  [1]{\csname bibitem#1\endcsname}%
\let\auto@bib@innerbib\@empty
\end{thebibliography}%


\begin{thebibliography}{3}%
\makeatletter
\providecommand \@ifxundefined [1]{%
 \@ifx{#1\undefined}
}%
\providecommand \@ifnum [1]{%
 \ifnum #1\expandafter \@firstoftwo
 \else \expandafter \@secondoftwo
 \fi
}%
\providecommand \@ifx [1]{%
 \ifx #1\expandafter \@firstoftwo
 \else \expandafter \@secondoftwo
 \fi
}%
\providecommand \natexlab [1]{#1}%
\providecommand \enquote  [1]{``#1''}%
\providecommand \bibnamefont  [1]{#1}%
\providecommand \bibfnamefont [1]{#1}%
\providecommand \citenamefont [1]{#1}%
\providecommand \href@noop [0]{\@secondoftwo}%
\providecommand \href [0]{\begingroup \@sanitize@url \@href}%
\providecommand \@href[1]{\@@startlink{#1}\@@href}%
\providecommand \@@href[1]{\endgroup#1\@@endlink}%
\providecommand \@sanitize@url [0]{\catcode `\\12\catcode `\$12\catcode
  `\&12\catcode `\#12\catcode `\^12\catcode `\_12\catcode `\%12\relax}%
\providecommand \@@startlink[1]{}%
\providecommand \@@endlink[0]{}%
\providecommand \url  [0]{\begingroup\@sanitize@url \@url }%
\providecommand \@url [1]{\endgroup\@href {#1}{\urlprefix }}%
\providecommand \urlprefix  [0]{URL }%
\providecommand \Eprint [0]{\href }%
\providecommand \doibase [0]{https://doi.org/}%
\providecommand \selectlanguage [0]{\@gobble}%
\providecommand \bibinfo  [0]{\@secondoftwo}%
\providecommand \bibfield  [0]{\@secondoftwo}%
\providecommand \translation [1]{[#1]}%
\providecommand \BibitemOpen [0]{}%
\providecommand \bibitemStop [0]{}%
\providecommand \bibitemNoStop [0]{.\EOS\space}%
\providecommand \EOS [0]{\spacefactor3000\relax}%
\providecommand \BibitemShut  [1]{\csname bibitem#1\endcsname}%
\let\auto@bib@innerbib\@empty
\bibitem [{\citenamefont {Suzuki}(1976)}]{Suzuki1976}%
  \BibitemOpen
  \bibfield  {author} {\bibinfo {author} {\bibfnamefont {M.}~\bibnamefont
  {Suzuki}},\ }\bibfield  {title} {\bibinfo {title} {Generalized trotter's
  formula and systematic approximants of exponential operators and inner
  derivations with applications to many-body problems},\ }\href
  {https://doi.org/10.1007/BF01609348} {\bibfield  {journal} {\bibinfo
  {journal} {Commun. Math. Phys.}\ }\textbf {\bibinfo {volume} {51}},\ \bibinfo
  {pages} {183} (\bibinfo {year} {1976})}\BibitemShut {NoStop}%
\bibitem [{\citenamefont {Gradshtein}\ and\ \citenamefont
  {Ryzhik}(1965)}]{Gradshtein}%
  \BibitemOpen
  \bibfield  {author} {\bibinfo {author} {\bibfnamefont {I.~S.}\ \bibnamefont
  {Gradshtein}}\ and\ \bibinfo {author} {\bibfnamefont {I.~M.}\ \bibnamefont
  {Ryzhik}},\ }\href@noop {} {\emph {\bibinfo {title} {Table of Integrals,
  Series, and Products}}}\ (\bibinfo  {publisher} {Academic Press, New York},\
  \bibinfo {year} {1965})\BibitemShut {NoStop}%
\bibitem [{\citenamefont {Kasprzak}\ \emph {et~al.}(2010)\citenamefont
  {Kasprzak}, \citenamefont {Reitzenstein}, \citenamefont {Muljarov},
  \citenamefont {Kistner}, \citenamefont {Schneider}, \citenamefont {Strauss},
  \citenamefont {H{\"o}fling}, \citenamefont {Forchel},\ and\ \citenamefont
  {Langbein}}]{KasprzakNMa10}%
  \BibitemOpen
  \bibfield  {author} {\bibinfo {author} {\bibfnamefont {J.}~\bibnamefont
  {Kasprzak}}, \bibinfo {author} {\bibfnamefont {S.}~\bibnamefont
  {Reitzenstein}}, \bibinfo {author} {\bibfnamefont {E.~A.}\ \bibnamefont
  {Muljarov}}, \bibinfo {author} {\bibfnamefont {C.}~\bibnamefont {Kistner}},
  \bibinfo {author} {\bibfnamefont {C.}~\bibnamefont {Schneider}}, \bibinfo
  {author} {\bibfnamefont {M.}~\bibnamefont {Strauss}}, \bibinfo {author}
  {\bibfnamefont {S.}~\bibnamefont {H{\"o}fling}}, \bibinfo {author}
  {\bibfnamefont {A.}~\bibnamefont {Forchel}},\ and\ \bibinfo {author}
  {\bibfnamefont {W.}~\bibnamefont {Langbein}},\ }\bibfield  {title} {\bibinfo
  {title} {Up on the jaynes{\textendash}cummings ladder of a
  quantum-dot/microcavity system},\ }\href {https://doi.org/10.1038/nmat2717}
  {\bibfield  {journal} {\bibinfo  {journal} {Nat. Mater.}\ }\textbf {\bibinfo
  {volume} {9}},\ \bibinfo {pages} {304} (\bibinfo {year} {2010})}\BibitemShut
  {NoStop}%
\end{thebibliography}%


\begin{thebibliography}{29}%
	\makeatletter
	\providecommand \@ifxundefined [1]{%
	 \@ifx{#1\undefined}
	}%
	\providecommand \@ifnum [1]{%
	 \ifnum #1\expandafter \@firstoftwo
	 \else \expandafter \@secondoftwo
	 \fi
	}%
	\providecommand \@ifx [1]{%
	 \ifx #1\expandafter \@firstoftwo
	 \else \expandafter \@secondoftwo
	 \fi
	}%
	\providecommand \natexlab [1]{#1}%
	\providecommand \enquote  [1]{``#1''}%
	\providecommand \bibnamefont  [1]{#1}%
	\providecommand \bibfnamefont [1]{#1}%
	\providecommand \citenamefont [1]{#1}%
	\providecommand \href@noop [0]{\@secondoftwo}%
	\providecommand \href [0]{\begingroup \@sanitize@url \@href}%
	\providecommand \@href[1]{\@@startlink{#1}\@@href}%
	\providecommand \@@href[1]{\endgroup#1\@@endlink}%
	\providecommand \@sanitize@url [0]{\catcode `\\12\catcode `\$12\catcode
	  `\&12\catcode `\#12\catcode `\^12\catcode `\_12\catcode `\%12\relax}%
	\providecommand \@@startlink[1]{}%
	\providecommand \@@endlink[0]{}%
	\providecommand \url  [0]{\begingroup\@sanitize@url \@url }%
	\providecommand \@url [1]{\endgroup\@href {#1}{\urlprefix }}%
	\providecommand \urlprefix  [0]{URL }%
	\providecommand \Eprint [0]{\href }%
	\providecommand \doibase [0]{https://doi.org/}%
	\providecommand \selectlanguage [0]{\@gobble}%
	\providecommand \bibinfo  [0]{\@secondoftwo}%
	\providecommand \bibfield  [0]{\@secondoftwo}%
	\providecommand \translation [1]{[#1]}%
	\providecommand \BibitemOpen [0]{}%
	\providecommand \bibitemStop [0]{}%
	\providecommand \bibitemNoStop [0]{.\EOS\space}%
	\providecommand \EOS [0]{\spacefactor3000\relax}%
	\providecommand \BibitemShut  [1]{\csname bibitem#1\endcsname}%
	\let\auto@bib@innerbib\@empty
	\bibitem [{\citenamefont {Hopfield}(1958)}]{HopfieldPR58}%
	  \BibitemOpen
	  \bibfield  {author} {\bibinfo {author} {\bibfnamefont {J.~J.}\ \bibnamefont
	  {Hopfield}},\ }\bibfield  {title} {\bibinfo {title} {Theory of the
	  contribution of excitons to the complex dielectric constant of crystals},\
	  }\href {https://doi.org/10.1103/PhysRev.112.1555} {\bibfield  {journal}
	  {\bibinfo  {journal} {Phys. Rev.}\ }\textbf {\bibinfo {volume} {112}},\
	  \bibinfo {pages} {1555} (\bibinfo {year} {1958})}\BibitemShut {NoStop}%
	\bibitem [{\citenamefont {Weisbuch}\ \emph {et~al.}(1992)\citenamefont
	  {Weisbuch}, \citenamefont {Nishioka}, \citenamefont {Ishikawa},\ and\
	  \citenamefont {Arakawa}}]{WeisbuchPRL92}%
	  \BibitemOpen
	  \bibfield  {author} {\bibinfo {author} {\bibfnamefont {C.}~\bibnamefont
	  {Weisbuch}}, \bibinfo {author} {\bibfnamefont {M.}~\bibnamefont {Nishioka}},
	  \bibinfo {author} {\bibfnamefont {A.}~\bibnamefont {Ishikawa}},\ and\
	  \bibinfo {author} {\bibfnamefont {Y.}~\bibnamefont {Arakawa}},\ }\bibfield
	  {title} {\bibinfo {title} {Observation of the coupled exciton-photon mode
	  splitting in a semiconductor quantum microcavity},\ }\href@noop {} {\bibfield
	   {journal} {\bibinfo  {journal} {Phys. Rev. Lett.}\ }\textbf {\bibinfo
	  {volume} {69}},\ \bibinfo {pages} {3314} (\bibinfo {year}
	  {1992})}\BibitemShut {NoStop}%
	\bibitem [{\citenamefont {Hood}\ \emph {et~al.}(1998)\citenamefont {Hood},
	  \citenamefont {Chapman}, \citenamefont {Lynn},\ and\ \citenamefont
	  {Kimble}}]{HoodPRL98}%
	  \BibitemOpen
	  \bibfield  {author} {\bibinfo {author} {\bibfnamefont {C.~J.}\ \bibnamefont
	  {Hood}}, \bibinfo {author} {\bibfnamefont {M.~S.}\ \bibnamefont {Chapman}},
	  \bibinfo {author} {\bibfnamefont {T.~W.}\ \bibnamefont {Lynn}},\ and\
	  \bibinfo {author} {\bibfnamefont {H.~J.}\ \bibnamefont {Kimble}},\ }\bibfield
	   {title} {\bibinfo {title} {Real-time cavity qed with single atoms},\ }\href
	  {https://doi.org/10.1103/PhysRevLett.80.4157} {\bibfield  {journal} {\bibinfo
	   {journal} {Phys. Rev. Lett.}\ }\textbf {\bibinfo {volume} {80}},\ \bibinfo
	  {pages} {4157} (\bibinfo {year} {1998})}\BibitemShut {NoStop}%
	\bibitem [{\citenamefont {Reithmaier}\ \emph {et~al.}(2004)\citenamefont
	  {Reithmaier}, \citenamefont {S\c{e}k}, \citenamefont {L{\"o}ffler},
	  \citenamefont {Hoffmann}, \citenamefont {Kuhn}, \citenamefont {Reitzenstein},
	  \citenamefont {Keldysh}, \citenamefont {Kulakovskii}, \citenamefont
	  {Reinecke},\ and\ \citenamefont {Forchel}}]{ReithmaierN04}%
	  \BibitemOpen
	  \bibfield  {author} {\bibinfo {author} {\bibfnamefont {J.~P.}\ \bibnamefont
	  {Reithmaier}}, \bibinfo {author} {\bibfnamefont {G.}~\bibnamefont {S\c{e}k}},
	  \bibinfo {author} {\bibfnamefont {A.}~\bibnamefont {L{\"o}ffler}}, \bibinfo
	  {author} {\bibfnamefont {C.}~\bibnamefont {Hoffmann}}, \bibinfo {author}
	  {\bibfnamefont {S.}~\bibnamefont {Kuhn}}, \bibinfo {author} {\bibfnamefont
	  {S.}~\bibnamefont {Reitzenstein}}, \bibinfo {author} {\bibfnamefont {L.~V.}\
	  \bibnamefont {Keldysh}}, \bibinfo {author} {\bibfnamefont {V.~D.}\
	  \bibnamefont {Kulakovskii}}, \bibinfo {author} {\bibfnamefont {T.~L.}\
	  \bibnamefont {Reinecke}},\ and\ \bibinfo {author} {\bibfnamefont
	  {A.}~\bibnamefont {Forchel}},\ }\bibfield  {title} {\bibinfo {title} {Strong
	  coupling in a single quantum dot-semiconductor microcavity system},\
	  }\href@noop {} {\bibfield  {journal} {\bibinfo  {journal} {Nature}\ }\textbf
	  {\bibinfo {volume} {432}},\ \bibinfo {pages} {197} (\bibinfo {year}
	  {2004})}\BibitemShut {NoStop}%
	\bibitem [{\citenamefont {Yoshie}\ \emph {et~al.}(2004)\citenamefont {Yoshie},
	  \citenamefont {Scherer}, \citenamefont {Hendrickson}, \citenamefont
	  {Khitrova}, \citenamefont {Gibbs}, \citenamefont {Rupper}, \citenamefont
	  {Ell}, \citenamefont {Shchekin},\ and\ \citenamefont {Deppe}}]{YoshieN04}%
	  \BibitemOpen
	  \bibfield  {author} {\bibinfo {author} {\bibfnamefont {T.}~\bibnamefont
	  {Yoshie}}, \bibinfo {author} {\bibfnamefont {A.}~\bibnamefont {Scherer}},
	  \bibinfo {author} {\bibfnamefont {J.}~\bibnamefont {Hendrickson}}, \bibinfo
	  {author} {\bibfnamefont {G.}~\bibnamefont {Khitrova}}, \bibinfo {author}
	  {\bibfnamefont {H.~M.}\ \bibnamefont {Gibbs}}, \bibinfo {author}
	  {\bibfnamefont {G.}~\bibnamefont {Rupper}}, \bibinfo {author} {\bibfnamefont
	  {C.}~\bibnamefont {Ell}}, \bibinfo {author} {\bibfnamefont {O.~B.}\
	  \bibnamefont {Shchekin}},\ and\ \bibinfo {author} {\bibfnamefont {D.~G.}\
	  \bibnamefont {Deppe}},\ }\bibfield  {title} {\bibinfo {title} {Vacuum rabi
	  splitting with a single quantum dot in a photonic crystal nanocavity},\
	  }\href {https://doi.org/10.1038/nature03119} {\bibfield  {journal} {\bibinfo
	  {journal} {Nature}\ }\textbf {\bibinfo {volume} {432}},\ \bibinfo {pages}
	  {200} (\bibinfo {year} {2004})}\BibitemShut {NoStop}%
	\bibitem [{\citenamefont {Fink}\ \emph {et~al.}(2008)\citenamefont {Fink},
	  \citenamefont {Goppl}, \citenamefont {Baur}, \citenamefont {Bianchetti},
	  \citenamefont {Leek}, \citenamefont {Blais},\ and\ \citenamefont
	  {Wallraff}}]{FinkN08}%
	  \BibitemOpen
	  \bibfield  {author} {\bibinfo {author} {\bibfnamefont {J.~M.}\ \bibnamefont
	  {Fink}}, \bibinfo {author} {\bibfnamefont {M.}~\bibnamefont {Goppl}},
	  \bibinfo {author} {\bibfnamefont {M.}~\bibnamefont {Baur}}, \bibinfo {author}
	  {\bibfnamefont {R.}~\bibnamefont {Bianchetti}}, \bibinfo {author}
	  {\bibfnamefont {P.~J.}\ \bibnamefont {Leek}}, \bibinfo {author}
	  {\bibfnamefont {A.}~\bibnamefont {Blais}},\ and\ \bibinfo {author}
	  {\bibfnamefont {A.}~\bibnamefont {Wallraff}},\ }\bibfield  {title} {\bibinfo
	  {title} {Climbing the jaynes-cummings ladder and observing its nonlinearity
	  in a cavity qed system},\ }\href {https://doi.org/10.1038/nature07112}
	  {\bibfield  {journal} {\bibinfo  {journal} {Nature}\ }\textbf {\bibinfo
	  {volume} {454}},\ \bibinfo {pages} {315} (\bibinfo {year}
	  {2008})}\BibitemShut {NoStop}%
	\bibitem [{\citenamefont {Bishop}\ \emph {et~al.}(2009)\citenamefont {Bishop},
	  \citenamefont {Chow}, \citenamefont {Koch}, \citenamefont {Houck},
	  \citenamefont {Devoret}, \citenamefont {Thuneberg}, \citenamefont {Girvin},\
	  and\ \citenamefont {Schoelkopf}}]{BishopNP08}%
	  \BibitemOpen
	  \bibfield  {author} {\bibinfo {author} {\bibfnamefont {L.~S.}\ \bibnamefont
	  {Bishop}}, \bibinfo {author} {\bibfnamefont {J.~M.}\ \bibnamefont {Chow}},
	  \bibinfo {author} {\bibfnamefont {J.}~\bibnamefont {Koch}}, \bibinfo {author}
	  {\bibfnamefont {A.~A.}\ \bibnamefont {Houck}}, \bibinfo {author}
	  {\bibfnamefont {M.~H.}\ \bibnamefont {Devoret}}, \bibinfo {author}
	  {\bibfnamefont {E.}~\bibnamefont {Thuneberg}}, \bibinfo {author}
	  {\bibfnamefont {S.~M.}\ \bibnamefont {Girvin}},\ and\ \bibinfo {author}
	  {\bibfnamefont {R.~J.}\ \bibnamefont {Schoelkopf}},\ }\bibfield  {title}
	  {\bibinfo {title} {Nonlinear response of the vacuum rabi resonance},\ }\href
	  {https://doi.org/10.1038/nphys1154} {\bibfield  {journal} {\bibinfo
	  {journal} {Nat. Phys.}\ }\textbf {\bibinfo {volume} {5}},\ \bibinfo {pages}
	  {105} (\bibinfo {year} {2009})}\BibitemShut {NoStop}%
	\bibitem [{\citenamefont {Birnbaum}\ \emph {et~al.}(2005)\citenamefont
	  {Birnbaum}, \citenamefont {Boca}, \citenamefont {Miller}, \citenamefont
	  {Boozer}, \citenamefont {Northup},\ and\ \citenamefont
	  {Kimble}}]{BirnbaumN05}%
	  \BibitemOpen
	  \bibfield  {author} {\bibinfo {author} {\bibfnamefont {K.~M.}\ \bibnamefont
	  {Birnbaum}}, \bibinfo {author} {\bibfnamefont {A.}~\bibnamefont {Boca}},
	  \bibinfo {author} {\bibfnamefont {R.}~\bibnamefont {Miller}}, \bibinfo
	  {author} {\bibfnamefont {A.~D.}\ \bibnamefont {Boozer}}, \bibinfo {author}
	  {\bibfnamefont {T.~E.}\ \bibnamefont {Northup}},\ and\ \bibinfo {author}
	  {\bibfnamefont {H.~J.}\ \bibnamefont {Kimble}},\ }\bibfield  {title}
	  {\bibinfo {title} {Photon blockade in an optical cavity with one trapped
	  atom},\ }\href {https://doi.org/10.1038/nature03804} {\bibfield  {journal}
	  {\bibinfo  {journal} {Nature}\ }\textbf {\bibinfo {volume} {436}},\ \bibinfo
	  {pages} {87} (\bibinfo {year} {2005})}\BibitemShut {NoStop}%
	\bibitem [{\citenamefont {Schuster}\ \emph {et~al.}(2008)\citenamefont
	  {Schuster}, \citenamefont {Kubanek}, \citenamefont {Fuhrmanek}, \citenamefont
	  {Puppe}, \citenamefont {Pinkse}, \citenamefont {Murr},\ and\ \citenamefont
	  {Rempe}}]{SchusterN08}%
	  \BibitemOpen
	  \bibfield  {author} {\bibinfo {author} {\bibfnamefont {I.}~\bibnamefont
	  {Schuster}}, \bibinfo {author} {\bibfnamefont {A.}~\bibnamefont {Kubanek}},
	  \bibinfo {author} {\bibfnamefont {A.}~\bibnamefont {Fuhrmanek}}, \bibinfo
	  {author} {\bibfnamefont {T.}~\bibnamefont {Puppe}}, \bibinfo {author}
	  {\bibfnamefont {P.~W.~H.}\ \bibnamefont {Pinkse}}, \bibinfo {author}
	  {\bibfnamefont {K.}~\bibnamefont {Murr}},\ and\ \bibinfo {author}
	  {\bibfnamefont {G.}~\bibnamefont {Rempe}},\ }\bibfield  {title} {\bibinfo
	  {title} {Nonlinear spectroscopy of photons bound to one atom},\ }\href
	  {https://doi.org/10.1038/nphys940} {\bibfield  {journal} {\bibinfo  {journal}
	  {Nat. Phys.}\ }\textbf {\bibinfo {volume} {4}},\ \bibinfo {pages} {382}
	  (\bibinfo {year} {2008})}\BibitemShut {NoStop}%
	\bibitem [{\citenamefont {Kasprzak}\ \emph {et~al.}(2010)\citenamefont
	  {Kasprzak}, \citenamefont {Reitzenstein}, \citenamefont {Muljarov},
	  \citenamefont {Kistner}, \citenamefont {Schneider}, \citenamefont {Strauss},
	  \citenamefont {H{\"o}fling}, \citenamefont {Forchel},\ and\ \citenamefont
	  {Langbein}}]{KasprzakNMa10}%
	  \BibitemOpen
	  \bibfield  {author} {\bibinfo {author} {\bibfnamefont {J.}~\bibnamefont
	  {Kasprzak}}, \bibinfo {author} {\bibfnamefont {S.}~\bibnamefont
	  {Reitzenstein}}, \bibinfo {author} {\bibfnamefont {E.~A.}\ \bibnamefont
	  {Muljarov}}, \bibinfo {author} {\bibfnamefont {C.}~\bibnamefont {Kistner}},
	  \bibinfo {author} {\bibfnamefont {C.}~\bibnamefont {Schneider}}, \bibinfo
	  {author} {\bibfnamefont {M.}~\bibnamefont {Strauss}}, \bibinfo {author}
	  {\bibfnamefont {S.}~\bibnamefont {H{\"o}fling}}, \bibinfo {author}
	  {\bibfnamefont {A.}~\bibnamefont {Forchel}},\ and\ \bibinfo {author}
	  {\bibfnamefont {W.}~\bibnamefont {Langbein}},\ }\bibfield  {title} {\bibinfo
	  {title} {Up on the jaynes{\textendash}cummings ladder of a
	  quantum-dot/microcavity system},\ }\href {https://doi.org/10.1038/nmat2717}
	  {\bibfield  {journal} {\bibinfo  {journal} {Nat. Mater.}\ }\textbf {\bibinfo
	  {volume} {9}},\ \bibinfo {pages} {304} (\bibinfo {year} {2010})}\BibitemShut
	  {NoStop}%
	\bibitem [{\citenamefont {Jaynes}\ and\ \citenamefont
	  {Cummings}(1963)}]{JaynesIEEE63}%
	  \BibitemOpen
	  \bibfield  {author} {\bibinfo {author} {\bibfnamefont {E.}~\bibnamefont
	  {Jaynes}}\ and\ \bibinfo {author} {\bibfnamefont {F.}~\bibnamefont
	  {Cummings}},\ }\bibfield  {title} {\bibinfo {title} {Comparison of quantum
	  and semiclassical radiation theory with application to the beam maser.},\
	  }\href@noop {} {\bibfield  {journal} {\bibinfo  {journal} {Proc. IEEE}\
	  }\textbf {\bibinfo {volume} {51}},\ \bibinfo {pages} {89} (\bibinfo {year}
	  {1963})}\BibitemShut {NoStop}%
	\bibitem [{\citenamefont {Faraon}\ \emph {et~al.}(2008)\citenamefont {Faraon},
	  \citenamefont {Fushman}, \citenamefont {Englund}, \citenamefont {Stoltz},
	  \citenamefont {Petroff},\ and\ \citenamefont {Vuckovic}}]{FaraonNPhy08}%
	  \BibitemOpen
	  \bibfield  {author} {\bibinfo {author} {\bibfnamefont {A.}~\bibnamefont
	  {Faraon}}, \bibinfo {author} {\bibfnamefont {I.}~\bibnamefont {Fushman}},
	  \bibinfo {author} {\bibfnamefont {D.}~\bibnamefont {Englund}}, \bibinfo
	  {author} {\bibfnamefont {N.}~\bibnamefont {Stoltz}}, \bibinfo {author}
	  {\bibfnamefont {P.}~\bibnamefont {Petroff}},\ and\ \bibinfo {author}
	  {\bibfnamefont {J.}~\bibnamefont {Vuckovic}},\ }\bibfield  {title} {\bibinfo
	  {title} {Coherent generation of non-classical light on a chip via
	  photon-induced tunnelling and blockade},\ }\href
	  {https://doi.org/10.1038/nphys1078} {\bibfield  {journal} {\bibinfo
	  {journal} {Nat. Phys.}\ }\textbf {\bibinfo {volume} {4}},\ \bibinfo {pages}
	  {859} (\bibinfo {year} {2008})}\BibitemShut {NoStop}%
	\bibitem [{\citenamefont {Mollow}(1969)}]{MollowPR69}%
	  \BibitemOpen
	  \bibfield  {author} {\bibinfo {author} {\bibfnamefont {B.~R.}\ \bibnamefont
	  {Mollow}},\ }\bibfield  {title} {\bibinfo {title} {Power spectrum of light
	  scattered by two-level systems},\ }\href
	  {https://doi.org/10.1103/physrev.188.1969} {\bibfield  {journal} {\bibinfo
	  {journal} {Phys. Rev.}\ }\textbf {\bibinfo {volume} {188}},\ \bibinfo {pages}
	  {1969} (\bibinfo {year} {1969})}\BibitemShut {NoStop}%
	\bibitem [{\citenamefont {Muller}\ \emph {et~al.}(2007)\citenamefont {Muller},
	  \citenamefont {Flagg}, \citenamefont {Bianucci}, \citenamefont {Wang},
	  \citenamefont {Deppe}, \citenamefont {Ma}, \citenamefont {Zhang},
	  \citenamefont {Salamo}, \citenamefont {Xiao},\ and\ \citenamefont
	  {Shih}}]{MullerPRL07}%
	  \BibitemOpen
	  \bibfield  {author} {\bibinfo {author} {\bibfnamefont {A.}~\bibnamefont
	  {Muller}}, \bibinfo {author} {\bibfnamefont {E.~B.}\ \bibnamefont {Flagg}},
	  \bibinfo {author} {\bibfnamefont {P.}~\bibnamefont {Bianucci}}, \bibinfo
	  {author} {\bibfnamefont {X.~Y.}\ \bibnamefont {Wang}}, \bibinfo {author}
	  {\bibfnamefont {D.~G.}\ \bibnamefont {Deppe}}, \bibinfo {author}
	  {\bibfnamefont {W.}~\bibnamefont {Ma}}, \bibinfo {author} {\bibfnamefont
	  {J.}~\bibnamefont {Zhang}}, \bibinfo {author} {\bibfnamefont {G.~J.}\
	  \bibnamefont {Salamo}}, \bibinfo {author} {\bibfnamefont {M.}~\bibnamefont
	  {Xiao}},\ and\ \bibinfo {author} {\bibfnamefont {C.~K.}\ \bibnamefont
	  {Shih}},\ }\bibfield  {title} {\bibinfo {title} {Resonance fluorescence from
	  a coherently driven semiconductor quantum dot in a cavity},\ }\href
	  {https://doi.org/10.1103/PhysRevLett.99.187402} {\bibfield  {journal}
	  {\bibinfo  {journal} {Phys. Rev. Lett.}\ }\textbf {\bibinfo {volume} {99}},\
	  \bibinfo {eid} {187402} (\bibinfo {year} {2007})}\BibitemShut {NoStop}%
	\bibitem [{\citenamefont {Ates}\ \emph {et~al.}(2009)\citenamefont {Ates},
	  \citenamefont {Ulrich}, \citenamefont {Reitzenstein}, \citenamefont
	  {L\"offler}, \citenamefont {Forchel},\ and\ \citenamefont
	  {Michler}}]{AtesPRL09}%
	  \BibitemOpen
	  \bibfield  {author} {\bibinfo {author} {\bibfnamefont {S.}~\bibnamefont
	  {Ates}}, \bibinfo {author} {\bibfnamefont {S.~M.}\ \bibnamefont {Ulrich}},
	  \bibinfo {author} {\bibfnamefont {S.}~\bibnamefont {Reitzenstein}}, \bibinfo
	  {author} {\bibfnamefont {A.}~\bibnamefont {L\"offler}}, \bibinfo {author}
	  {\bibfnamefont {A.}~\bibnamefont {Forchel}},\ and\ \bibinfo {author}
	  {\bibfnamefont {P.}~\bibnamefont {Michler}},\ }\bibfield  {title} {\bibinfo
	  {title} {Post-selected indistinguishable photons from the resonance
	  fluorescence of a single quantum dot in a microcavity},\ }\href
	  {https://doi.org/10.1103/PhysRevLett.103.167402} {\bibfield  {journal}
	  {\bibinfo  {journal} {Phys. Rev. Lett.}\ }\textbf {\bibinfo {volume} {103}},\
	  \bibinfo {pages} {167402} (\bibinfo {year} {2009})}\BibitemShut {NoStop}%
	\bibitem [{\citenamefont {Vamivakas}\ \emph {et~al.}(2009)\citenamefont
	  {Vamivakas}, \citenamefont {Zhao}, \citenamefont {Lu},\ and\ \citenamefont
	  {Atatüre}}]{VamivakasNP09}%
	  \BibitemOpen
	  \bibfield  {author} {\bibinfo {author} {\bibfnamefont {A.~N.}\ \bibnamefont
	  {Vamivakas}}, \bibinfo {author} {\bibfnamefont {Y.}~\bibnamefont {Zhao}},
	  \bibinfo {author} {\bibfnamefont {C.-Y.}\ \bibnamefont {Lu}},\ and\ \bibinfo
	  {author} {\bibfnamefont {M.}~\bibnamefont {Atatüre}},\ }\bibfield  {title}
	  {\bibinfo {title} {Spin-resolved quantum-dot resonance fluorescence},\ }\href
	  {https://doi.org/10.1038/nphys1459} {\bibfield  {journal} {\bibinfo
	  {journal} {Nat. Phys.}\ }\textbf {\bibinfo {volume} {5}},\ \bibinfo {pages}
	  {925} (\bibinfo {year} {2009})}\BibitemShut {NoStop}%
	\bibitem [{\citenamefont {Flagg}\ \emph {et~al.}(2009)\citenamefont {Flagg},
	  \citenamefont {Muller}, \citenamefont {Robertson}, \citenamefont {Founta},
	  \citenamefont {Deppe}, \citenamefont {Xiao}, \citenamefont {Ma},
	  \citenamefont {Salamo},\ and\ \citenamefont {Shih}}]{FlaggNP09}%
	  \BibitemOpen
	  \bibfield  {author} {\bibinfo {author} {\bibfnamefont {E.~B.}\ \bibnamefont
	  {Flagg}}, \bibinfo {author} {\bibfnamefont {A.}~\bibnamefont {Muller}},
	  \bibinfo {author} {\bibfnamefont {J.~W.}\ \bibnamefont {Robertson}}, \bibinfo
	  {author} {\bibfnamefont {S.}~\bibnamefont {Founta}}, \bibinfo {author}
	  {\bibfnamefont {D.~G.}\ \bibnamefont {Deppe}}, \bibinfo {author}
	  {\bibfnamefont {M.}~\bibnamefont {Xiao}}, \bibinfo {author} {\bibfnamefont
	  {W.}~\bibnamefont {Ma}}, \bibinfo {author} {\bibfnamefont {G.~J.}\
	  \bibnamefont {Salamo}},\ and\ \bibinfo {author} {\bibfnamefont {C.~K.}\
	  \bibnamefont {Shih}},\ }\bibfield  {title} {\bibinfo {title} {Resonantly
	  driven coherent oscillations in a solid-state quantum emitter},\ }\href
	  {https://doi.org/10.1038/nphys1184} {\bibfield  {journal} {\bibinfo
	  {journal} {Nat. Phys.}\ }\textbf {\bibinfo {volume} {5}},\ \bibinfo {pages}
	  {203} (\bibinfo {year} {2009})}\BibitemShut {NoStop}%
	\bibitem [{\citenamefont {Astafiev}\ \emph {et~al.}(2010)\citenamefont
	  {Astafiev}, \citenamefont {Zagoskin}, \citenamefont {Abdumalikov},
	  \citenamefont {Pashkin}, \citenamefont {Yamamoto}, \citenamefont {Inomata},
	  \citenamefont {Nakamura},\ and\ \citenamefont {Tsai}}]{AstafievS10}%
	  \BibitemOpen
	  \bibfield  {author} {\bibinfo {author} {\bibfnamefont {O.}~\bibnamefont
	  {Astafiev}}, \bibinfo {author} {\bibfnamefont {A.~M.}\ \bibnamefont
	  {Zagoskin}}, \bibinfo {author} {\bibfnamefont {A.~A.}\ \bibnamefont
	  {Abdumalikov}}, \bibinfo {author} {\bibfnamefont {Y.~A.}\ \bibnamefont
	  {Pashkin}}, \bibinfo {author} {\bibfnamefont {T.}~\bibnamefont {Yamamoto}},
	  \bibinfo {author} {\bibfnamefont {K.}~\bibnamefont {Inomata}}, \bibinfo
	  {author} {\bibfnamefont {Y.}~\bibnamefont {Nakamura}},\ and\ \bibinfo
	  {author} {\bibfnamefont {J.~S.}\ \bibnamefont {Tsai}},\ }\bibfield  {title}
	  {\bibinfo {title} {Resonance fluorescence of a single artificial atom},\
	  }\href {https://doi.org/10.1126/science.1181918} {\bibfield  {journal}
	  {\bibinfo  {journal} {Science}\ }\textbf {\bibinfo {volume} {327}},\ \bibinfo
	  {pages} {840} (\bibinfo {year} {2010})}\BibitemShut {NoStop}%
	\bibitem [{\citenamefont {Laussy}\ \emph {et~al.}(2006)\citenamefont {Laussy},
	  \citenamefont {Glazov}, \citenamefont {Kavokin}, \citenamefont {Whittaker},\
	  and\ \citenamefont {Malpuech}}]{LaussyPRB06}%
	  \BibitemOpen
	  \bibfield  {author} {\bibinfo {author} {\bibfnamefont {F.~P.}\ \bibnamefont
	  {Laussy}}, \bibinfo {author} {\bibfnamefont {M.~M.}\ \bibnamefont {Glazov}},
	  \bibinfo {author} {\bibfnamefont {A.}~\bibnamefont {Kavokin}}, \bibinfo
	  {author} {\bibfnamefont {D.~M.}\ \bibnamefont {Whittaker}},\ and\ \bibinfo
	  {author} {\bibfnamefont {G.}~\bibnamefont {Malpuech}},\ }\bibfield  {title}
	  {\bibinfo {title} {Statistics of excitons in quantum dots and their effect on
	  the optical emission spectra of microcavities},\ }\href
	  {https://doi.org/10.1103/PhysRevB.73.115343} {\bibfield  {journal} {\bibinfo
	  {journal} {Phys. Rev. B}\ }\textbf {\bibinfo {volume} {73}},\ \bibinfo
	  {pages} {115343} (\bibinfo {year} {2006})}\BibitemShut {NoStop}%
	\bibitem [{\citenamefont {del Valle}\ \emph {et~al.}(2009)\citenamefont {del
	  Valle}, \citenamefont {Laussy},\ and\ \citenamefont {Tejedor}}]{VallePRB09}%
	  \BibitemOpen
	  \bibfield  {author} {\bibinfo {author} {\bibfnamefont {E.}~\bibnamefont {del
	  Valle}}, \bibinfo {author} {\bibfnamefont {F.~P.}\ \bibnamefont {Laussy}},\
	  and\ \bibinfo {author} {\bibfnamefont {C.}~\bibnamefont {Tejedor}},\
	  }\bibfield  {title} {\bibinfo {title} {Luminescence spectra of quantum dots
	  in microcavities. ii. fermions},\ }\href
	  {https://doi.org/10.1103/PhysRevB.79.235326} {\bibfield  {journal} {\bibinfo
	  {journal} {Phys. Rev. B}\ }\textbf {\bibinfo {volume} {79}},\ \bibinfo
	  {pages} {235326} (\bibinfo {year} {2009})}\BibitemShut {NoStop}%
	\bibitem [{\citenamefont {del Valle}\ and\ \citenamefont
	  {Laussy}(2010)}]{VallePRL10}%
	  \BibitemOpen
	  \bibfield  {author} {\bibinfo {author} {\bibfnamefont {E.}~\bibnamefont {del
	  Valle}}\ and\ \bibinfo {author} {\bibfnamefont {F.~P.}\ \bibnamefont
	  {Laussy}},\ }\bibfield  {title} {\bibinfo {title} {Mollow triplet under
	  incoherent pumping},\ }\href {https://doi.org/10.1103/PhysRevLett.105.233601}
	  {\bibfield  {journal} {\bibinfo  {journal} {Phys. Rev. Lett.}\ }\textbf
	  {\bibinfo {volume} {105}},\ \bibinfo {pages} {233601} (\bibinfo {year}
	  {2010})}\BibitemShut {NoStop}%
	\bibitem [{\citenamefont {Langbein}\ and\ \citenamefont
	  {Patton}(2006)}]{LangbeinOL06}%
	  \BibitemOpen
	  \bibfield  {author} {\bibinfo {author} {\bibfnamefont {W.}~\bibnamefont
	  {Langbein}}\ and\ \bibinfo {author} {\bibfnamefont {B.}~\bibnamefont
	  {Patton}},\ }\bibfield  {title} {\bibinfo {title} {Heterodyne spectral
	  interferometry for multidimensional nonlinear spectroscopy of individual
	  quantum systems},\ }\href@noop {} {\bibfield  {journal} {\bibinfo  {journal}
	  {Opt. Lett.}\ }\textbf {\bibinfo {volume} {31}},\ \bibinfo {pages} {1151}
	  (\bibinfo {year} {2006})}\BibitemShut {NoStop}%
	\bibitem [{\citenamefont {Kasprzak}\ \emph {et~al.}(2013)\citenamefont
	  {Kasprzak}, \citenamefont {Sivalertporn}, \citenamefont {Albert},
	  \citenamefont {Schneider}, \citenamefont {H{\"o}fling}, \citenamefont {Kamp},
	  \citenamefont {Forchel}, \citenamefont {Reitzenstein}, \citenamefont
	  {Muljarov},\ and\ \citenamefont {Langbein}}]{KasprzakNJP13}%
	  \BibitemOpen
	  \bibfield  {author} {\bibinfo {author} {\bibfnamefont {J.}~\bibnamefont
	  {Kasprzak}}, \bibinfo {author} {\bibfnamefont {K.}~\bibnamefont
	  {Sivalertporn}}, \bibinfo {author} {\bibfnamefont {F.}~\bibnamefont
	  {Albert}}, \bibinfo {author} {\bibfnamefont {C.}~\bibnamefont {Schneider}},
	  \bibinfo {author} {\bibfnamefont {S.}~\bibnamefont {H{\"o}fling}}, \bibinfo
	  {author} {\bibfnamefont {M.}~\bibnamefont {Kamp}}, \bibinfo {author}
	  {\bibfnamefont {A.}~\bibnamefont {Forchel}}, \bibinfo {author} {\bibfnamefont
	  {S.}~\bibnamefont {Reitzenstein}}, \bibinfo {author} {\bibfnamefont {E.~A.}\
	  \bibnamefont {Muljarov}},\ and\ \bibinfo {author} {\bibfnamefont
	  {W.}~\bibnamefont {Langbein}},\ }\bibfield  {title} {\bibinfo {title}
	  {Coherence dynamics and quantum-to-classical crossover in an exciton-cavity
	  system in the quantum strong coupling regime},\ }\href@noop {} {\bibfield
	  {journal} {\bibinfo  {journal} {New J. Phys.}\ }\textbf {\bibinfo {volume}
	  {15}},\ \bibinfo {pages} {045013} (\bibinfo {year} {2013})}\BibitemShut
	  {NoStop}%
	\bibitem [{\citenamefont {Albert}\ \emph {et~al.}(2013)\citenamefont {Albert},
	  \citenamefont {Sivalertporn}, \citenamefont {Kasprzak}, \citenamefont
	  {Strau\ss}, \citenamefont {Schneider}, \citenamefont {H{\"o}fling},
	  \citenamefont {Kamp}, \citenamefont {Forchel}, \citenamefont {Reitzenstein},
	  \citenamefont {Muljarov},\ and\ \citenamefont {Langbein}}]{AlbertNC13}%
	  \BibitemOpen
	  \bibfield  {author} {\bibinfo {author} {\bibfnamefont {F.}~\bibnamefont
	  {Albert}}, \bibinfo {author} {\bibfnamefont {K.}~\bibnamefont
	  {Sivalertporn}}, \bibinfo {author} {\bibfnamefont {J.}~\bibnamefont
	  {Kasprzak}}, \bibinfo {author} {\bibfnamefont {M.}~\bibnamefont {Strau\ss}},
	  \bibinfo {author} {\bibfnamefont {C.}~\bibnamefont {Schneider}}, \bibinfo
	  {author} {\bibfnamefont {S.}~\bibnamefont {H{\"o}fling}}, \bibinfo {author}
	  {\bibfnamefont {M.}~\bibnamefont {Kamp}}, \bibinfo {author} {\bibfnamefont
	  {A.}~\bibnamefont {Forchel}}, \bibinfo {author} {\bibfnamefont
	  {S.}~\bibnamefont {Reitzenstein}}, \bibinfo {author} {\bibfnamefont {E.~A.}\
	  \bibnamefont {Muljarov}},\ and\ \bibinfo {author} {\bibfnamefont
	  {W.}~\bibnamefont {Langbein}},\ }\bibfield  {title} {\bibinfo {title}
	  {Microcavity controlled coupling of excitonic qubits},\ }\href@noop {}
	  {\bibfield  {journal} {\bibinfo  {journal} {Nat. Comm.}\ }\textbf {\bibinfo
	  {volume} {4}},\ \bibinfo {pages} {1747} (\bibinfo {year} {2013})}\BibitemShut
	  {NoStop}%
	\bibitem [{\citenamefont {Groll}\ \emph {et~al.}(2020)\citenamefont {Groll},
	  \citenamefont {Wigger}, \citenamefont {J\"urgens}, \citenamefont {Hahn},
	  \citenamefont {Schneider}, \citenamefont {Kamp}, \citenamefont {H\"ofling},
	  \citenamefont {Kasprzak},\ and\ \citenamefont {Kuhn}}]{GrollPRB20}%
	  \BibitemOpen
	  \bibfield  {author} {\bibinfo {author} {\bibfnamefont {D.}~\bibnamefont
	  {Groll}}, \bibinfo {author} {\bibfnamefont {D.}~\bibnamefont {Wigger}},
	  \bibinfo {author} {\bibfnamefont {K.}~\bibnamefont {J\"urgens}}, \bibinfo
	  {author} {\bibfnamefont {T.}~\bibnamefont {Hahn}}, \bibinfo {author}
	  {\bibfnamefont {C.}~\bibnamefont {Schneider}}, \bibinfo {author}
	  {\bibfnamefont {M.}~\bibnamefont {Kamp}}, \bibinfo {author} {\bibfnamefont
	  {S.}~\bibnamefont {H\"ofling}}, \bibinfo {author} {\bibfnamefont
	  {J.}~\bibnamefont {Kasprzak}},\ and\ \bibinfo {author} {\bibfnamefont
	  {T.}~\bibnamefont {Kuhn}},\ }\bibfield  {title} {\bibinfo {title} {Four-wave
	  mixing dynamics of a strongly coupled quantum-dot--microcavity system driven
	  by up to 20 photons},\ }\href {https://doi.org/10.1103/PhysRevB.101.245301}
	  {\bibfield  {journal} {\bibinfo  {journal} {Phys. Rev. B}\ }\textbf {\bibinfo
	  {volume} {101}},\ \bibinfo {pages} {245301} (\bibinfo {year}
	  {2020})}\BibitemShut {NoStop}%
	\bibitem [{\citenamefont {Allcock}\ \emph {et~al.}(2021)\citenamefont
	  {Allcock}, \citenamefont {Langbein},\ and\ \citenamefont {Muljarov}}]{SI}%
	  \BibitemOpen
	  \bibfield  {author} {\bibinfo {author} {\bibfnamefont {T.}~\bibnamefont
	  {Allcock}}, \bibinfo {author} {\bibfnamefont {W.}~\bibnamefont {Langbein}},\
	  and\ \bibinfo {author} {\bibfnamefont {E.~A.}\ \bibnamefont {Muljarov}},\
	  }\href@noop {} {\emph {\bibinfo {title} {Supplementary Information}}}
	  (\bibinfo {year} {2021})\BibitemShut {NoStop}%
	\bibitem [{\citenamefont {Wiseman}\ and\ \citenamefont
	  {Milburn}(2009)}]{WisemanBook09}%
	  \BibitemOpen
	  \bibfield  {author} {\bibinfo {author} {\bibfnamefont {H.~M.}\ \bibnamefont
	  {Wiseman}}\ and\ \bibinfo {author} {\bibfnamefont {G.~J.}\ \bibnamefont
	  {Milburn}},\ }\href {https://doi.org/10.1017/cbo9780511813948} {\emph
	  {\bibinfo {title} {Quantum Measurement and Control}}}\ (\bibinfo  {publisher}
	  {Cambridge University Press},\ \bibinfo {year} {2009})\BibitemShut {NoStop}%
	\bibitem [{\citenamefont {Glauber}(1963)}]{GlauberPR63}%
	  \BibitemOpen
	  \bibfield  {author} {\bibinfo {author} {\bibfnamefont {R.~J.}\ \bibnamefont
	  {Glauber}},\ }\bibfield  {title} {\bibinfo {title} {Coherent and incoherent
	  states of the radiation field},\ }\href
	  {https://doi.org/10.1103/physrev.131.2766} {\bibfield  {journal} {\bibinfo
	  {journal} {Phys. Rev.}\ }\textbf {\bibinfo {volume} {131}},\ \bibinfo {pages}
	  {2766} (\bibinfo {year} {1963})}\BibitemShut {NoStop}%
	\bibitem [{\citenamefont {Morreau}\ and\ \citenamefont
	  {Muljarov}(2019)}]{MorreauPRB19}%
	  \BibitemOpen
	  \bibfield  {author} {\bibinfo {author} {\bibfnamefont {A.}~\bibnamefont
	  {Morreau}}\ and\ \bibinfo {author} {\bibfnamefont {E.~A.}\ \bibnamefont
	  {Muljarov}},\ }\bibfield  {title} {\bibinfo {title} {Phonon-induced dephasing
	  in quantum-dot--cavity qed},\ }\href
	  {https://doi.org/10.1103/PhysRevB.100.115309} {\bibfield  {journal} {\bibinfo
	   {journal} {Phys. Rev. B}\ }\textbf {\bibinfo {volume} {100}},\ \bibinfo
	  {pages} {115309} (\bibinfo {year} {2019})}\BibitemShut {NoStop}%
	\end{thebibliography}

%

\end{document}


\title{Supplementary information to\\ ``Quantum Mollow quadruplet in non-linear cavity-QED''}
\date{\today}	
\pacs{} 			
\keywords{} 	

\author{Thomas Allcock}
\author{Wolfgang Langbein}
\author{Egor A. Muljarov}
\affiliation{School\,of\,Physics\,and\,Astronomy,\,Cardiff\,University,\,The Parade,\,Cardiff\,CF24\,3AA,\,United\,Kingdom}

\maketitle 

\section{The effect of a pulsed optical excitation on the density matrix of a quantum dot-cavity system: Analytic solution}\label{sec:excitation}

Let us consider an excitation of a quantum dot (QD)-cavity system
by a sequence of ultrashort optical pulses or by an extended finite wave packet of light. The master equation  describing the time evolution of the density matrix (DM) is given by Eq.\,(1),
\be i\dot{\rho}(t)= \left[\hL+\hLc(t)\right]\rho(t)\,,
\label{eqn:ME}\ee
with the time-independent Lindblad operator $\hL$ defined by Eq.\,(2) and
a time-dependent operator $\hLc(t)$ defined as
\be \hLc(t)\rho(t)= [ V(t),\rho(t)]\,,
\label{eqn:L}\ee
where $V(t)$ is given by Eq.\,(4) as
\be V(t)=-\bmu\cdot \Ec(t) \ad - \bmu^\ast\!\!\cdot \Ec^\ast(t) a\,.
\label{equ:V} \ee
The formal solution of \Eq{eqn:ME} can be written as
\be \rho(t)=T\exp\left\{ -i\int_{t_0}^t\left[\hL+\hLc(\tau)\right]d\tau\right\}  \rho(t_0)
= T\prod_{j=0}^{J-1} Q_j \rho(t_0)\,,
\label{equ:split} \ee
where $T$ is the standard time-ordering operator. In the second part of \Eq{equ:split}, the full time evolution of the DM, between $t_0$ and $t$, is split into a time-ordered product of $J$ operators
$$ Q_j=T\exp\left\{ -i\int_{t_j}^{t_{j+1}}\left[\hL+\hLc(\tau)\right]d\tau\right\}\,,$$
obtained by dividing the full time interval (from $t_0$ to $t$) into $J$ pieces, which are not necessarily equal:
$$ t_0<t_1<\dots <t_j<t_{j+1}<\dots < t_J=t\,.$$
Assuming that the time steps $\Delta t_j=t_{j+1}-t_j$ are small enough, these operators may be approximated as
\be Q_j\approx T\exp\left\{ -i\int_{t_j}^{t_{j+1}}\hL d\tau \right\} T\exp\left\{-i\int_{t_j}^{t_{j+1}} \hLc(\tau) d\tau \right\}\,,
\label{eqn:Q}\ee
with an error scaling as $(\Delta t_j)^2$~\cite{Suzuki1976}. While the first operator in \Eq{eqn:Q} can be written as $e^{-i\hL\Delta t_j}$ due to the time-independent $\hL$, the second operator requires integration of the time-dependent field $\Ec(t)$ exciting the system. Using its definition \Eq{eqn:L}, the action of the second operator in \Eq{eqn:Q} on the DM can be evaluated as
\be T\exp\left\{ -i\int_{t_j}^{t_{j+1}}\hLc(\tau)d\tau\right\}\rho(t_j)= U_j \rho(t_j) U^\dagger_j\,,
\label{equ:pulse}\ee
where
$$ U_j=T\exp\left\{ -i\int_{t_j}^{t_{j+1}}V(\tau)d\tau\right\} $$
is the standard evolution operator due to a time-dependent interaction $V(t)$.
Using the explicit form of $V(t)$, given by \Eq{equ:V}, from which it follows in particular that the commutator
$[V(t),V(t')]=0$ vanishes for any $t$ and $t'$, we obtain
\be U_j=e^{i(E_j\ad +E_j^\ast a)} \,,
\label{equ:U}\ee
where
\be E_j= \int_{t_j}^{t_{j+1}} \bmu\cdot \Ec(\tau) d\tau\,.
\label{equ:Ej}\ee
Combining the results, we find
\be Q_j\rho(t_j)\approx e^{-i\hL\Delta t_j}U_j \rho(t_j) U^\dagger_j\,,
\label{equ:Qj}\ee
where $U_j$ is given by \Eq{equ:U}.

Note that the full time evolution of the DM described by Eqs.\,(\ref{equ:split}) and (\ref{equ:U})--(\ref{equ:Qj}) becomes exact if the excitation field is represented by a sequence of $\delta$ pulses, given by Eq.\,(5):
\be \bmu\cdot \Ec(t)=\sum_j E_j \delta(t-t_j)\,.
\label{eqn:delta}\ee
In fact, \Eq{eqn:delta} is equivalent to the rectangular rule of numerical integration of a finite wave packet
$$ \int_{-\infty}^\infty \bmu\cdot \Ec(t) dt = \sum_j E_j=\sum_j \bmu\cdot \Ec(t_j) \Delta t_j\,,$$
where  $E_j$ are the pulse areas corresponding to the time intervals  $\Delta t_j$\,.

\bigskip Let us now consider the effect of a single $\delta$ pulse on the DM, which is given by \Eq{equ:pulse}\,. Dropping the index $j$ for brevity, we first transform the evolution operator
$$ U(E)=e^{i(E\ad +E^\ast a)}=e^{-|E|^2/2}e^{iE\ad}e^{iE^\ast a}\,,$$
using the fact~\cite{Suzuki1976} that $e^{A+B}=e^A e^B e^C$, if $C=-\frac{1}{2}[A,B]$ commutes with both operators $A$ and $B$, which is true in the present case. When acting on the ground state $|0\rangle$ of the optical cavity (with a single cavity mode), this operator generates a Glauber coherent states $|\alpha \rangle$ with the eigenvalue $\alpha=iE$. In fact,
$$ |\alpha \rangle= U(E)|0 \rangle =  e^{-|E|^2/2}e^{iE\ad} |0 \rangle =  e^{-|E|^2/2} \sum_{n=0}^\infty \frac{(iE)^n (a^\dagger)^n}{n!} |0 \rangle =  e^{-|E|^2/2} \sum_{n=0}^\infty \frac{(iE)^n}{\sqrt{n!}} |n \rangle\,,$$
so that
$$ a|\alpha \rangle = e^{-|E|^2/2} \sum_{n=1}^\infty \frac{(iE)^n\sqrt{n}}{\sqrt{n!}} |n-1 \rangle
 = e^{-|E|^2/2} \sum_{n=0}^\infty \frac{(iE)^{n+1}}{\sqrt{n!}} |n \rangle = iE|\alpha \rangle\,.$$
This result is useful if the system is initially in its ground state, so that the density matrix before the $\delta$ pulse is given by $|0 \rangle \langle 0|$. In general, this is not the case, and the density matrix before the pulsed excitation is given by Eq.\,(8), or \Eq{equ:rhofull} in \Sec{Sec:Lindblad} below. We therefore need to evaluate the effect of a $\delta$ pulse on an arbitrary state $|m \rangle$ of the cavity, which is given by a matrix $U_{nm}(E)$ defined by
$$ U(E)|m \rangle =\sum_{n=0}^\infty |n \rangle U_{nm}(E)\,, $$
where
\bea U_{nm}(E)= \langle n| U(E)|m \rangle &=& e^{-|E|^2/2}\sum_{k=0}^\infty \langle n| e^{iE\ad}|k \rangle \langle k| e^{iE^\ast a}|m \rangle \label{equ:Unm}\\
&=& e^{-|E|^2/2}\sum_{k=0}^l \frac{(iE)^{n-k}}{(n-k)!}\sqrt{\frac{n!}{k!}} \frac{(iE^\ast)^{m-k}}{(m-k)!} \sqrt{\frac{m!}{k!}}\,,\nonumber\eea
with $l=\min(n,m)$. Introducing the phase $\varphi$ of the excitation pulse, via $E=|E|e^{i\varphi}$, \Eq{equ:Unm} becomes
\be U_{nm}(E)=i^{n-m} e^{i\varphi(n-m)} |E|^{n-m}\sqrt{\frac{m!}{n!}}  e^{-|E|^2/2}
\sum_{k=0}^m\frac{ (-|E|^2)^{m-k} n!}{(n-k)! (m-k)! k!}
\label{equ:Unm1}\ee
for $n\geqslant m$, and
\be U_{nm}(E)=i^{m-n} e^{i\varphi(n-m)} |E|^{m-n}\sqrt{\frac{n!}{m!}}  e^{-|E|^2/2}
\sum_{k=0}^n\frac{ (-|E|^2)^{n-k} m!}{(n-k)! (m-k)! k!}
\label{equ:Unm2}\ee
for $n\leqslant m$. Comparing the series in \Eqs{equ:Unm1}{equ:Unm2} with the associated Laguerre polynomials~\cite{Gradshtein}, given by a series
$$ L^\alpha_p(x)=\sum_{j=0}^p \frac{(-x)^j (p+\alpha)!}{(p-j)!(\alpha+j)!j!} $$
for $p\geqslant0$, we find that for any values of $n$ and $m$\,,
$$ U_{nm}(E) =e^{i\varphi(n-m)} C_{nm}(|E|) $$
with $C_{nm}(|E|)$ expressed in terms of the Laguerre polynomials:
\be C_{nm}(|E|)=i^\alpha |E|^\alpha\sqrt{\frac{p!}{(p+\alpha)!}}L^\alpha_{p}(|E|^2) e^{-|E|^2/2}\,,
\label{Cnm}\ee
where $\alpha=|n-m|$ and $p=\min(n,m)$. Using the property
$$ L_m^{n-m}(x)=L_n^{m-n}(x)\frac{n!}{m!} (-x)^{m-n}\,,$$
\Eq{Cnm} can be written more explicitly as Eq.\,(11)
$$ C_{nm}(|E|)=i^{n-m} |E|^{n-m} \sqrt{\frac{m!}{n!}}L^{n-m}_{m}(|E|^2) e^{-|E|^2/2}\,. $$

Finally, applying the operators $U(E)$ and $U^\dagger(E)$, respectively,  on the left and right hand sides of the DM, in accordance with \Eq{equ:pulse}, we arrive at Eq\,(10) of the main text.

\section{Analytic Diagonalization of the Lindblad Operator for the QD-cavity system}
\label{Sec:Lindblad}

The evolution of the QD-cavity system between and after excitation pulses is described by the master equation
\be i\dot{\rho}= \hL\rho\,.
\label{equ:ME}\ee
The action of the Lindblad operator on the DM can be conveniently expressed as
\be \hat{L}\rho=H\rho-\rho H^\ast +2i\gamma_xd\rho \dd+2i\gamma_ca\rho \ad
\label{equ:Lindblad2}\ee
where the Jaynes-Cummings (JC) Hamiltonian and its complex conjugate are, respectively, given by
\be \begin{split}
H&=\omega_X\dd d+\omega_C\ad a+g(\ad d+\dd a)\,,\\
H^\ast&=\omega^\ast_X\dd d+\omega^\ast_C\ad a+g(\ad d+\dd a)\,.
\end{split} \nonumber \ee
Here, $\omega_X=\Omega_X-i\gamma_X$ and $\omega_C=\Omega_C-i\gamma_C$ are the complex frequencies of the QD exciton and the cavity mode, respectively. Note that \Eq{equ:Lindblad2} is equivalent to Eq.\,(2).

In the basis of Fock states of the QD-cavity system, the full DM is given by Eq.\,(8):
\be \rho=\sum_{\nu\nu' n n'}\rho^{\nu\nu'}_{nn'}|\nu,n\rangle \langle\nu',n'|\,,
\label{equ:rhofull}\ee
where $\nu$ and $\nu'$ refer to the exciton and $n$ and $n'$ to the cavity indices. Let us consider a group of elements of the DM describing the coherence between rungs $N+\DN$ and $N$ of the JC ladder, where $\DN$ is the separation between rungs. These are the elements with
$$ \nu+n=N+\DN\equiv \ND\ \ \ {\rm and} \ \ \  \nu'+n'=N $$
in \Eq{equ:rhofull}. The corresponding part of the DM is given by
\bea
\rho(\ND;N)&=&\phantom{+}\rho_1^{(N)} |1,\ND-1 \rangle \langle 1, N-1| \nonumber\\
&&+\rho_2^{(N)} |1,\ND-1 \rangle \langle 0, N| \nonumber\\
&&+\rho_3^{(N)} |0,\ND \rangle \langle 1, N-1| \nonumber\\
&&+\rho_4^{(N)} |0,\ND \rangle \langle 0, N|\,,
\label{equ:rhoN}
\eea
where for convenience we have introduced new notations for the DM elements: $\rho_1^{(N)}=\rho^{11}_{\ND-1,N-1}$, $\rho_2^{(N)}=\rho^{10}_{\ND-1,N}$, $\rho_3^{(N)}=\rho^{01}_{\ND,N-1}$, and $\rho_4^{(N)}=\rho^{00}_{\ND,N}$. For the elements involving the ground state ($N=0$ or $\ND=0$), the DM reduces to only two elements:
\be \rho(\DN;0)=\rho_1^{(0)} |1,\DN-1 \rangle \langle 0, 0| +\rho_2^{(0)} |0,\DN \rangle \langle 0, 0|
\label{equ:rho0}\ee
with $\DN>0$ taken for definiteness. We use these new notations, in order to form a vector $\vec{\rho}$ consisting of the elements of the DM which appear in \Eqs{equ:rhoN}{equ:rho0}, for all rungs and a fixed $\DN$:
\be
\vec{\rho}=
\begin{bmatrix}
\vec{\rho}^{\:(0)}\\
\vec{\rho}^{\:(1)}\\
\vec{\rho}^{\:(2)}\\
\vdots
\end{bmatrix}
,
\ \ \ {\rm where}\ \ \
\vec{\rho}^{\:(0)}=
\begin{bmatrix}
\rho_1^{(0)}\\
\rho_2^{(0)}
\end{bmatrix}
, \ \mbox{and} \ \
\vec{\rho}^{\:(N)}=
\begin{bmatrix}
\rho_1^{(N)}\\
\rho_2^{(N)}\\
\rho_3^{(N)}\\
\rho_4^{(N)}
\end{bmatrix}
\ \mbox{for} \ N>0\, .
\label{equ:vecs}\ee

The master equation (\ref{equ:ME}) then takes the matrix form $i\dot{\vec{\rho}}= \hL\vec{\rho}$, where  $\hat{L}$ is a matrix consisting of the blocks
\begin{align}\label{equ:L}
\hat{L}&= \begin{bmatrix}
L_0 & M_{01} & \zero & \hdots \\
\zero & L_1 & M_{12} & \hdots\\
\zero & \zero & L_2 & \hdots\\
\vdots & \vdots & \vdots & \ddots \end{bmatrix}
\,,\end{align}
where $\zero$ denotes blocks of zero elements. It is convenient at this point to introduce a $2\times2$ matrix of the $N$-th rung of the JC Hamiltonian, as in Eq.\,(13),
\begin{align}\label{equ:HN}
H_N&= \begin{bmatrix}
\omega_X+(N-1)\omega_C & \sqrt{N}g \\
\sqrt{N}g & N\omega_C \end{bmatrix}
\,.\end{align}
The diagonal blocks of $\hL$
are produced by the first two terms of the Lindblad operator \Eq{equ:Lindblad2}
and are given by
\bea
\label{equ:L0}
L_0&=& H_{\DN}\,,\\
L_N&=& G_{\ND}- F_N^\ast\ \ \ \mbox{for} \ N>0\,,
\label{equ:LN}
\eea
where
\be  G_{N}= \begin{bmatrix}
\omega_X+(N-1)\omega_C & 0 & \sqrt{N}g & 0 \\
0 & \omega_X+(N-1)\omega_C & 0 & \sqrt{N}g \\
\sqrt{N}g & 0 & N\omega_C & 0 \\
0 & \sqrt{N}g & 0 & N\omega_C \end{bmatrix}
\label{equ:G} \ee
consists of the four elements of $H_N$, contributing twice, one time distributed over the first and third rows and columns of $G_{N}$, the other over the second and fourth rows and columns. The other matrix,
$F_{N}^\ast$, is the complex conjugate of
\be F_N=\begin{bmatrix}
H_{N}& \zero\\
\zero &H_{N} \end{bmatrix}\,,
\label{equ:F}\ee
which in turn consists of $2\times 2$ diagonal blocks $H_{N}$, given by \Eq{equ:HN}, and $2\times 2$ zero matrices $\zero$ occupying its off-diagonal blocks. The off-diagonal blocks of $\hL$ are due to the last two terms of the Lindblad operator \Eq{equ:Lindblad2} and take the form
\bea M_{01}&= &\begin{bmatrix}
0 & 2i\gamma_C \sqrt{\DN}& 0 & 0 \\
2i\gamma_X & 0 & 0 & 2i\gamma_C \sqrt{\DN+1} \end{bmatrix} \,,
\label{equ:Mnn1}
\\
M_{N,N+1}&= &\begin{bmatrix}
2i\gamma_C \sqrt{\ND N}& 0 & 0 &0 \\
0& 2i\gamma_C \sqrt{\ND (N+1)} & 0 &0 \\
0& 0 & 2i\gamma_C \sqrt{(\ND+1)N} & 0 \\
2i\gamma_X & 0 & 0 & 2i\gamma_C \sqrt{(\ND+1)(N+1)}
\nonumber
\end{bmatrix}\,.
\eea

An analytic diagonalization of the matrix $\hL$ presented below is based on the eigenvalues and eigenvectors of the Hamiltonian matrix $H_N$ of the $N$-th rung of the JC ladder. This $2\times2$ matrix, playing the role of a building block for the diagonalization of $\hL$, is diagonalized as
\begin{equation}
\label{equ:evN}
H_N Y_N=Y_N\Lambda_N\,,
\end{equation}
where  the transformation matrix $Y_N$ and the eigenvalue matrix $\Lambda_N$ are given by
\begin{align}
Y_N&= \begin{bmatrix}
\alpha_N & \beta_N \\
-\beta_N & \alpha_N \end{bmatrix}
\quad\text{and}\quad\Lambda_N=
\begin{bmatrix}
\lambda^-_N & 0 \\
0 & \lambda^+_N \end{bmatrix},
\end{align}
respectively, with
\begin{align}
\label{equ:lam}
\lambda^\pm_N&=N\omega_C+\delta/2\pm\Delta_N \,,\\
\alpha_N&=\frac{\Delta_N-\delta/2}{D^-_N}=\frac{\sqrt{N}g}{D^+_N} \,,\ \ \ \ \beta_N=\frac{\sqrt{N}g}{D^-_N}=\frac{\Delta_N+\delta/2}{D^+_N}\,,\nonumber\\
\Delta_N&=\sqrt{(\delta/2)^2+Ng^2}\,,\ \ \ \ \ \ \ \ \ \, D^\pm_N=\sqrt{(\Delta_N\pm\delta/2)^2+Ng^2}\,,
\label{equ:Dgam}
\end{align}
where $\delta=\omega_X-\omega_C$ is the complex frequency detuning, and constants $D^\pm_N$ are normalizing the eigenvectors of $H_N$ in such a way that
\be
\label{equ:norm}
\alpha_N^2+\beta_N^2=1\,.
\ee
Note that $\Delta_N$ and $D^\pm_N$ are also complex-valued and expressed by \Eq{equ:Dgam} in terms of square roots, each having two values, or two branches. The choice of the sign (i.e. the square root branch) can be arbitrary in each case. However, this choice has to be used consistently in all the equations containing $\Delta_N$ and $D^\pm_N$, with the sign of $\Delta_N$ being independent from those of $D^\pm_N$, while the signs of $D^+_N$ and $D^-_N$ are linked together (however, only one of these two constants, either $D^+_N$ or $D^-_N$, is required in calculations). Owing to the normalization \Eq{equ:norm}, the transformation matrix $Y_N$ is orthogonal, i.e.
$$ Y^{-1}_N=Y^{\rm T}_N\,,$$
where $Y^{\rm T}_N$ is the transpose of $Y_N$, so that \Eq{equ:evN} can also be written as $Y^{\rm T}_N H_N Y_N=\Lambda_N$.

The diagonal block $L_0$ of the Lindblad matrix $\hL$, which is given by \Eq{equ:L0}, is identical to $H_{\DN}$ and is thus diagonalized by $Y_{\DN}$:
\be Y_{\DN}^{\rm T}L_0 Y_{\DN}=\Omega_0=\begin{bmatrix}
\lambda^-_{\DN} & 0 \\
0 & \lambda^+_{\DN} \end{bmatrix}\,.
\label{equ:Omega0}\ee
To diagonalize any other diagonal block $L_N$ with $N>0$, which is given by \Eq{equ:LN}, we introduce two $4\times4$ matrices
\be A_N=\begin{bmatrix}
\alpha_N & 0 & \beta_N & 0 \\
0 & \alpha_N & 0 & \beta_N \\
-\beta_N & 0 & \alpha_N & 0 \\
0 & -\beta_N & 0 & \alpha_N \end{bmatrix}
\,,\ \ \ B_N= \begin{bmatrix}
\alpha_N & \beta_N & 0 & 0 \\
-\beta_N & \alpha_N & 0 & 0 \\
0 & 0 & \alpha_N & \beta_N \\
0 & 0 & -\beta_N & \alpha_N \end{bmatrix}\,.
\label{equ:AB}\ee
Clearly, matrix $B_N$ is block-diagonal, consisting of two identical blocks of $Y_N$. Matrix $A_N$  can be obtained from $B_N$ by simultaneous swapping the 2nd and 3rd rows and columns. Note that exactly the same link exists between matrices $G_N$ and $F_N$ contributing to $L_N$ and consisting of zero elements and the elements of $H_N$, see \Eqs{equ:G}{equ:F}. Consequently, matrices $A_N$ and $B_N$ are orthogonal, i.e. $A_N^{-1}=A_N^{\rm T}$ and $B_N^{-1}=B_N^{\rm T}$, and diagonalize matrices $G_N$ and $F_N$, respectively. At the same time, owing to the structure of these matrices, the following commutation relations hold:
\be [A_{\ND},B^\ast_N]=[A_{\ND},F^\ast_N]=[G_{\ND},B^\ast_N]=0\,.
\label{equ:com}\ee
Owing to the above properties, the matrix
\be S_N=A_{\ND} B^\ast_N
\label{equ:SAB}\ee
is also orthogonal, $S_N^{-1}=S_N^{\rm T}$, and diagonalizes $L_N$, a diagonal block of the Lindblad matrix $\hL$:
\be S_N^{\rm T}L_N S_N=\Omega_N\,.
\label{equ:SLS}\ee
In fact, matrix $B^\ast_N$ diagonalized $F^\ast_N$ while keeping $G_{\ND}$ untouched, due to \Eq{equ:com}. Similarly, $A_\ND$ diagonalizes $G_{\ND}$ while keeping $F^\ast_N$ untouched. The diagonal matrix $\Omega_N$ of the eigenvalues of $L_N$ then takes the form:
\be \Omega_N=\begin{bmatrix}
\lambda^-_{\ND}-{\lambda^-_N}^\ast & 0 & 0 & 0 \\
0 & \lambda^-_{\ND}-{\lambda^+_N}^\ast & 0 & 0 \\
0 & 0 & \lambda^+_{\ND}-{\lambda^-_N}^\ast & 0 \\
0 & 0 & 0 & \lambda^+_{\ND}-{\lambda^+_N}^\ast \end{bmatrix}\,,
\label{equ:OmegaN}\ee
where ${\lambda^\pm_N}$ are given by \Eq{equ:lam}. The eigenvalues $\Omega_N$ are considered in more detail in \Sec{Sec:frequencies}, where limiting cases of large and zero detuning, and of large rung number $N$ are analyzed.

Let us now diagonalize the full matrix $\hat{L}$, finding matrices $\hat{U}$ and $\hat{V}$ of right and left eigenvectors, respectively:
\be \hat{L}\hat{U}=\hat{U}\hat{\Omega}\,, \qquad \hat{V}\hat{L}=\hat{\Omega}\hat{V}\,.
\label{equ:ev}\ee
Due to the block form of $\hat{L}$, the diagonal matrix  $\hat{\Omega}$  consists of the eigenvalue matrices $\Omega_N$ found above, and $\hat{U}$ and $\hat{V}$ are the block-triangular matrices:
\be \hat{\Omega}= \begin{bmatrix}
\Omega_0 & \zero & \zero & \hdots \\
\zero & \Omega_1 & \zero &  \hdots \\
\zero & \zero & \Omega_2 &  \hdots \\
\vdots & \vdots & \vdots &  \ddots \end{bmatrix}
\,,\ \ \
\hat{U}=\begin{bmatrix}
U_{00} & U_{01} & U_{02}  & \hdots \\
\zero & U_{11} & U_{12} & \hdots \\
\zero& \zero & U_{22}  & \hdots \\
 \vdots & \vdots & \vdots & \ddots \end{bmatrix}
 \,,\ \ \
\hat{V}=\begin{bmatrix}
V_{00} & V_{01} & V_{02}  & \hdots \\
\zero & V_{11} & V_{12} & \hdots \\
\zero& \zero & V_{22}  & \hdots \\
\vdots & \vdots & \vdots & \ddots \end{bmatrix}
\,.
\label{equ:OUV}\ee
Here, $\Omega_0$, $U_{00}$ and $V_{00}$ are $2\times 2$ blocks, $U_{0N}$ and $V_{0N}$ with $N>0$ are
$2\times 4$ blocks, and $U_{NK}$ and $V_{NK}$ with both $N,\,K>0$ are $4\times 4$ matrices. Substituting $\hat{U}$ and $\hat{V}$ into the eigenvalue equations (\ref{equ:ev}), we find series of recursive relations for all blocks $U_{NK}$ and $V_{NK}$ and explicit analytic expressions for their elements.

Let us first consider the right eigenvectors $\hat{U}$. Substituting $\hat{\Omega}$ and $\hat{U}$ from \Eq{equ:OUV} and $\hat{L}$ from \Eq{equ:L} into the first eigenvalue equation (\ref{equ:ev}), we obtain
for any fixed $N$ the matrix equation $ L_N U_{NN}=U_{NN} \Omega_N$, so that $U_{NN}=S_N$, having the explicit form given by \Eqs{equ:AB}{equ:SAB}. For any $0\leqslant K<N$, we then find a matrix equation linking $U_{KN}$ to $U_{K+1,N}$:
$$ L_K U_{KN}+M_{K,K+1}U_{K+1,N}=U_{KN}\Omega_N\,.$$
Multiplying this equation from the left with $S_K^{\rm T}$, and using $S_K^{\rm T}L_K=\Omega_K S_K^{\rm T}$, we obtain
\be \Omega_K \tilde{U}_{KN}+\tilde{M}_{K,K+1}\tilde{U}_{K+1,N}=\tilde{U}_{KN}\Omega_N\,,
\label{equ:left}\ee
where
\be \tilde{U}_{KN}=S_K^{\rm T}U_{KN}\,,\ \ \ \tilde{M}_{K,K+1}= S_K^{\rm T}M_{K,K+1} S_{K+1}\,.
\label{equ:Utilde}\ee
As $\Omega_K$ is a diagonal matrix, \Eq{equ:left} results in the following explicit form of the matrix elements of
$\tilde{U}_{KN}$:
\be (\tilde{U}_{KN})_{ij}=\frac{(\tilde{M}_{K,K+1}\tilde{U}_{K+1,N})_{ij}}{(\Omega_N)_{jj}-(\Omega_K)_{ii}}\,.
\label{equ:UKN}\ee
For each $N$, we use $\tilde{U}_{NN}=S_N^{\rm T}U_{NN} =S_N^{\rm T}S_N =\one$ (here $\one$ is the identity matrix) as a start point and calculate $\tilde{U}_{KN}$ from \Eq{equ:UKN} sequentially, for $K=N-1,N-2,...,0$. Note that the index $i$ ($j$) of the matrix elements takes the values of 1 or 2 for $K=0$ ($N=0$) and 1,\,2,\,3 or 4 for $K>0$ ($N>0$), due to the sizes of the corresponding blocks.

\begin{figure}
	\centering
	\includegraphics[height=6cm,clip]{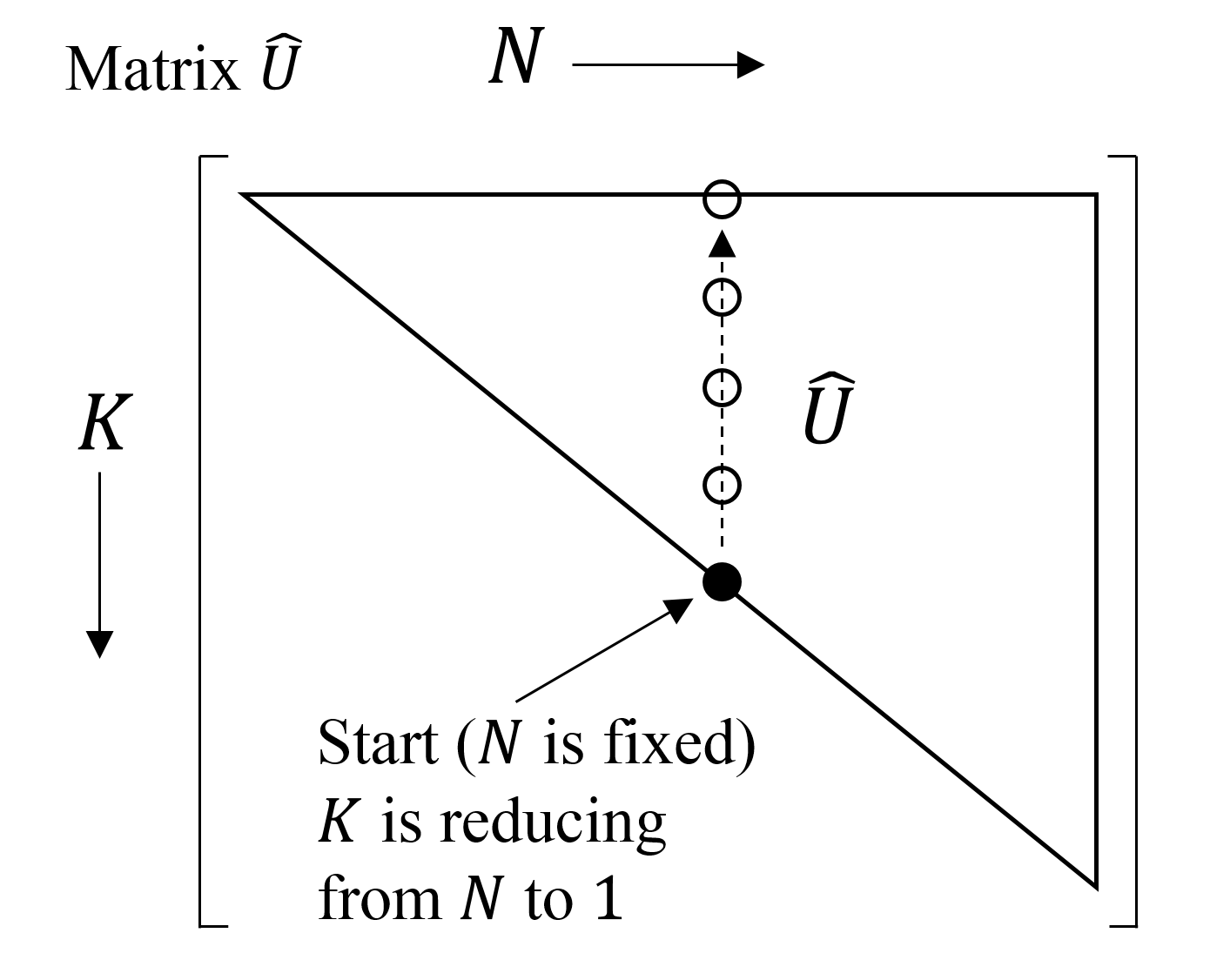}
	\caption{Scheme illustrating the algorithm of the analytic calculation of matrix $\hat{U}$ of right eigenvectors.  The diagonal blocks of $\hat{U}$ are found by iterating over $N$ which changes from 0 to $\infty$. Nonzero off-diagonal blocks are found by fixing $N$ and iterating over $K$ which changes from the diagonal value $K=N$ to $K=0$.  }
\label{fig:U}
\end{figure}

Finally the blocks of the right eigenvector matrix $\hat{U}$ are found from the matrix multiplication $U_{KN}=S_K\tilde{U}_{KN}$, which is the inverse transformation compared to \Eq{equ:Utilde}. Figure~\ref{fig:U} illustrates the above algorithm.

\begin{figure}
\centering
\includegraphics[height=6cm,clip]{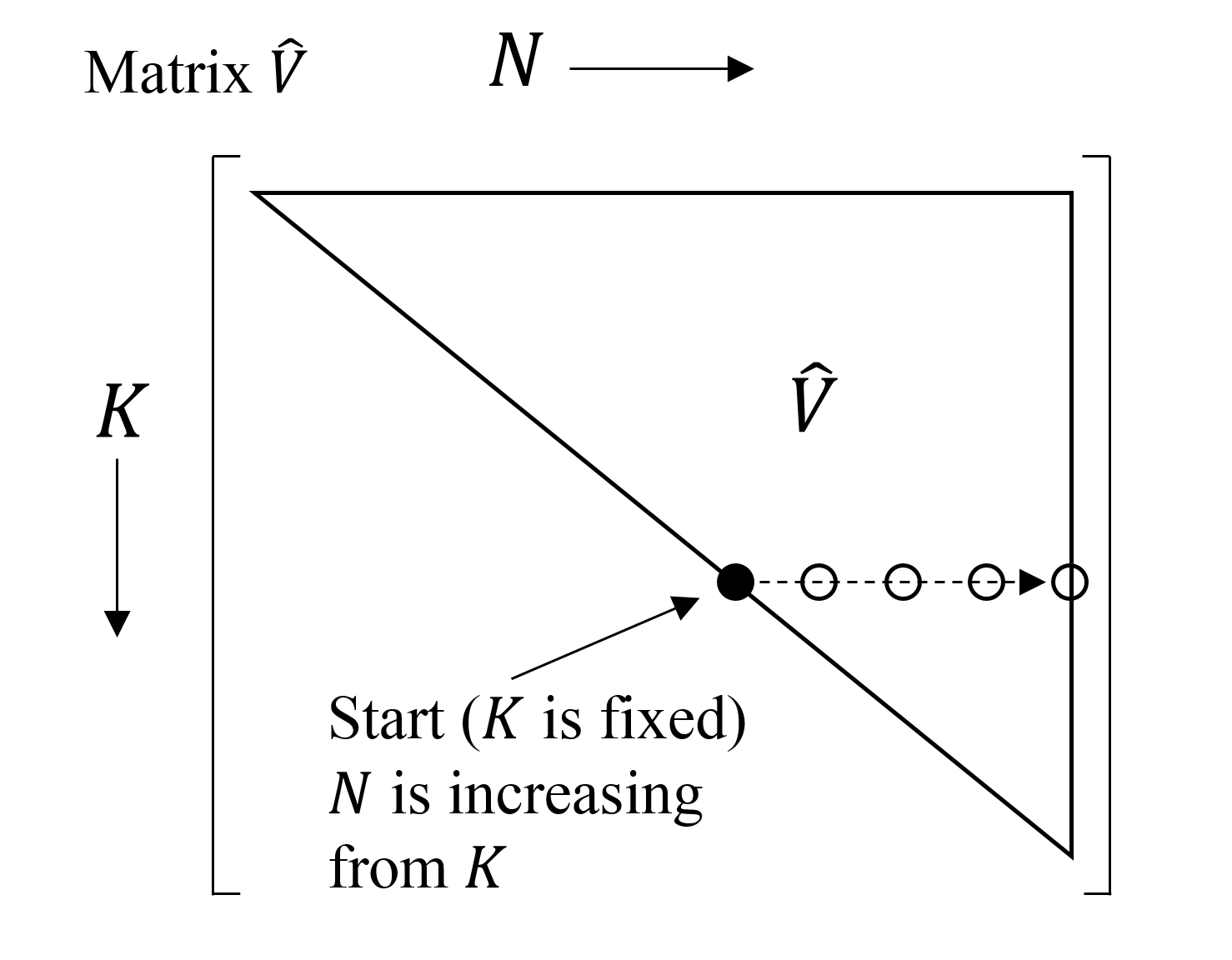}
\caption{As \Fig{fig:U} but for matrix $\hat{V}$ of left eigenvectors.  }
\label{fig:V}
\end{figure}

The procedure of finding the left eigenvector matrix $\hat{V}$ is similar. We first obtain matrix equations $V_{NN}L_N = \Omega_N V_{NN}$ for the diagonal blocks of $\hat{V}$, concluding that $V_{NN}=S_N^{\rm T}$\,. Then, for any $K>N$, we have
$$
V_{NK}L_K +V_{N,K-1}M_{K-1,K}=\Omega_N V_{NK}\,.
$$
Multiplying this equation with matrix $S_K$ from the right, and using the fact that $L_K S_K=S_K\Omega_K$, we find
$$
\tilde{V}_{NK}\Omega_K+\tilde{V}_{N,K-1}\tilde{M}_{K-1,K}=\Omega_N\tilde{V}_{NK}\,,
$$
where $\tilde{V}_{NK}=V_{NK}S_K$ and $\tilde{M}_{K-1,K}$ is defined in \Eq{equ:Utilde}. This again allows us to obtain an explicit form of the matrix elements:
\be (\tilde{V}_{NK})_{ij}=\frac{(\tilde{V}_{N,K-1}\tilde{M}_{K-1,K})_{ij}}{(\Omega_N)_{ii}-(\Omega_K)_{jj}}\,.
\label{equ:VNK}\ee
For a given fixed $N$,  one can generate sequentially, starting from $\tilde{V}_{NN}=\one$ and using \Eq{equ:VNK},  all the matrices $\tilde{V}_{NK}$  for $K=N+1,N+2,...$. Matrices $V_{NK}$ can then be found, using the inverse transformation, as $V_{NK}=\tilde{V}_{NK} S_K^{\rm T}$. This algorithm of reconstructing the full matrix $\hat{V}$ is illustrated by Figure~\ref{fig:V}.

Note that  the left and right eigenvectors are orthogonal,
$$ \hat{V}\hat{U}=\hat{U}\hat{V}=\hat{\mathbb{1}}\,,$$
which results for $N'\geqslant N$ in the relations
\be \sum_{K=N}^{N'}V_{NK}U_{KN'}=\sum_{K=N}^{N'}U_{NK}V_{KN'}=\mathbb{1}\delta_{NN'}\,,
\label{equ:orth}\ee
where $\mathbb{1}$ is the identity matrix and $\delta_{NN'}$ is the Kronecker delta. Equation (\ref{equ:orth}) can also be written as
$$\sum_{K=N}^{N'}\tilde{V}_{NK}\tilde{U}_{KN'}=\sum_{K=N}^{N'}\tilde{U}_{NK}\tilde{V}_{KN'}=\mathbb{1}\delta_{NN'}\,.
$$

\bigskip

To conclude this section, let us illustrate the analytic diagonalization presented above on a $6\times6$ Lindblad matrix corresponding to the lowest order of the standard FWM polarization treated in~\cite{KasprzakNMa10}:
$$ \hat{L}= \begin{bmatrix}
L_0 & M_{01} \\
\mathbb{0} & L_1 \end{bmatrix}\,,
\quad
\hat{U}= \begin{bmatrix}
S_0 & U_{01} \\
\mathbb{0} & S_1 \end{bmatrix}\,,
\quad
\hat{V}= \begin{bmatrix}
S^{\rm T}_0 & V_{01} \\
\mathbb{0} & S^{\rm T}_1 \end{bmatrix}\,.$$
We find $U_{01}=S_0 \tilde{U}_{01}$ and $V_{01}=\tilde{V}_{01}S^{\rm T}_2$, where
\be (\tilde{U}_{01})_{ij}=\frac{(\tilde{M}_{01})_{ij}}{(\Omega_1)_{jj}-(\Omega_0)_{ii}} \ , \quad
(\tilde{V}_{01})_{ij}=\frac{(\tilde{M}_{01})_{ij}}{(\Omega_0)_{ii}-(\Omega_1)_{jj}}=-(\tilde{U}_{01})_{ij}\,,
\label{equ:UeqV} \ee
and $\tilde{M}_{01}=S^{\rm T}_0 M_{01}S_1$.
The orthogonality of $\hat{V}$ and $\hat{U}$ can then be verified:
$$ \hat{U}\hat{V}=\begin{bmatrix}
S_0 & U_{01} \\
\mathbb{0} & S_1 \end{bmatrix}
\begin{bmatrix}
S^{\rm T}_0 & V_{01} \\
\mathbb{0} & S^{\rm T}_1 \end{bmatrix}
=\begin{bmatrix}
S_0S^{\rm T}_0 & S_0V_{01}+U_{01}S^{\rm T}_1 \\
\mathbb{0} & S_1 S^{\rm T}_1 \end{bmatrix}
=\hat{\mathbb{1}}\,,$$
using
$$ S_0V_{01}+U_{01}S^{\rm T}_1=S_0\tilde{V}_{01}S^{\rm T}_1+
S_0\tilde{U}_{01}S^{\rm T}_1
=S_0(\tilde{V}_{01}+\tilde{U}_{01})S^{\rm T}_1=\mathbb{0}\,, $$
which follows from \Eq{equ:UeqV}\,.

\section{Transition frequencies}
\label{Sec:frequencies}

By fixing the distance $\DN$ between rungs of the JC ladder, contributing to the left and right parts of the DM, we isolate a specific component of the coherent dynamics of the QD-cavity system, corresponding to a selected phase combination of the optical pulses exciting it. The time dependence of this polarization is given by Eq.\,(14) which contains the transition frequencies $\omega_r$ between rungs. These frequencies are the eigenvalues of the reduced Lindblad matrix $\hat{L}$, isolated from the full Lindblad operator by fixing the $\DN$. The eigenvalues are given by  the diagonal matrix $\hat{\Omega}$, \Eq{equ:OUV}, which consists of blocks $\Omega_N$ described by \Eqs{equ:Omega0}{equ:OmegaN}. For $N>0$, these diagonal blocks can be written as
\be
\Omega_N= (\bar{\Omega} -i\gamma_N)\one +
\begin{bmatrix}
-\Delta^i_N & 0 & 0 & 0 \\
0 & -\Delta^o_N & 0 & 0 \\
0 & 0 & \Delta^o_N & 0 \\
0 & 0 & 0 & \Delta^i_N \end{bmatrix}\,,
\label{equ:four}\ee
where
\be
\Delta^{o,i}_N=\sqrt{(\delta/2)^2+\ND g^2}\pm\sqrt{(\delta^\ast/2)^2+Ng^2}\,.
\label{equ:delpm}\ee
The complex frequencies of all four transitions, which occur between the two pairs of quantum levels of rungs $N$ and $\ND=N+\DN$, have the same dominant contribution, described by the first term of \Eq{equ:four}.  It consists of the average frequency distance between the rungs,
\be \bar{\Omega}=\DN\Omega_C\,,
\label{equ:Om}\ee
which is the same for all pairs of rungs separated by $\DN$, and
the average damping,
\be \gamma_N=(N+\ND-1)\gamma_C+\gamma_X\,,
\label{equ:gam}\ee
showing a linear increase with $N$, as the dampings of both rungs add up.

The second term in \Eq{equ:four} describes a fine structure of the transitions, given by the splittings $\Delta^{o,i}_N$, which depend on the detuning $\delta=\omega_X-\omega_C$, the coupling constant $g$, and the rung number $N$. Below we analyze this fine structure in more detail, providing simple asymptotic expressions for limiting cases of (i) large and (ii) small or zero detuning, in the latter case paying attention to the limit of large $N$.

\subsection{Large detuning}
Assuming $|\delta/2|\gg \sqrt{\ND}g$ and $|\delta/2|\gg \sqrt{N}g$, we find from \Eq{equ:delpm}
$$ \Delta^{o,i}_N\approx\frac{\delta\pm\delta^\ast}{2}+g^2\left(\frac{\ND}{\delta}\pm\frac{N}{\delta^\ast}\right)\,,$$
so that
$$ \Delta^o_N\approx \delta' + g^2\frac{(N+\ND)\delta'-i\DN\delta''}{|\delta|^2}\,,\\
\quad \Delta^i_N\approx i\delta'' + g^2\frac{\DN\delta'-i(N+\ND)\delta''}{|\delta|^2}\,,$$
where the complex detuning $\delta=\delta'+i\delta''$ is split into the real and imaginary parts, given by $\delta'=\Omega_X-\Omega_C$ and $\delta''=\gamma_C-\gamma_X$, respectively. Furthermore, for $|\delta'|\gg|\delta''|$, the above equations simplify to
$$ \Delta^o_N\approx \delta' +g^2\frac{N+\ND}{|\delta|}\ \ \ \ {\rm and}\ \ \ \
\Delta^i_N\approx i\delta''+g^2\frac{\DN}{|\delta|}\,,$$
giving approximate frequencies $\pm\Delta^o_N$ and $\pm\Delta^i_N$ of, respectively, ``outer'' and  ``inner'' transition doublets, to be considered on top of the same for all four transitions lead frequency and damping, described by \Eqs{equ:Om}{equ:gam}, respectively. We see that the splitting between the outer transitions is dominated by $2\delta'$, with a correction proportional to $g^2$ and growing linearly with $N$. At the same time, the inner transitions have a small splitting $2g^2\DN/|\delta|$, which is independent of $N$. The damping of both outer transitions is given by $\gamma_N$. For the inner transitions ($\bar{\Omega}\pm\Delta^i_N$), the dampings are different, $\gamma_N\mp\delta'' $, so that the lower-frequency transition is broader than the higher-frequency one for $\gamma_C>\gamma_X$.

\subsection{Small detuning}
Assuming $|\delta/2|\ll \sqrt{\ND}g$ and $|\delta/2|\ll\sqrt{N}g$, we find from \Eq{equ:delpm}

$$ \Delta^{o,i}_N\approx \left(\sqrt{\ND}\pm \sqrt{N} \right)g+\frac{\delta^2}{8\ND g^2}\pm
\frac{{\delta^\ast}^2}{8Ng^2}\,.$$
For $N\gg\DN$, this result further simplifies to
$$\Delta^o_N\approx  2\sqrt{N}g+\frac{\DN}{2\sqrt{N}} g +\frac{(\delta')^2-(\delta'')^2}{4Ng^2} \ , \quad
\Delta^i_N\approx \frac{\DN}{2\sqrt{N}} g + i\frac{\delta'\delta''}{2Ng^2}\,.$$
Finally, for zero detuning, $\Omega_X=\Omega_C$, we obtain in leading order of $\DN/N$
$$ \Delta^o_N\approx 2\sqrt{N}g \ \ \ \ {\rm and} \ \ \ \ \Delta^i_N\approx \frac{\DN}{2\sqrt{N}} g \,,$$
from where we find also the change of the transition frequencies of outer and inner doublets with rung number $N$,
\be \Delta^o_{N+1}-\Delta^o_N\approx \frac{g}{\sqrt{N}} \ \ \ \ {\rm and} \ \ \ \
\Delta^i_{N+1}-\Delta^i_N\approx -\frac{\DN g}{4N\sqrt{N}} \,.
\label{equ:freq}\ee

\section{Degenerate ${\cal N}$-wave mixing in the low-damping limit}\label{sec:lowdamp}
In this section, we consider all possible phase channels in the optical polarization when the system is excited by two laser pulses of arbitrary strength. We focus on the situation when both pulses arrive simultaneously (i.e. the time delay is zero) and call the optical response on this excitation  {\it degenerate} ${\cal N}$-wave mixing (${\cal N}$WM) polarization, where ${\cal N}$ determines the detected phase channel. We first treat rigorously the change of the DM due to the pulsed excitation, concentrating on two important special cases when the pulse area of one of the two pulses small enough to be accounted for in the lowest order while the pulse area of the other pulse can be arbitrarily large.  Then we consider  the coherent dynamics after the pulses in the limit of small exciton and cavity damping, so that the off-diagonal blocks of the Lindblad matrix can be neglected. Finally, we treat analytically the limit of a large average number of photons $n_{\rm ph}$ excited in the cavity, $n_{\rm ph}\gg 1$, corresponding to a large pulse area of one of the pulses, which allows us to develop a closed-form solution for the ${\cal N}$WM polarization in both time and frequency domain.

\subsection{Two-pulse excitation}

The so-called ${\cal N}$WM mentioned above describes a mixing of ${\cal N}$ waves which produce an optical response of the system with a phase
\be \Phi=\DN_1 \varphi_1+\DN_2 \varphi_2\,,
\label{equ:phase}\ee
where $\varphi_j=\arg(E_j)$ and
\be |\DN_1|+|\DN_2|+1={\cal N}\,.
\label{equ:N}\ee
For example, the standard FWM corresponds to $\DN_1=-1$ and $\DN_2=2$; therefore ${\cal N}=4$ and
$\Phi=2\varphi_2-\varphi_1$\,. Starting from the DM of a fully relaxed system before the excitation,
\be
\rho(0_-)=|0 \rangle\langle 0|\,,
\label{equ:DM0}
\ee
where $|0\rangle$ is its absolute ground state and $0_-$ is a negative infinitesimal, we consider    below the effect on the DM of the two pulses both arriving at $t=0$ and having pulse areas $E_1$ and $E_2$, focusing on the two limiting cases mentioned above.

\subsubsection*{Case 1: $E_1$ is small, $E_2$ is arbitrary.}

While we are assuming that the pulses arrive simultaneously at $t=0$, it is convenient to consider first the effect of the smaller pulse. Using the general Eq.\,(10) with the QD exciton indices dropped for brevity and the unexcited DM in the form of \Eq{equ:DM0}, the DM straight after the first pulse takes the form
$$ \rho_{kk'}(0)= \left[\hat{X}(E_1) \rho(0_-)\right]_{kk'}=e^{i\varphi_1(k-k')} C_{k0}(|E_1|) C^\ast_{k'0}(|E_1|)\,.$$
From \Eq{equ:phase} it follows that $k-k'=\DN_1$. Then, concentrating on the lowest-order response, we find the following two options for $k$ and $k'$:
$$ \begin{array}{cllllll}
(i)&k=\DN_1 &{\rm and} & k'=0 && {\rm for} & \DN_1\geqslant 0\,,\\
(ii)&k=0 &{\rm and} & k'=-\DN_1 && {\rm for} & \DN_1\leqslant 0\,.
\end{array}$$
Since the pulse area $E_2$ of the second pulse can be arbitrarily large, we take into account its effect rigorously in all orders, which results in the following DM after the pulses:
\bea \rho_{nn'}(0_+)= \left[\hat{X}(E_2) \rho(0)\right]_{nn'}&=&e^{i\varphi_2(n-n'-\DN_1)} C_{nk}(|E_2|) C^\ast_{n'k'}(|E_2|) \rho_{kk'}(0)
\nonumber\\
&=& e^{i\Phi} C_{nk}(|E_2|) C^\ast_{n'k'}(|E_2|) C_{k0}(|E_1|) C^\ast_{k'0}(|E_1|) \,,
\nonumber\eea
according to Eq.\,(10). From \Eq{equ:phase} we find $ n-n'=\DN_1+\DN_2=\DN$, and then from Eq.\,(11) obtain
\be \rho_{n+\DN N,n}(0_+)=e^{i\Phi} i^{\DN} |E_1|^{|\DN_1|}  |E_2|^{|\DN_2|} R_n\,,
\label{equ:rhonn}\ee
where
\be R_n=\frac{\lambda^n e^{-\lambda}}{\sqrt{n!(n+\DN)!}} \tilde{R}_n
\label{equ:Rn}\ee
with $\lambda=|E_2|^2$ and
\be
\tilde{R}_n = \left\{
\begin{array}{rlll}
\medskip
\lambda^m L_{\DN_1}^{n+\DN_2}(\lambda) && {\rm for} & \DN_1\geqslant 0\,,\\
\lambda^{m+\DN_1} L_{-\DN_1}^{n+\DN_2}(\lambda) && {\rm for} & \DN_1\leqslant 0\,.
\end{array}
\right.
\label{equ:Rtilde}
\ee
Here $L^k_p(x)$ are the Laguerre polynomials, and
\be m = \left\{\begin{array}{clll}\medskip
0 && {\rm for} & \DN_2\geqslant 0\,,\\
\DN_2&& {\rm for} & \DN_2\leqslant 0\,.\end{array}\right.
\label{equ:m}\ee

For ${\cal N}$WM,  we have in particular  $\DN_1=1-{\cal N}/2$ and $\DN_2={\cal N}/2$, so that $\DN=1$, in accordance with \Eq{equ:N}, and \Eqss{equ:rhonn}{equ:m} reduce to
\be \rho_{n+1,n}(0_+)=e^{i\Phi} i  |E_1|^{|\DN_1|}  |E_2|^{|\DN_2|} R_n\,,
\label{equ:rhonnN}\ee
where
\be R_n=\frac{\lambda^n e^{-\lambda}}{n!\sqrt{n+1}} \tilde{R}_n
\label{equ:RnN}\ee
and
\be \tilde{R}_n= \lambda^{1-{\cal N}/2} L_{{\cal N}/2-1}^{n+1-{\cal N}/2}(\lambda)\,.
\label{equ:NWM1} \ee
The last equation simplifies to
\be
\tilde{R}_n= \frac{1}{\lambda} L_{1}^{n-1}(\lambda)
\label{equ:FWM1}
\ee
for the standard FWM, in which case ${\cal N}=4$,   $\DN_1=-1$, and $\DN_2=2$.

\subsubsection*{Case 2: $E_2$ is small, $E_1$ is arbitrary.}

To address this case we use the fact that for simultaneous pulses, the pulse operators $\hat{X}(E_1)$ and $\hat{X}(E_2)$ commute:
\be
\hat{X}(E_1)\hat{X}(E_2)\rho=\hat{X}(E_2)\hat{X}(E_1)\rho\,.
\ee
Technically, this is easy to see from the definition of $\hat{X}(E)$, Eq.\,(7), and the fact that the commutator
$[E_1\ad +E_1^\ast a,E_2\ad +E_2^\ast a]= E_1^\ast E_2-E_1 E_2^\ast $ is a constant.
Physically, this means that for an infinitesimal delay between the pulses exciting the cavity, the time-ordering of the pulses does not matter. It would matter, however, if the QD-cavity system was excited via the QD, since the QD exciton is described by Fermionic operators obeying anti-commutation relations, and therefore the corresponding pulse operators do not commute.

The result obtained for {\it Case 1}, given by \Eqss{equ:rhonn}{equ:m}, can therefore be used for {\it Case 2} by swapping $E_1\leftrightarrow E_2$ and $\DN_1\leftrightarrow\DN_2$. Then the DM after the pulses is described by the same \Eqs{equ:rhonn}{equ:Rn} with $\lambda=|E_1|^2$ and $\tilde{R}_n$ now given by
\be
\tilde{R}_n = \left\{\begin{array}{rlll}\medskip
\lambda^m L_{\DN_2}^{n+\DN_1}(\lambda) && {\rm for} & \DN_2\geqslant 0\,,\\
\lambda^{m+\DN_2} L_{-\DN_2}^{n+\DN_1}(\lambda) && {\rm for} & \DN_2\leqslant 0\,,\end{array}\right.
\label{equ:Rtilde2}\ee
where
\be
m = \left\{\begin{array}{clll}\medskip
0 && {\rm for} & \DN_1\geqslant 0\,,\\
\DN_1&& {\rm for} & \DN_1\leqslant 0\,.\end{array}\right.
\label{equ:m2}\ee
Again, for ${\cal N}$WM, \Eqs{equ:rhonnN}{equ:RnN} remain the same as in {\it Case 1}, while
\Eqs{equ:Rtilde2}{equ:m2} simplify to
\be
\tilde{R}_n= \lambda^{1-{\cal N}/2} L_{{\cal N}/2}^{n+1-{\cal N}/2}(\lambda)\,,
\label{equ:NWM2}
\ee
which reduces for the standard FWM with  $\DN_1=-1$ and $\DN_2=2$ to
\be \tilde{R}_n= \frac{1}{\lambda} L_{2}^{n-1}(\lambda)\,.
\label{equ:FWM2}\ee

Note that in the ${\cal N}$WM, the difference between the two {\it Cases} is only in the lower index of the Laguerre polynomials; compare \Eqs{equ:NWM1}{equ:NWM2} and similarly \Eqs{equ:FWM1}{equ:FWM2}.

\subsection{Coherent dynamics after the pulses}
Now, omitting the factor
\be e^{i\Phi} i^{\DN} |E_1|^{|\DN_1|}  |E_2|^{|\DN_2|}
\label{equ:factor} \ee
in \Eq{equ:rhonn}, which is common for all elements of the DM, we write the initial DM straight after the pulses in vector form,
$$ \vec{\rho}^{\:(0)}(0_+)=\begin{bmatrix}
0\\
R_0 \end{bmatrix}
, \ \ \
\vec{\rho}^{\:(n)}(0_+)= \begin{bmatrix}
0\\
0\\
0\\
R_n \end{bmatrix}\,,$$
where $n\geqslant 1$ and the exciton components have been restored; for definition of the basis, see \Eqss{equ:rhoN}{equ:vecs} in \Sec{Sec:Lindblad}.

In the limit of small damping of both the QD exciton and the cavity mode, $\gamma_X,\gamma_C\ll g$, one can neglect the off-diagonal blocks $M_{n,n+1}$ of the Lindblad matrix $\hat{L}$, see \Eq{equ:Mnn1}. The remaining diagonal blocks of $\hat{L}$ are diagonalized according to \Eq{equ:SLS},
\be L_n=S_n\Omega_n S_n^{\rm T}\,,
\label{equ:SOS}\ee
where matrices $S_n$ and $\Omega_n$ are given, respectively, by \Eqs{equ:SAB}{equ:OmegaN}. The time evolution is then described as
$$ \vec{\rho}^{\:(n)}(t) = e^{-iL_n t}\vec{\rho}^{\:(n)}(0_+)=S_ne^{-i\Omega_n t}S_n^{\rm T}\vec{\rho}^{\:(n)}(0_+)\,.$$
Using the general form Eq.\,(9) of the optical polarization, we find
$$ P(t)= \sum_{n=0}^\infty \vec{a}^{\:(n)}\cdot \vec{\rho}^{\:(n)}(t)\,,$$
where $\vec{a}^{\:(n)}$ is the vector representation of the photon annihilation operator $a$:
$$ \vec{a}^{\:(0)}= \begin{bmatrix}
0\\
1 \end{bmatrix}
, \ \ \
\vec{a}^{\:(n)}= \begin{bmatrix}
\sqrt{n}\\
0\\
0\\
\sqrt{n+1} \end{bmatrix}\,,$$
in accordance with the basis defined in \Eqs{equ:rhoN}{equ:rho0}. Now, using the explicit form of the matrices $S_n$ and $\Omega_n$ provided in \Eqs{equ:AB}{equ:SAB}, and \Eqs{equ:four}{equ:delpm}, respectively, we find
$$P(t)= e^{-i\bar{\Omega}t} \sum_{\sigma=i,o} \sum_{s=\pm} P_{\sigma s}(t)\,,$$
where
\be P_{\sigma s}(t)= \sum_{n=0}^\infty  R_n C_n^{\sigma s} e^{-i(s\Delta_n^\sigma -i\gamma_n) t}\,.
\label{equ:Pss}\ee
The frequencies $\Delta_n^i$ and $\Delta_n^o$ of, respectively, the inner and outer transitions are given by \Eq{equ:delpm}, and the damping $\gamma_n$ by \Eq{equ:gam}.
Using the matrices $A_{n+1}$ and $B_n^\ast$ [\Eq{equ:AB}] forming the transformation matrix $S_n$, we find the coefficients $C_n^{\sigma s}$ for an arbitrary detuning:
\bea
C_n^{i+}&=&\alpha_n^\ast \alpha_{n+1} \left(\alpha_n^\ast \alpha_{n+1} \sqrt{n+1} +\beta_n^\ast \beta_{n+1}\sqrt{n}\right)\,,\nonumber \\
C_n^{i-}&=&\beta_n^\ast \beta_{n+1} \left(\beta_n^\ast \beta_{n+1} \sqrt{n+1} +\alpha_n^\ast \alpha_{n+1}\sqrt{n}\right)\,,\nonumber \\
C_n^{o+}&=&\beta_n^\ast \alpha_{n+1} \left(\beta_n^\ast \alpha_{n+1} \sqrt{n+1} -\alpha_n^\ast \beta_{n+1}\sqrt{n}\right)\,,\nonumber \\
C_n^{o-}&=&\alpha_n^\ast \beta_{n+1} \left(\alpha_n^\ast \beta_{n+1} \sqrt{n+1} -\beta_n^\ast \alpha_{n+1}\sqrt{n}\right)\,.
\label{equ:CC}
\eea
For a detuning much smaller than the energy splitting of the $n$-th rung, $|\delta|\ll \sqrt{n}g$,
which is relevant to the case of large excitation pulse area treated below, they
take approximate forms
\be
C_n^{i\pm}=\frac{1}{4} \left(\sqrt{n+1} +\sqrt{n}\right) \quad \mbox{and} \quad
C_n^{o\pm}=\frac{1}{4} \left(\sqrt{n+1} -\sqrt{n}\right)
\label{equ:Cio}\ee
for $n\geqslant1$, as well as $C_0^{i+}=C_0^{o-}=1/2$ and $C_0^{o+}=C_0^{i-}=0$,  using $\alpha_n=\beta_n=1/\sqrt{2}$ for $n\geqslant1$, and $\alpha_0=1$ and $\beta_0=0$. For zero detuning, $\delta=0$, \Eq{equ:Cio} is exact. The general property  $C_n^{\sigma -}= (C_n^{\sigma +})^\ast$ is fulfilled for \Eq{equ:CC} only approximately but becomes strict at zero detuning, since all the coefficients in \Eq{equ:Cio} are real.

\subsection{Large pulse area}
In the limit of large pulse area ($\lambda=|E_2|^2\gg 1$ in {\it Case 1} or $\lambda=|E_1|^2\gg 1$ in {\it Case 2}), the excited system contains a large number of photons, $n_{\rm ph}\approx \lambda\gg1$. The Poisson distribution in \Eq{equ:RnN} then becomes Gaussian, with the mean rung number $\langle n \rangle = \lambda$ and the mean square deviation $\langle n^2 -  \langle n \rangle ^2\rangle = \lambda$. To achieve this limit mathematically, we replace in \Eq{equ:RnN}
$$ n!\approx \sqrt{2\pi n} e^{-n} n^n$$
and, introducing a small quantity $\varepsilon\ll 1$, which is defined in such a way that
$ n=\lambda(1+\varepsilon)\,, $
we further approximate
\bea e^{-n} n^n&=& e^{-n} \lambda^n (1+\varepsilon)^{\lambda(1+\varepsilon)}=
e^{-n} \lambda^n e^{\lambda(1+\varepsilon)\ln(1+\varepsilon)}\nonumber\\\nonumber
&\approx &   e^{-n} \lambda^n e^{\lambda(1+\varepsilon)(\varepsilon- \varepsilon^2/2)}
\approx  e^{-\lambda-\lambda\varepsilon} \lambda^n e^{\lambda(\varepsilon+ \varepsilon^2/2)}
=e^{-\lambda} \lambda^n e^{\lambda \varepsilon^2/2}\,.\eea
Equation (\ref{equ:RnN}) then becomes
\be R_n=\frac{\lambda^n e^{-\lambda}}{n!\sqrt{n+1}} \tilde{R}_n\approx
\frac{\lambda^n e^{-\lambda}}{\sqrt{2\pi \lambda}e^{-\lambda} \lambda^n e^{\lambda \varepsilon^2/2}\sqrt{\lambda}} \tilde{R}_n=\frac{e^{-z^2}}{\sqrt{2\pi} \lambda}\tilde{R}_n\,,
\label{equ:RnNa} \ee
where we have introduced for convenience a new variable
\be z=\frac{n-\lambda}{\sqrt{2\lambda}}\,,
\label{equ:zdef}\ee
such that $\langle z \rangle =0$ and $\langle z^2 \rangle =1/2$. The Lagguere polynomials in \Eqs{equ:NWM1}{equ:NWM2} for $\tilde{R}_n$ are approximated as
\be L^{n-m}_m(\lambda) \approx \frac{1}{m!} \left(\frac{\lambda}{2}\right)^{\frac{m}{2}} H_m(z)\,,
\label{equ:LH}\ee
where $ H_m(z)$ are Hermite polynomials. To prove \Eq{equ:LH}, we use the recursive relation~\cite{Gradshtein}
\be m L^{n-m}_m(\lambda)=(n-\lambda) L^{n-m+1}_{m-1}(\lambda) -n L^{n-m+2}_{m-2}(\lambda)+n L^{n-m+2}_{m-3}(\lambda)\,.
\label{equ:Lrec}\ee
The first few polynomials in this sequence have the following form:
\bea
L^{n}_{0}(\lambda) &=&1\,, \nonumber\\
L^{n-1}_{1}(\lambda) &=&n-\lambda= \left(\frac{\lambda}{2}\right)^{\frac{1}{2}} 2z \,, \nonumber\\
L^{n-2}_{2}(\lambda) &=&\frac{1}{2} \left(-\lambda+(n-\lambda)^2\right)= \frac{1}{2}\,\frac{\lambda}{2}(4z^2-2)\,,
\label{equ:L123}\eea
demonstrating the strict validity of \Eq{equ:LH} for $m=0$, 1, and 2. To prove \Eq{equ:LH} for higher $m$, we note that  $L^{n-m}_m(\lambda) \sim \lambda^{\frac{m}{2}}$, which is clear from \Eq{equ:L123} and the recursive formula \Eq{equ:Lrec}. In fact, all terms in \Eq{equ:Lrec} except the last one are of order $\lambda^{\frac{m}{2}}$, while the last term is of order $\lambda^{\frac{m-1}{2}}$ and thus can be neglected for large $\lambda$. For the same reason, $L^{n-m}_m(\lambda)\approx L^{n+1-m}_m(\lambda)$, so that \Eq{equ:LH} can be used for both {\it Cases} in the ${\cal N}$WM, described by \Eqs{equ:NWM1}{equ:NWM2}. Finally, substituting \Eq{equ:LH} into \Eq{equ:Lrec} and dropping the last term in \Eq{equ:Lrec}, in accordance with the above discussion, results in a recursive relation
\be H_m(z)=2z H_{m-1}(z) - 2(m-1) H_{m-2}(z) \,,
\label{equ:Hrec}\ee
which generates the Hermite polynomials~\cite{Gradshtein}, starting from $H_0(z)=1$ and $H_1(z)=2z$. Note that the latter are the two lowest-order polynomials which appear in \Eq{equ:L123}.

We further approximate the eigenfrequencies $\Delta_n^\sigma$, the damping $\gamma_n$, and the transition amplitudes $C_n^{\sigma s}$ in \Eq{equ:Pss} for large excitation pulse area ($\lambda\gg1$):
\begin{align}
&\Delta_n^o\approx 2\sqrt{\lambda}g+\sqrt{2}g z\,,
&&\Delta_n^i\approx \frac{g}{2\sqrt{\lambda}}-\frac{g}{2\sqrt{2}\lambda} z\,,\nonumber\\
&\gamma_n\approx(2\lambda+1)\gamma+2\sqrt{2\lambda}\gamma z\,,\nonumber\\
&C_n^{o \pm} \approx \frac{1}{8\sqrt{\lambda}}\,,
&&C_n^{i \pm} \approx \frac{\sqrt{\lambda}}{2}\,,
\label{equ:Capp}
\end{align}
with $z$ defined by \Eq{equ:zdef}.
Again, the approximation is valid for relatively small ($|\delta|\ll \sqrt{\lambda} g$) or zero detuning ($\delta=0$, so that $\gamma_X=\gamma_C=\gamma$).

Finally, switching in \Eq{equ:Pss} from summation to integration,
\be
\sum_{n=0}^\infty \to \sqrt{2\lambda} \int_{-\infty}^\infty dz\,,
\ee
and using the approximations \Eqsss{equ:RnNa}{equ:LH}{equ:Capp}, we obtain
\be
P_{\sigma s}(t)\approx \frac{i^m}{2} A^{(m)}_\sigma e^{-i\omega_{\sigma s} t}
\frac{1}{\sqrt{\pi}} \int_{-\infty}^\infty dz e^{-z^2}  e^{-i\gamma_{\sigma s} tz} H_m(z)\,,
\label{equ:Pss2}
\ee
where
\begin{align}
&\omega_{o s}= s 2\sqrt{\lambda}g -i(2\lambda+1)\gamma\,,
&&\omega_{i s}=s\frac{g}{2\sqrt{\lambda}}-i(2\lambda+1)\gamma\,,\nonumber\\
&\gamma_{o s}= s\sqrt{2}g -i 2\sqrt{2\lambda }\gamma\,,
&&\gamma_{i s}= -s\frac{g}{2\sqrt{2}\lambda} -i2\sqrt{2\lambda }\gamma\,,\nonumber\\
&A^{(m)}_o =\frac{(-i)^m}{4 m! (\sqrt{2\lambda})^m}\,,
&&A^{(m)}_i =4\lambda A^{(m)}_o \,,
\label{equ:Adef}
\end{align}
and $s=\pm1$. The amplitudes $A^{(m)}_\sigma$ of the ${\cal N}$WM polarization are given in \Eq{equ:Adef} for {\it Case 2}, for which $m={\cal N}/2$. Note, however, that \Eqs{equ:Pss2}{equ:Adef}  describe also the ${\cal N}$WM polarization in {\it Case 1}, provided that all $P_{\sigma s}(t)$ are divided by $\lambda$ and $m={\cal N}/2-1$ is used.

Now, performing the integration in \Eq{equ:Pss2} we find
\be P_{\sigma s}(t)\approx \frac{1}{2} A^{(m)}_\sigma(\gamma_{\sigma s} t)^m \exp\left\{{-i\omega_{\sigma s} t}-(\gamma_{\sigma s} t)^2/4\right\} \,,
\label{equ:Pss3}\ee
using the analytic integral
\bea I_m(p)&=&\int_{-\infty}^\infty e^{ipz}H_m(z) e^{-z^2} dz \nonumber\\
&=&\int_{-\infty}^\infty e^{ipz}\left[2z H_{m-1}(z) - 2(m-1) H_{m-2}(z)\right] e^{-z^2} dz \nonumber\\
&=&\int_{-\infty}^\infty e^{ipz}\left[ip H_{m-1}(z) + H'_{m-1}(z)- 2(m-1) H_{m-2}(z)\right] e^{-z^2} dz  \nonumber\\
&=& ip I_{m-1}=(ip)^m\sqrt{\pi}  e^{-p^2/4}\,.
\label{equ:Im}\eea
Note that in deriving \Eq{equ:Im} we have used the recursive relation \Eq{equ:Hrec}, integration by parts, the property of Hermite polynomials
$$ H'_m(z)=2m H_{m-1}(z)\,,$$
where the prime indicates the derivative versus the argument, and the Fourier transform of the Gaussian function
$$ I_0(p)=\int_{-\infty}^\infty e^{ipz} e^{-z^2} dz =\sqrt{\pi}  e^{-p^2/4}\,.$$

Finally, to obtain the ${\cal N}$WM spectrum, using $\bar{\Omega}$ as zero of frequency for convenience, we Fourier transform the time-dependent optical polarization:
\bea
\tilde{P}_{\sigma s}(\omega)&= &\int_0^\infty e^{i\omega t} P_{\sigma s}(t) dt\nonumber\\
&=&A^{(m)}_\sigma \frac{1}{2} \int_0^\infty (\gamma_{\sigma s} t)^m \exp\left\{{i(\omega-\omega_{\sigma s}) t}-(\gamma_{\sigma s} t)^2/4\right\} dt\nonumber\\
&=& \frac{A^{(m)}_\sigma}{\gamma_{\sigma s}} w_m\left(\frac{\omega- \omega_{\sigma s}}{\gamma_{\sigma s}}\right)\,,
\label{equ:Pw}
\eea
for $|\arg(\gamma_{\sigma s})|<\pi/4$. Otherwise, $\gamma_{\sigma s}$ must be replaced with $-\gamma_{\sigma s}$ and a sign factor $(-1)^m$ be added, see below for more details. This is actually the case of $o-$ and $i+$ transitions, for which Re\,$\gamma_{o-}<0$ and Re\,$\gamma_{i+}<0$. However, this can be conveniently dealt with by using the spectral symmetry:
\be \tilde{P}(\omega)=\sum_{\sigma=i,o} \sum_{s=\pm} \tilde{P}_{\sigma s}(\omega)=\bar{P}(\omega)+\bar{P}^\ast(-\omega)\,,
\label{equ:Pw0}\ee
where
\be \bar{P}(\omega)=\tilde{P}_{o+}(\omega)+\tilde{P}_{i-}(\omega)=
A^{(m)}_o\left[\frac{1}{\gamma_{o+}} w_m\left(\frac{\omega- \omega_{o+}}{\gamma_{o+}}\right)
+\frac{4\lambda}{\gamma_{i-}} w_m\left(\frac{\omega- \omega_{i-}}{\gamma_{i-}}\right) \right]\,.
\label{equ:Pw1}\ee
The function $w_m(z)$ in \Eqs{equ:Pw}{equ:Pw1} is defined as
\be w_m(z)= \frac{1}{2} \int_0^\infty t^m e^{izt}e^{-t^2/4} dt\,,
\label{equ:wn}\ee
%
and  can be  expressed in terms of the Faddeeva function, $w(z)=2w_0(z)/\sqrt{\pi}$, via its $m$-th  derivative
$$ w_m(z)=(-i)^m\frac{d^m}{dz^m} w_0(z)\,. $$
It is, however, more practical to use a recursive formula
which can be obtained integrating \Eq{equ:wn} by parts, which gives
\be w_m(z)=2izw_{m-1}(z) + 2(m-1) w_{m-2}(z)
\label{equ:wn2}\ee
for $m\geqslant 2$,
$$ w_1(z)=1+2izw_0(z)$$
for $m=1$, and
\be w_0(z)=G(z)+iD(z) =\frac{\sqrt{\pi}}{2} w(z)
\label{equ:w0}\ee
for $m=0$. Here, $G(z)$ is the Gaussian function,
$$ G(z)= \frac{\sqrt{\pi}}{2} e^{-z^2}\,,$$
$D(z)$ is the standard Dawson's integral,
$$ D(z)= \frac{1}{2} \int_0^\infty e^{-t^2/4}\sin(zt) dt = e^{-z^2} \int_0^z e^{t^2} dt\,,$$
and $w(z)$ is the Faddeeva function. The latter is well-know due to its real part, describing a Voigt (Gaussian) profile for complex (real) $z$.

The integral $w_m(z)$ in \Eq{equ:wn} can also be written explicitly using the Faddeeva function, Hermite polynomials and associated polynomials $Q_m(z)$ satisfying the recursive relation \Eq{equ:Hrec} of Hermite polynomials,
\be Q_m(z)=2z Q_{m-1}(z) - 2(m-1) Q_{m-2}(z)\,,
\label{equ:Qrec}\ee
but starting from $Q_1(z)=1$ and $Q_2(z)=2z$ instead. Functions $w_m(z)$ take the form
$$ w_m(z)=i^m H_m(z) w_0(z) +i^{m-1} Q_m(z)$$
with $w_0(z)$ given by \Eq{equ:w0} and polynomials
\begin{align}
&H_0(z)=1\,, && Q_0(z)=0\,, \nonumber\\
&H_1(z)=2z\,, && Q_1(z)=1\,, \nonumber\\
&H_2(z)=4z^2-2\,, && Q_2(z)=2z\,, \nonumber\\
&H_3(z)=8z^3-12z\,, && Q_3(z)=4z^2-4\,, \nonumber\\
&H_4(z)=16z^4-48z^2+12\,, && Q_4(z)=8z^3-20z\,, \nonumber\\
&H_5(z)=32z^5-160z^3+120z\,, && Q_5(z)=16z^4-72z^2+32\,,
\nonumber\end{align}
listed above for the first few $m$.

Note also that we have reduced the integral in \Eq{equ:Pw} to the Faddeeva function in the following way
\bea \int_0^\infty e^{iat}e^{-(bt)^2/4} dt &=&  e^{-(a/b)^2} \int_0^\infty e^{-b^2(t-t_0)^2/4} dt \nonumber\\
&=& e^{-(a/b)^2} \left[ \int_0^{t_0} e^{-b^2t^2/4} dt + \int_0^\infty e^{-b^2 t^2/4} dt  \right] \nonumber\\
&=& e^{-(a/b)^2} \left[\frac{2i}{b} \int_0^{a/b} e^{t^2} dt + \frac{\sqrt{\pi}}{b}  \right] =\frac{2}{b} w_0(a/b)\,,
\label{equ:b}\eea
where $t_0=2ia/b$. While the initial integral is invariant with respect to the sign change of $b$ and only requires Re\,$(b^2)>0$ for convergence, the Gaussian term in the last line of \Eq{equ:b}, containing the factor ${\sqrt{\pi}}/{b}$, is valid only if $|\arg(b)|<\pi/4$. This leads to the requirement introduced above that $|\arg(\gamma_{\sigma s})|<\pi/4$\,, otherwise $\gamma_{\sigma s}$ should be taken with the opposite sign.

\begin{figure}
	\centering
	\includegraphics[width=\columnwidth]{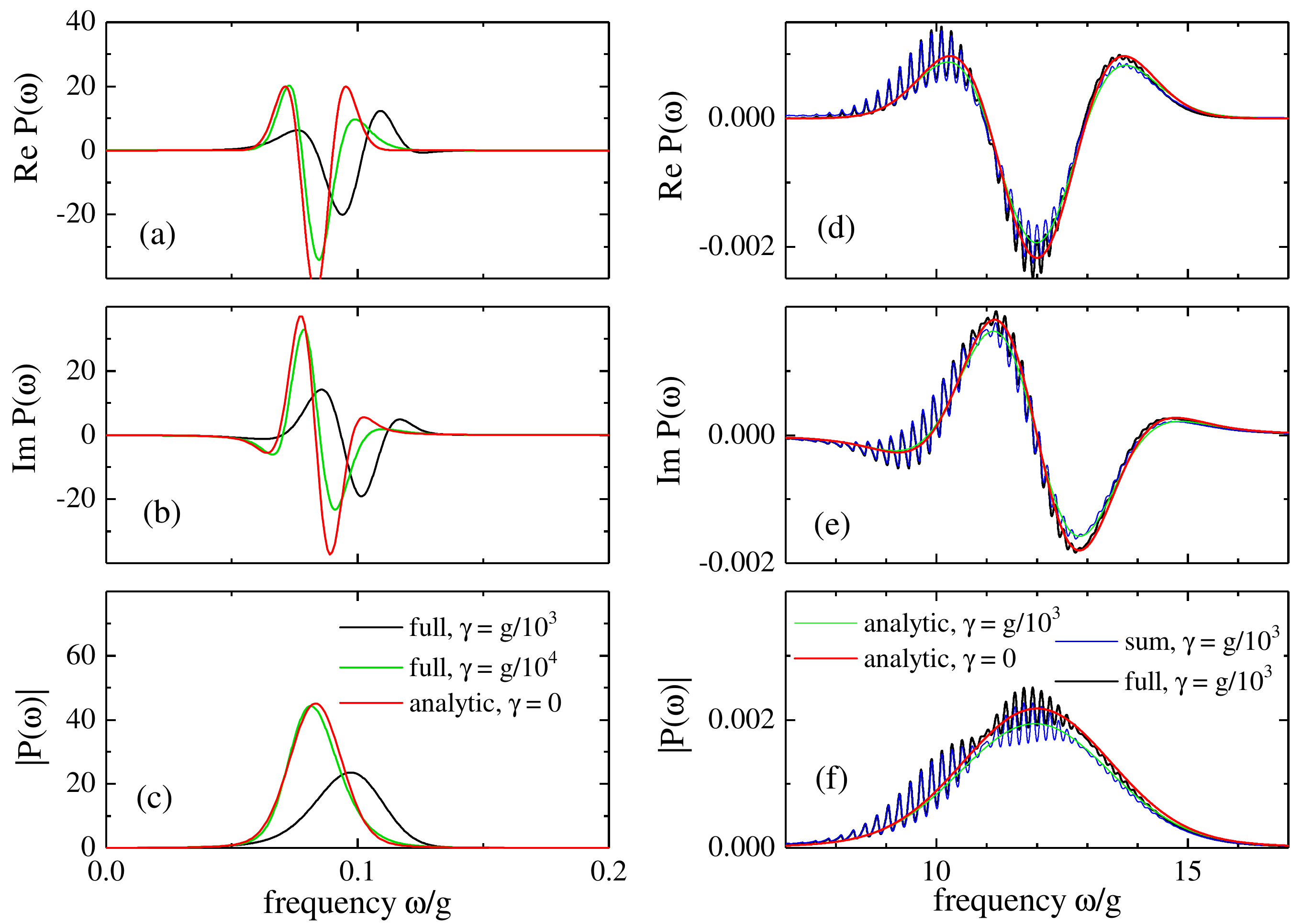}
	\caption{ Exact FWM spectrum (black and green lines) for $|E_1|=6$, $|E_2|=0.001$, zero detuning $\delta=0$, so that $\gamma_C=\gamma_X=\gamma$, with the values of $\gamma$ as given, in comparison with the analytic approximation \Eqs{equ:Pw0}{equ:Pw1} (red lines), and the full sum \Eq{equ:Pss} (blue lines). Left and right panels show the spectral regions of, respectively, inner and outer transitions (for positive frequencies). The spectra are shown without the factor \Eq{equ:factor}. }
\label{fig:FWMcomp}
\end{figure}

\begin{figure}
	\centering
	\includegraphics[width=0.85\columnwidth]{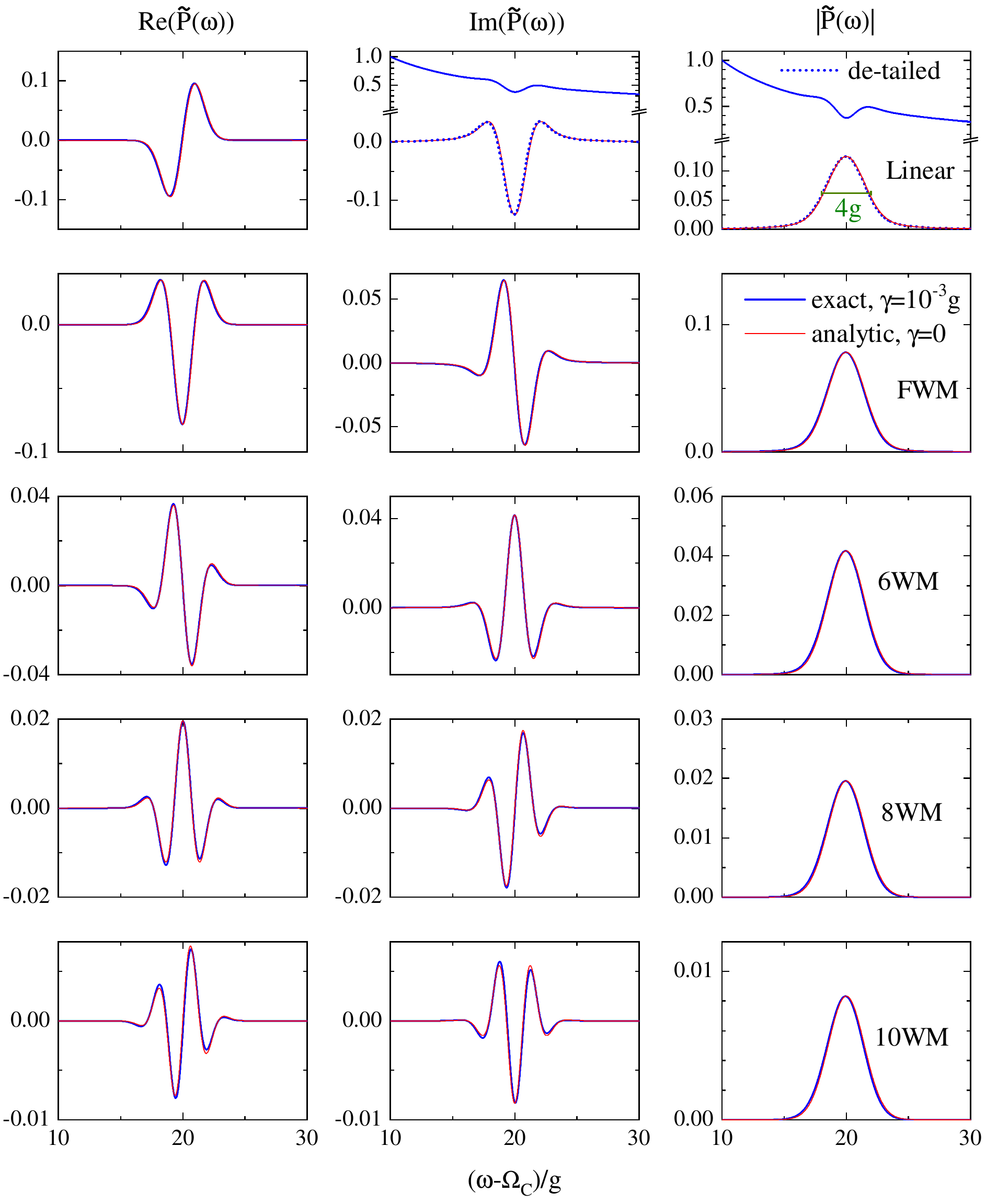}
	\caption{  Analytic approximation \Eqs{equ:Pw0}{equ:Pw1} (red lines) for the outer-transition sideband  of the ${\cal N}$WM spectrum with ${\cal N}=2,\,4,\,6,\,8,$ and 10,  $|E_1|=10$, $|E_2|=0.001$, zero detuning, and $\gamma=\gamma_C=\gamma_X=0$, in comparison with the exact calculation with $\gamma=0.001g$ (blue lines). The horizontal bars show the spectral linewidth of $4g$. The left, middle, and right panels show, respectively the real, imaginary part, and the absolute value of $\tilde P(\omega)$. All spectra are shown without the factor \Eq{equ:factor} and are multiplied with $|E_1|^{{\cal N}/2}$. The 2WM contains the linear response, creating a spectral tail of the inner doublet. The dotted lines show the exact result minus this tail $10ig/\omega$.}
\label{fig:NWM12}
\end{figure}

Figure~\ref{fig:FWMcomp} illustrates a comparison of FWM spectra, calculated using the exact solution,  given by the analytic formulas \Eqs{equ:Pw0}{equ:Pw1}, and the sum \Eq{equ:Pss} without converting it to an integral, with coefficients taken in the form of \Eq{equ:Cio}. For a damping of $\gamma=0.001g$, the sideband (right panels) shows the contributions of individual outer transitions, both in the sum and in the full spectrum. This is visible because the difference between the transition frequencies, $g/\sqrt{n}$ [see \Eq{equ:freq}], is larger than the damping $\gamma_n=(2n+1)\gamma$ (here $n\sim 36$). The pattern of oscillations seen in the spectral profile can be understood from the modulation of the Poisson distribution by the Lagguere polynomial $L_2^{n-1}(\lambda)$ specific to this nonlinearity channel, see \Eq{equ:FWM2}. In fact, $L_2^{n-1}(\lambda)$ presents a parabola, which is clearly seen in the amplitude of oscillations, having knots at around $\omega/g=11$ and 13. The frequency difference between the neighboring inner transitions, $-g/(4n\sqrt{n})$, is in turn much smaller than the damping, so that similar oscillations in the peak of the central band (left panel) are not seen even for a 10 times smaller damping. The analytic approximation (red curves) shows no oscillations, since the conversion of a sum into an integral used in its derivation effectively introduces a continuum of transitions. Interestingly, the analytic approximation shows somewhat better agreement with the full calculation when it is taken with zero damping, instead of using the correct $\gamma=0.001g$.

We further look at the spectral profile for higher non-linearities, concentrating on the outer transitions. Figure~\ref{fig:NWM12} shows the real and imaginary parts, as well as the absolute value of the ${\cal N}$WM spectrum, for all even ${\cal N}$ from 2 to 10. The number of oscillations in the real and imaginary parts grows linearly with ${\cal N}$, the real part of the ${\cal N}$WM spectrum being very similar to the imaginary part of the $({\cal N}-2)$WM spectrum, which is the property of the generalized Faddeeva function $w_m$ determining the spectra. The absolute value, however, shows no oscillations, and the same linewidth of around $4g$, essentially independent of ${\cal N}$. The right panels demonstrate excellent agreement between the analytic approximation and the exact calculation, for all spectra, apart from the case ${\cal N}=2$ which contains the linear response. Here, an extended spectral tail scaling with the inverse frequency remains in the full calculation, which is not reproduced by the analytic solution. Subtracting this tail, a good agreement is found.

\section{FWM power versus pulse area}

\begin{figure}
	\centering
	\includegraphics[width=0.7\columnwidth]{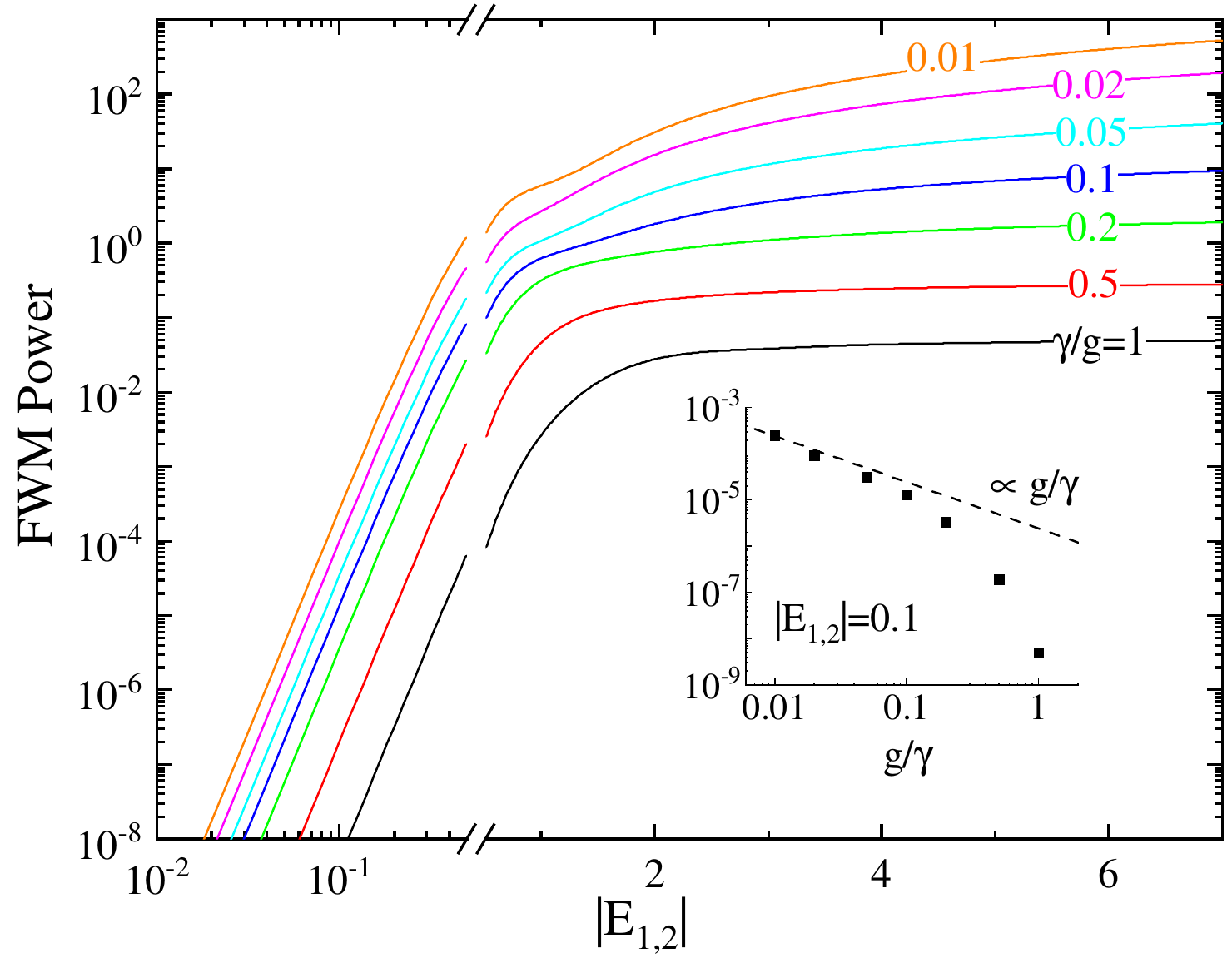}
	\caption{Power of FWM response for varying $|E|=|E_2|=|E_1|$, with $\delta=0$ and various $\gamma=\gamma_X=\gamma_C$ as indicated. Inset: the FWM power for $|E|=0.1$ versus $\gamma/g$. The scaling $\propto g/\gamma$ is given as dashed line.}
	\label{fig:E12powGcomp}
\end{figure}

\begin{figure}
	\centering
	\includegraphics[width=0.7\columnwidth]{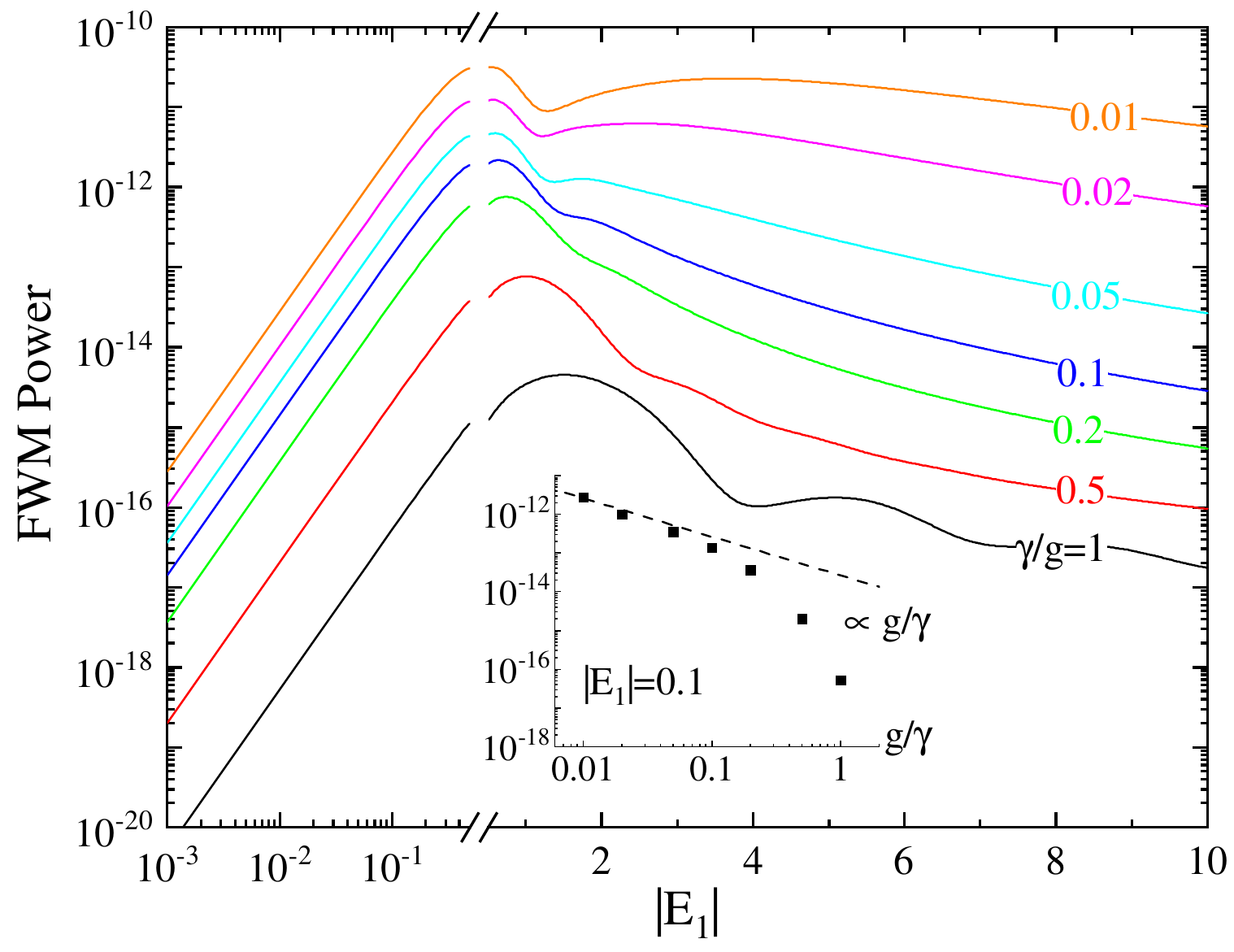}
	\caption{As \Fig{fig:E12powGcomp}, but for varying $|E_1|$, with $|E_2|=0.001$. }
	\label{fig:E1powGcomp}
\end{figure}

\begin{figure}
	\centering
	\includegraphics[width=0.7\columnwidth]{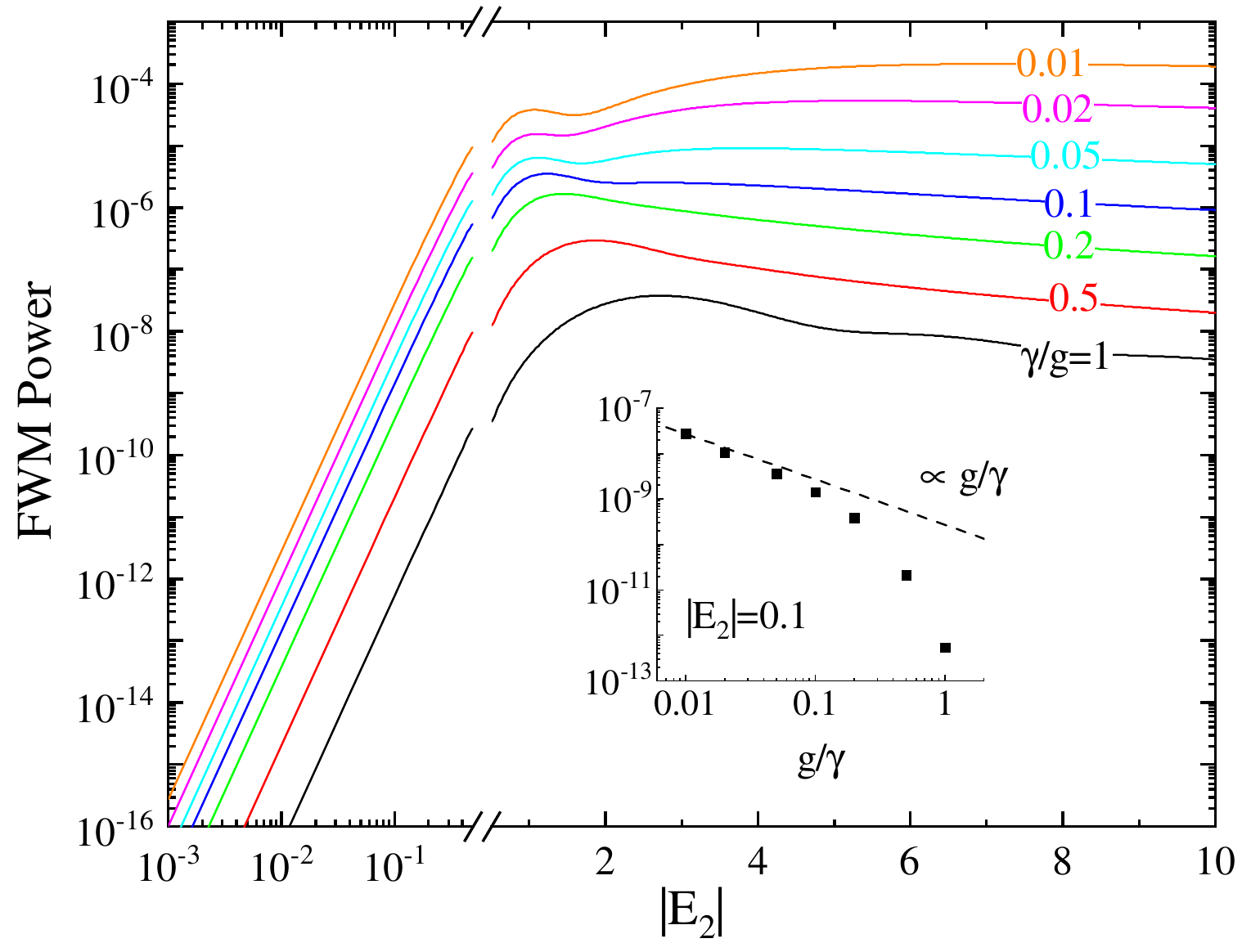}
	\caption{As \Fig{fig:E12powGcomp}, but for varying $|E_2|$, with $|E_1|=0.001$.}
	\label{fig:E2powGcomp}
\end{figure}

The scaling of the FWM power, $\int_0^\infty |P(t)|^2 dt$,  versus pulse area $|E|=|E_2|=|E_1|$ is shown in \Fig{fig:E12powGcomp} for zero detuning and zero delay between pulses. In the perturbative (i.e. low-excitation) regime, the expected scaling $\propto |E|^6$ is observed, and in the high pulse area regime a saturation of the power is seen.  The scaling of the FWM power versus pulse area $|E_1|$ with a fixed $|E_2|=0.001$ is shown in \Fig{fig:E1powGcomp}. In the perturbative regime, the expected scaling $\propto |E_1|^2$ is observed. In the high pulse area regime, a reduction of the power is observed, and for $\gamma/g$ of the order of one, Rabi-oscillations are seen.
The scaling of the FWM power versus pulse area $|E_2|$ with a fixed $|E_1|=0.001$ is shown in \Fig{fig:E2powGcomp}. In the perturbative regime, the expected scaling $\propto |E_2|^4$ is observed.
In the high pulse area regime, the behaviour is somewhat similar to the case of changing $|E_1|$. However, almost no Rabi oscillations are seen, which is different from the previous case.

\clearpage
\section{More results on the FWM spectra}\label{sec:moreresults}

This section contains a collection of results similar to Fig.\,1 of the main text, investigating the effect of varying system parameters on transition amplitudes and FWM polarization, presented to provide the reader with a broader picture of possible responses. We vary, in particular, the exciton and cavity dampings, which are assumed equal, $\gamma_X=\gamma_C$, and taking values of $g$, $g/5$, and $g/20$. We also vary the detuning $\delta =\Omega_X-\Omega_C$ between the exciton and cavity-mode transition frequencies: results are shown for $\delta=0$ and $g$. Finally, for each parameters set, we vary the pulse strength $|E_1|$ while keeping $|E_2|$ small, or $|E_2|$ while keeping $|E_1|$ small, or both using $|E_1|=|E_2|$.



The resulting 18 figures listed in \Tab{tab:sim} show a number of effects on the FWM polarization.
By varying the excitation pulse strength within each figure, we demonstrate a formation of the QMQ for each excitation condition, damping and detuning. By reducing the damping we demonstrate a fine structure, both in the inner and outer doublets. By increasing the damping up to the values of critical damping $\gamma_C=g$, we show how the QMQ gradually disappears. This happens not only because of the spectral broadening but also due to a population relaxation down the ladder, which reduces the outer doublet splitting. Varying the detuning from 0 to $g$ results in spectral asymmetry: the low-intensity strong-coupling doublet shifts towards the positive frequency, and the shape of the QMQ changes.




\setlength{\tabcolsep}{10pt} 

\begin{table}
		\renewcommand*{\arraystretch}{1.1}
	\begin{tabular}{l|l|l|l|l}
		\hline
		Figure & $\delta/g$ & $\gamma_C/g$ & $|E_1|$ & $|E_2|$ \\
		\hline
		\Fig{fig:e1d0g1} & 0 & 1 & 0-10 & 0.001\\
		Fig\,1 & 0 & 1/2 & 0-10 & 0.001\\
		\Fig{fig:e1d0g5} & 0 & 1/5 & 0-10 & 0.001\\
		\Fig{fig:e1d0g20} & 0 & 1/20 & 0-10 & 0.001\\
		\hline
		\Fig{fig:e1d1g1} & 1 & 1 & 0-10 & 0.001\\
		\Fig{fig:e1d1g5} & 1 & 1/5 & 0-10 & 0.001\\
		\Fig{fig:e1d1g20} & 1 & 1/20 & 0-10 & 0.001\\
		\hline
		\Fig{fig:e2d0g1} & 0 & 1  & 0.001 & 0-10\\
		\Fig{fig:e2d0g5} & 0 & 1/5  & 0.001 & 0-10\\
		\Fig{fig:e2d0g20} & 0 & 1/20  & 0.001 & 0-10\\
		\hline
		\Fig{fig:e2d1g1} & 1 & 1  & 0.001 & 0-10\\
		\Fig{fig:e2d1g5} & 1 & 1/5  & 0.001 & 0-10\\
		\Fig{fig:e2d1g20} & 1 & 1/20  & 0.001 & 0-10\\
		\hline
		\Fig{fig:e12d0g1} & 0 & 1  & 0-10 & 0-10\\
		\Fig{fig:e12d0g5} & 0 & 1/5  & 0-10 & 0-10\\
		\Fig{fig:e12d0g20} & 0 & 1/20  & 0-10 & 0-10\\
		\hline
		\Fig{fig:e12d1g1} & 1 & 1  & 0-10 & 0-10\\
		\Fig{fig:e12d1g5} & 1 & 1/5  & 0-10 & 0-10\\
		\Fig{fig:e12d1g20} & 1 & 1/20  & 0-10 & 0-10\\
		\hline
	\end{tabular}
	\caption{Overview of available simulation results.}
	\label{tab:sim}
\end{table}


\begin{figure}[H]
	\centering
	\includegraphics[width=0.8\textwidth]{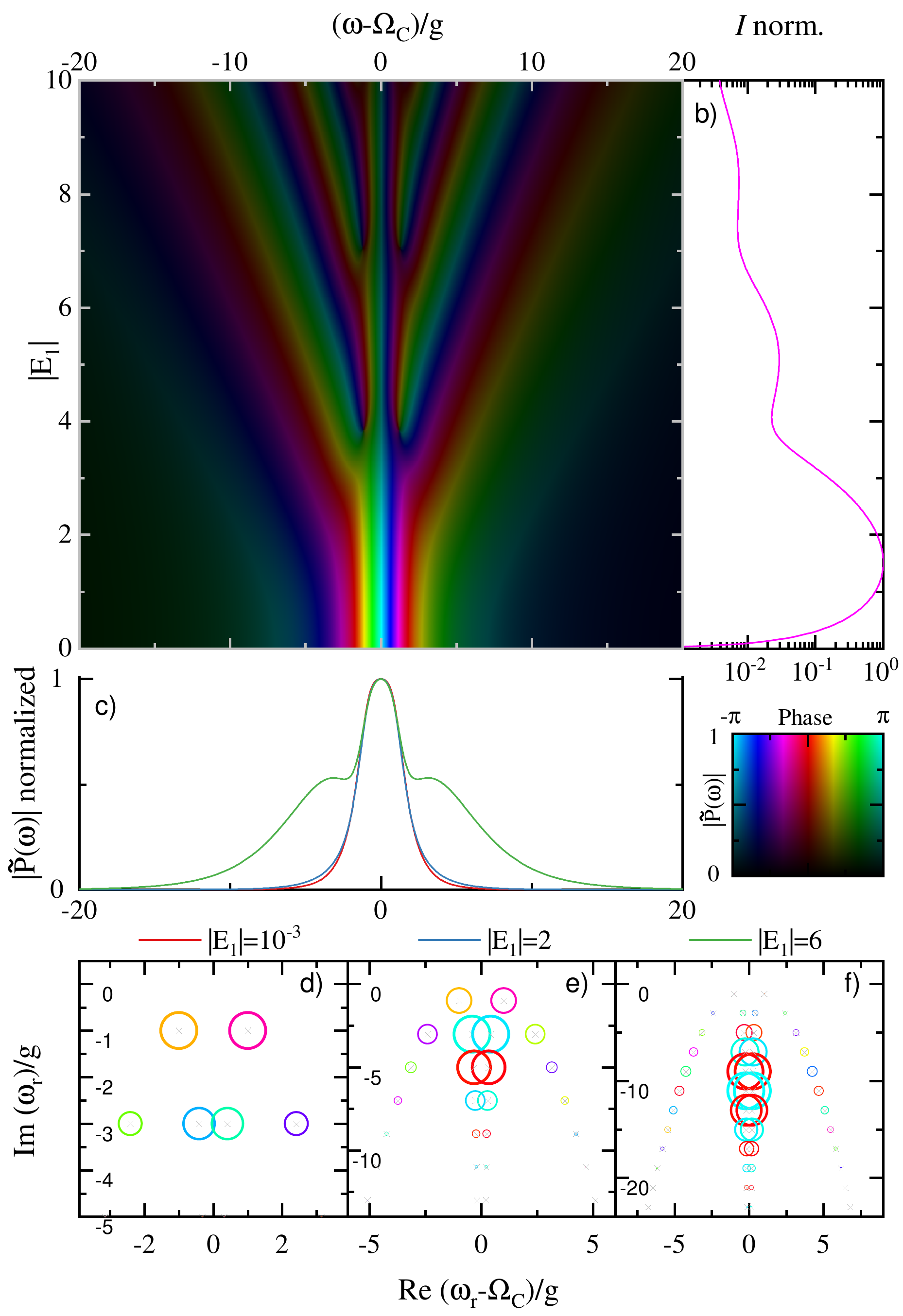}
	\caption{As Fig.\,1, with alternate parameters $\delta=0$, $\gamma_C=g$.}
	\label{fig:e1d0g1}
\end{figure}

\begin{figure}[H]
	\centering
	\includegraphics[width=0.8\textwidth]{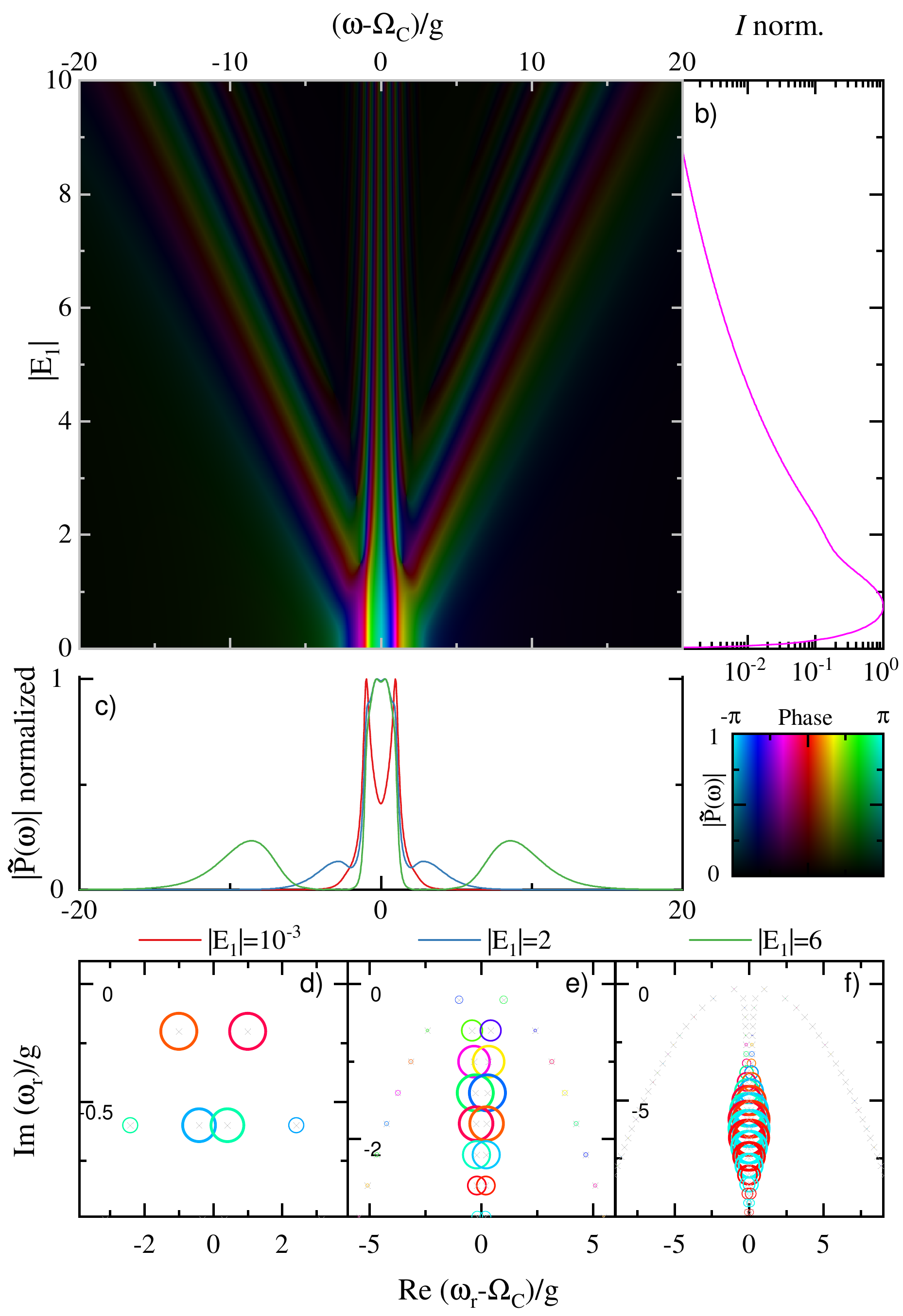}
	\caption{As Fig.\,1, with alternate parameters $\delta=0$, $\gamma_C=g/5$.}
	\label{fig:e1d0g5}
\end{figure}

\begin{figure}[H]
	\centering
	\includegraphics[width=0.8\textwidth]{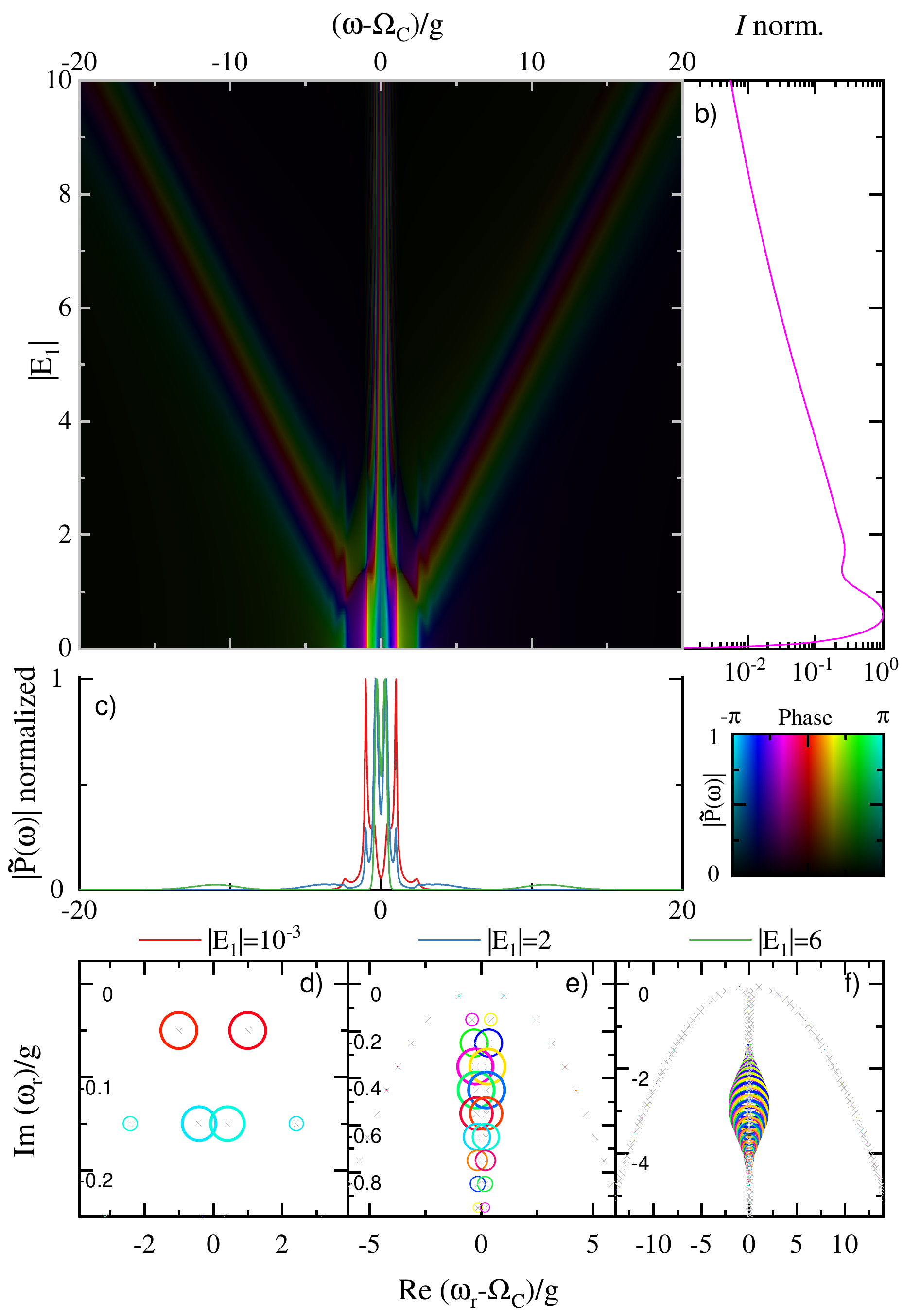}
	\caption{As Fig.\,1, with alternate parameters $\delta=0$, $\gamma_C=g/20$.}
	\label{fig:e1d0g20}
\end{figure}

\begin{figure}[H]
	\centering
	\includegraphics[width=0.8\textwidth]{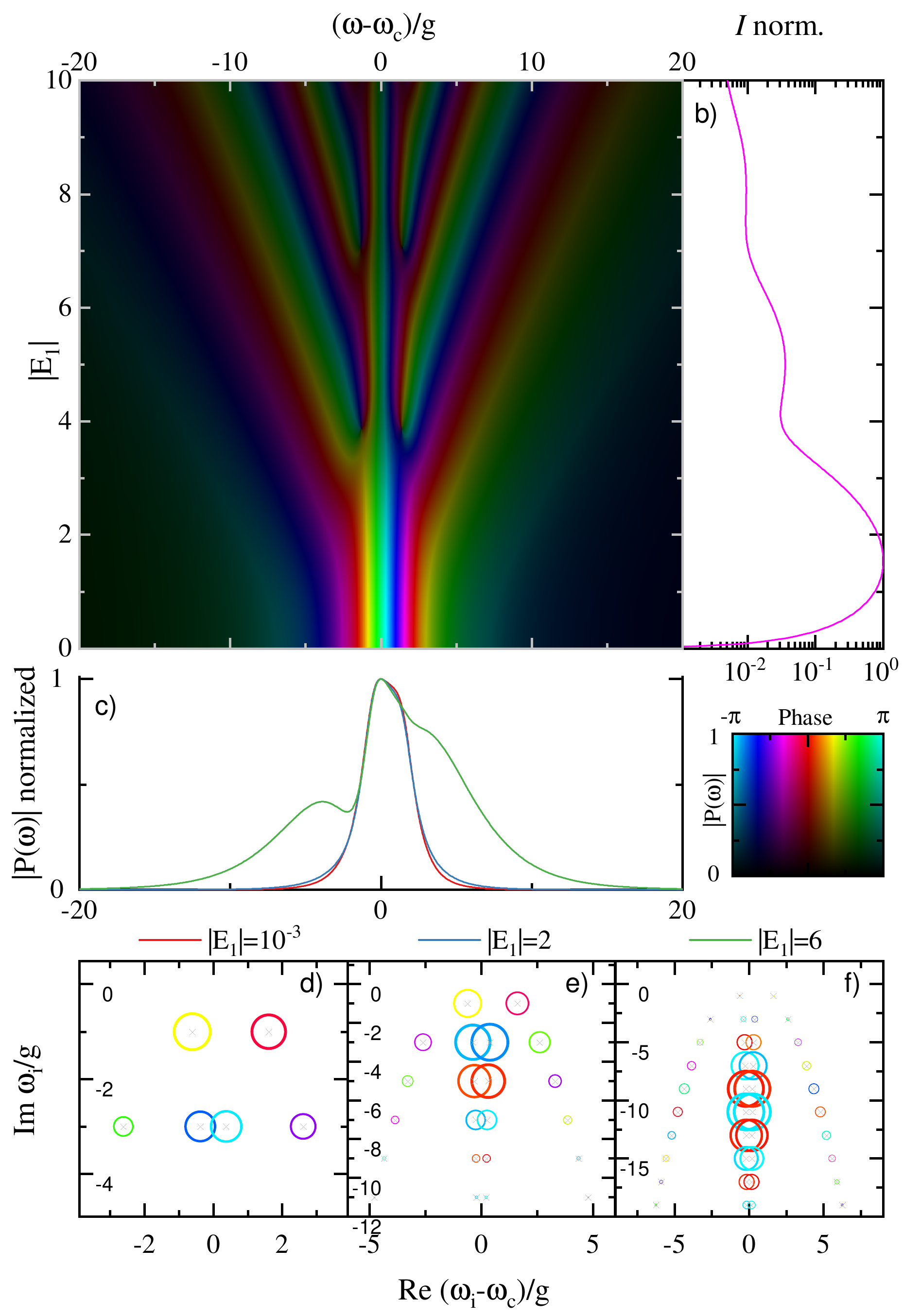}
	\caption{As Fig.\,1, with alternate parameters $\delta=g$, $\gamma_C=g$.}
	\label{fig:e1d1g1}
\end{figure}

\begin{figure}[H]
	\centering
	\includegraphics[width=0.8\textwidth]{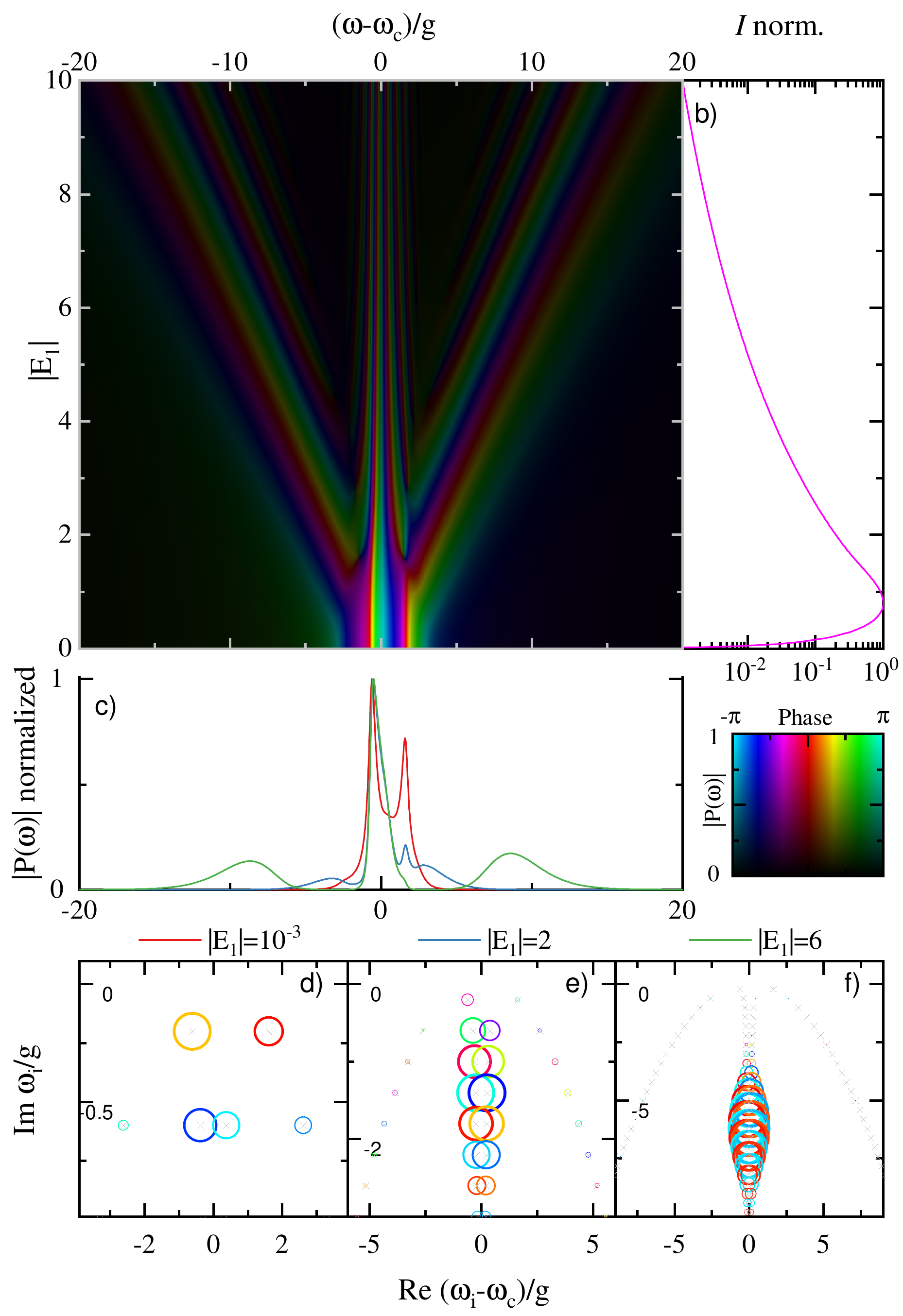}
	\caption{As Fig.\,1, with alternate parameters $\delta=g$, $\gamma_C=g/5$.}
	\label{fig:e1d1g5}
\end{figure}

\begin{figure}[H]
	\centering
	\includegraphics[width=0.8\textwidth]{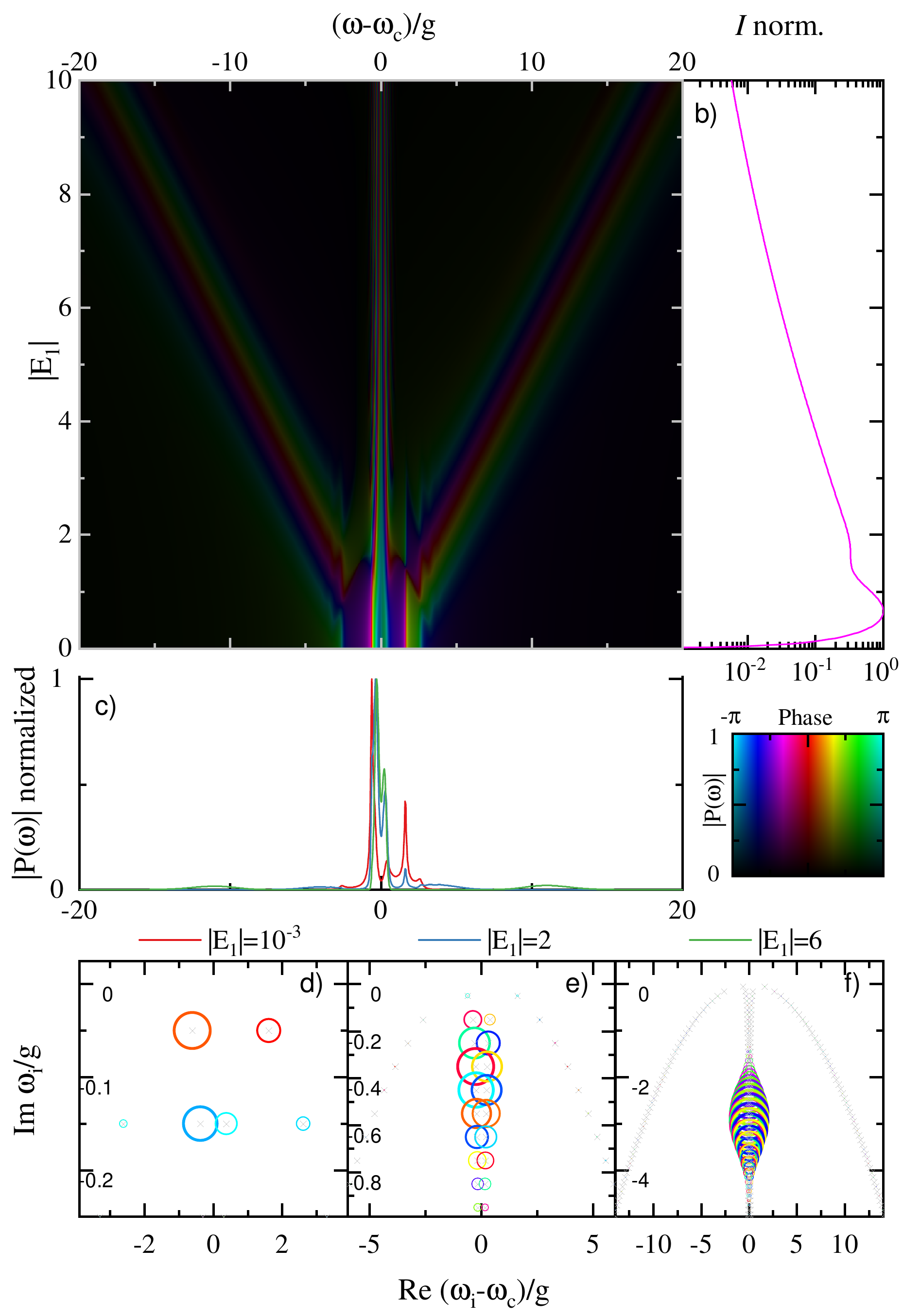}
	\caption{As Fig.\,1, with alternate parameters $\delta=g$, $\gamma_C=g/20$.}
	\label{fig:e1d1g20}
\end{figure}

\begin{figure}[H]
	\centering
	\includegraphics[width=0.8\textwidth]{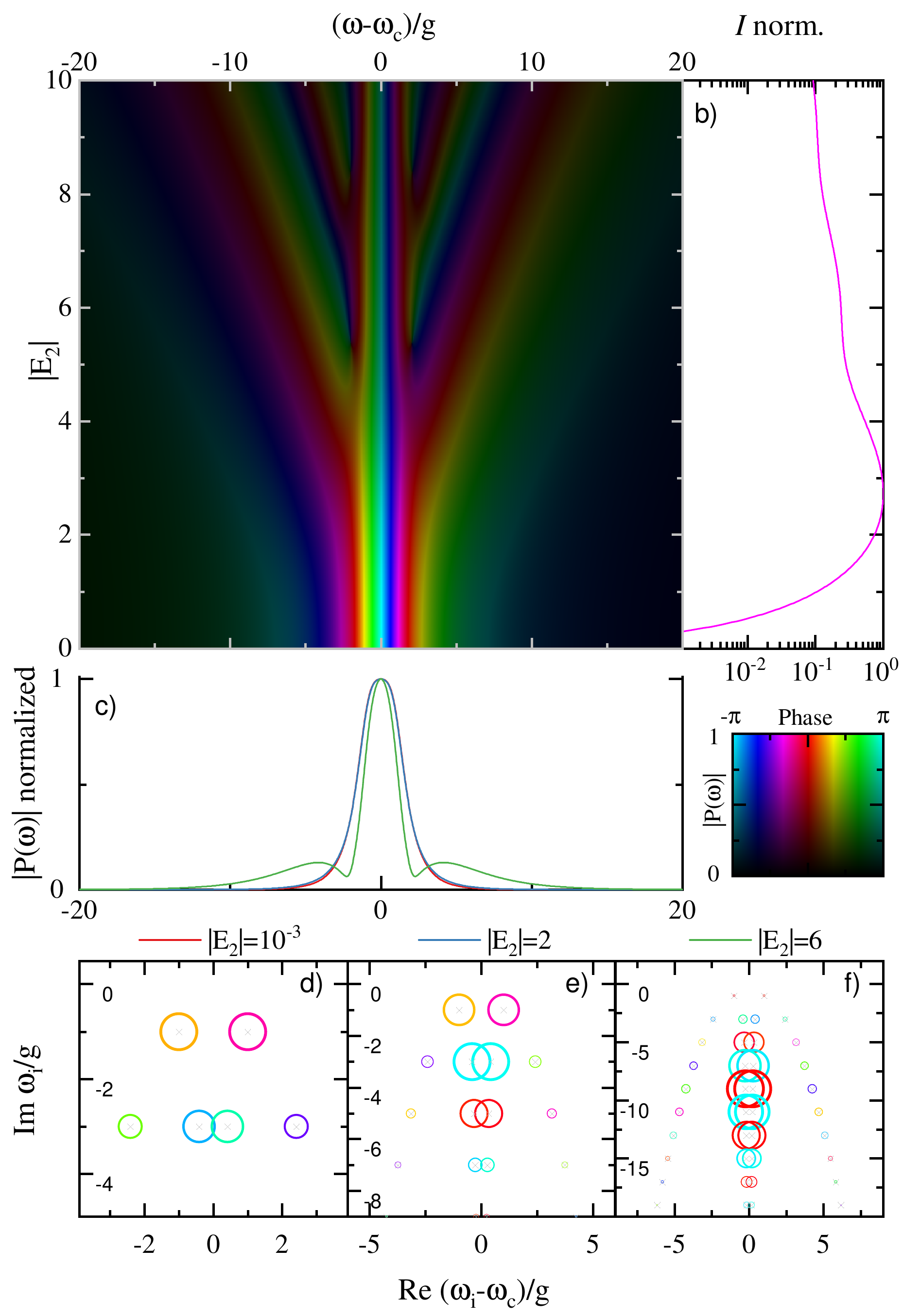}
	\caption{As Fig.\,1 with alternate parameters $\delta=0$, $\gamma_C=g$, $|E_1|=0.001$ and varying $|E_2|$.}
	\label{fig:e2d0g1}
\end{figure}

\begin{figure}[H]
	\centering
	\includegraphics[width=0.8\textwidth]{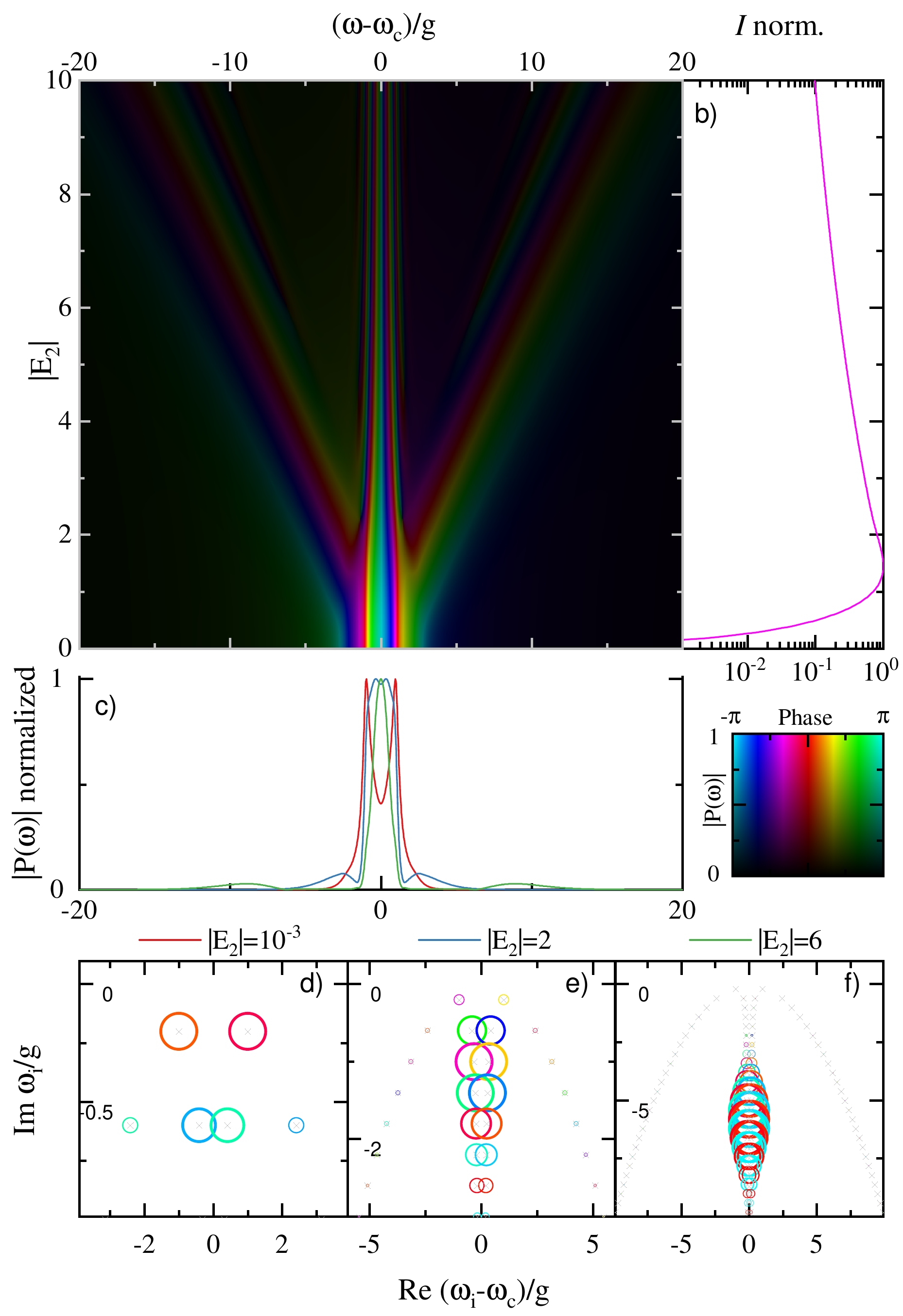}
	\caption{As \Fig{fig:e2d0g1} with alternate parameters  $\delta=0$, $\gamma_C=g/5$.}
	\label{fig:e2d0g5}
\end{figure}

\begin{figure}[H]
	\centering
	\includegraphics[width=0.8\textwidth]{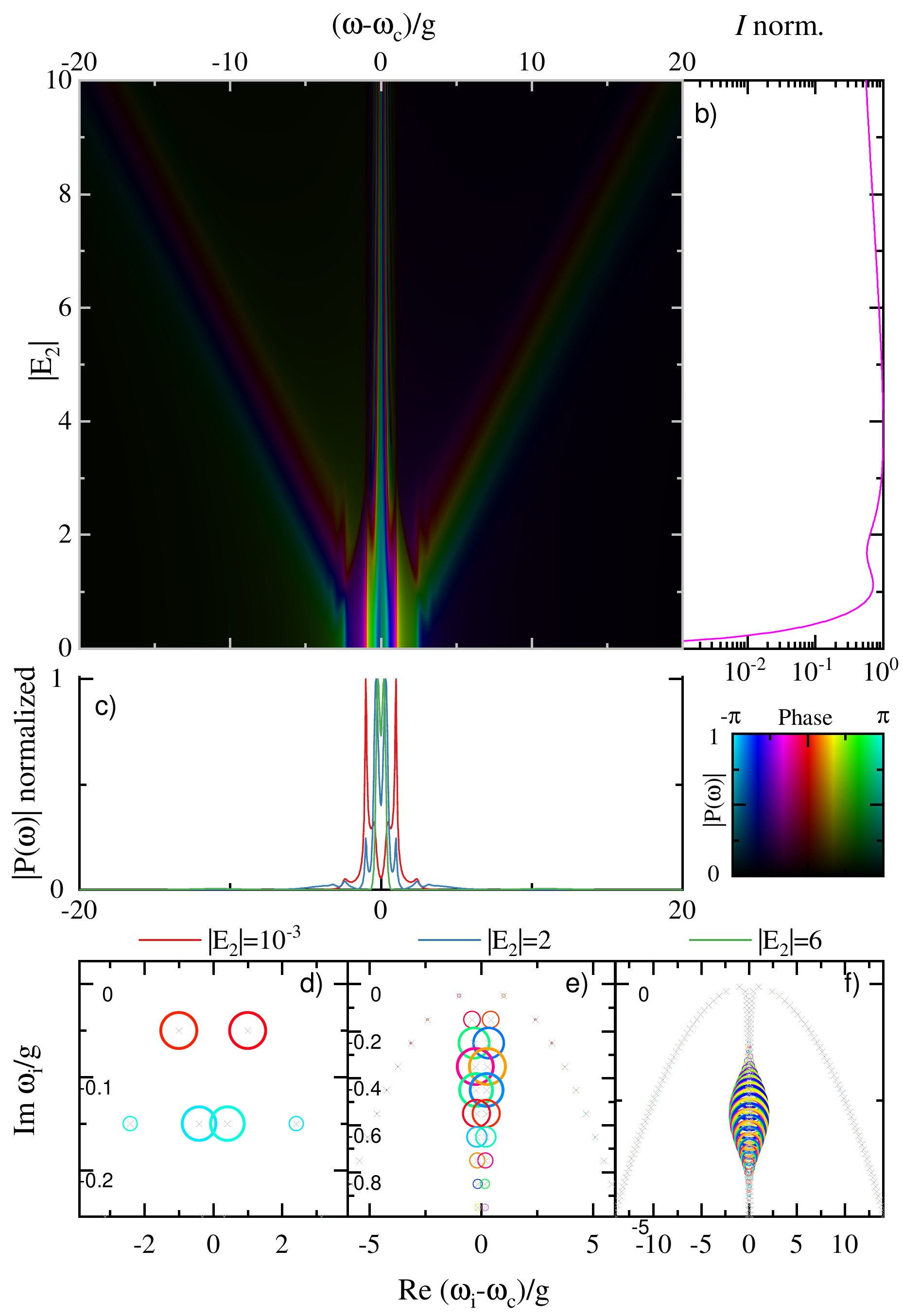}
	\caption{As \Fig{fig:e2d0g1} with alternate parameters $\delta=0$, $\gamma_C=g/20$.}
	\label{fig:e2d0g20}
\end{figure}

\begin{figure}[H]
	\centering
\includegraphics[width=0.8\textwidth]{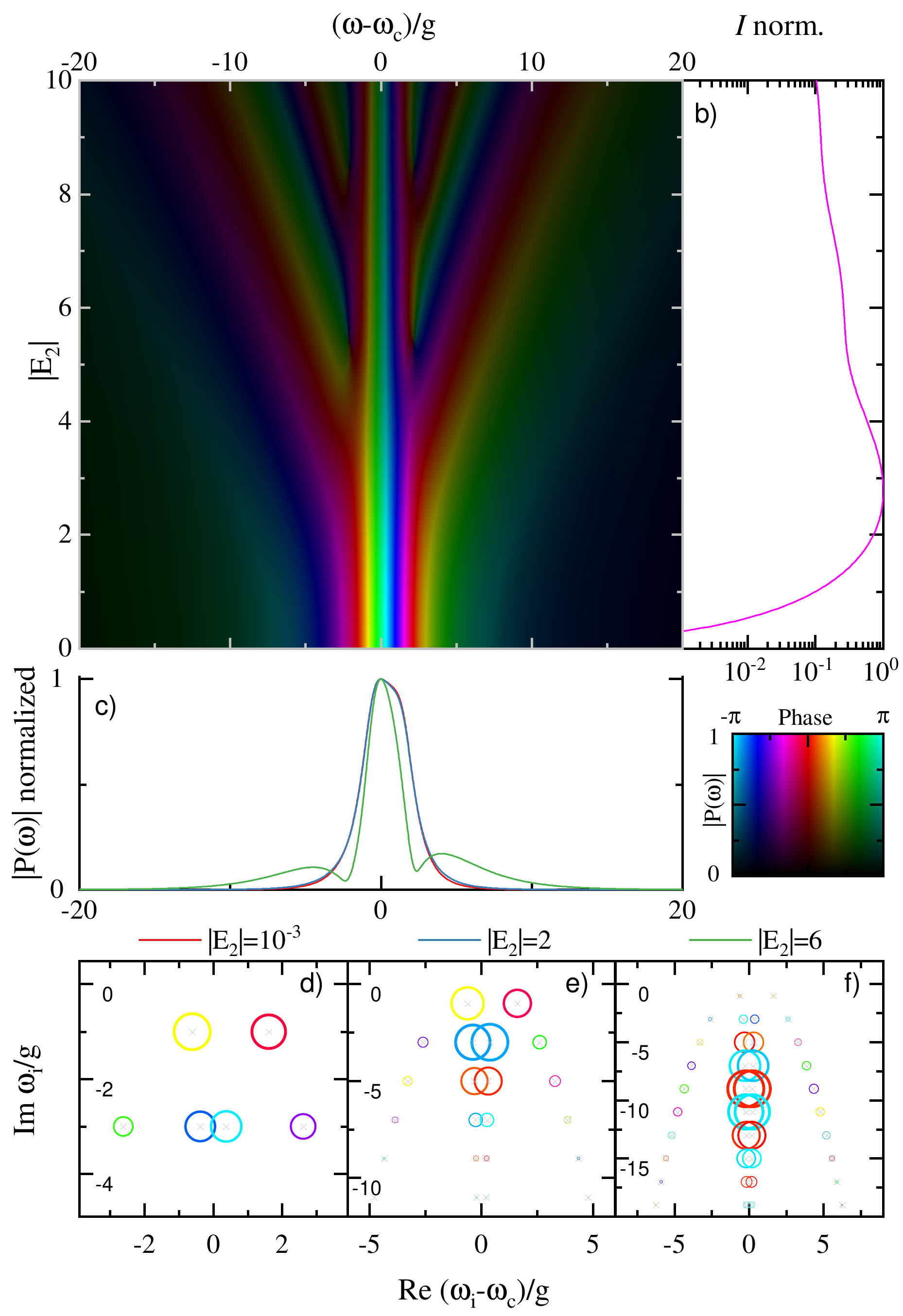}
	\caption{As \Fig{fig:e2d0g1} with alternate parameters $\delta=g$, $\gamma_C=g$.}
\label{fig:e2d1g1}
\end{figure}

\begin{figure}[H]
	\centering
	\includegraphics[width=0.8\textwidth]{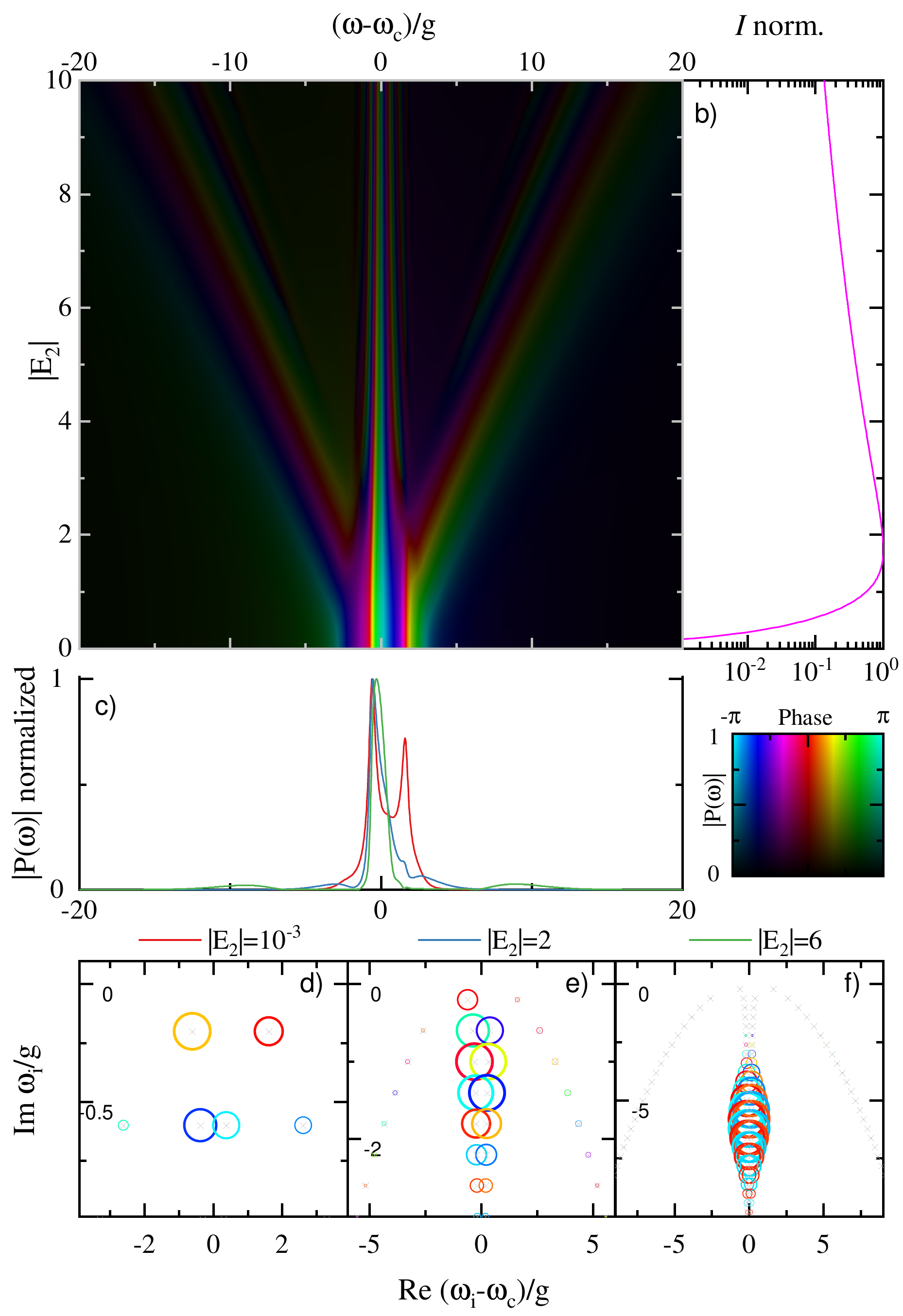}
	\caption{As \Fig{fig:e2d0g1} with alternate parameters $\delta=g$, $\gamma_C=g/5$.}
	\label{fig:e2d1g5}
\end{figure}

\begin{figure}[H]
	\centering
	\includegraphics[width=0.8\textwidth]{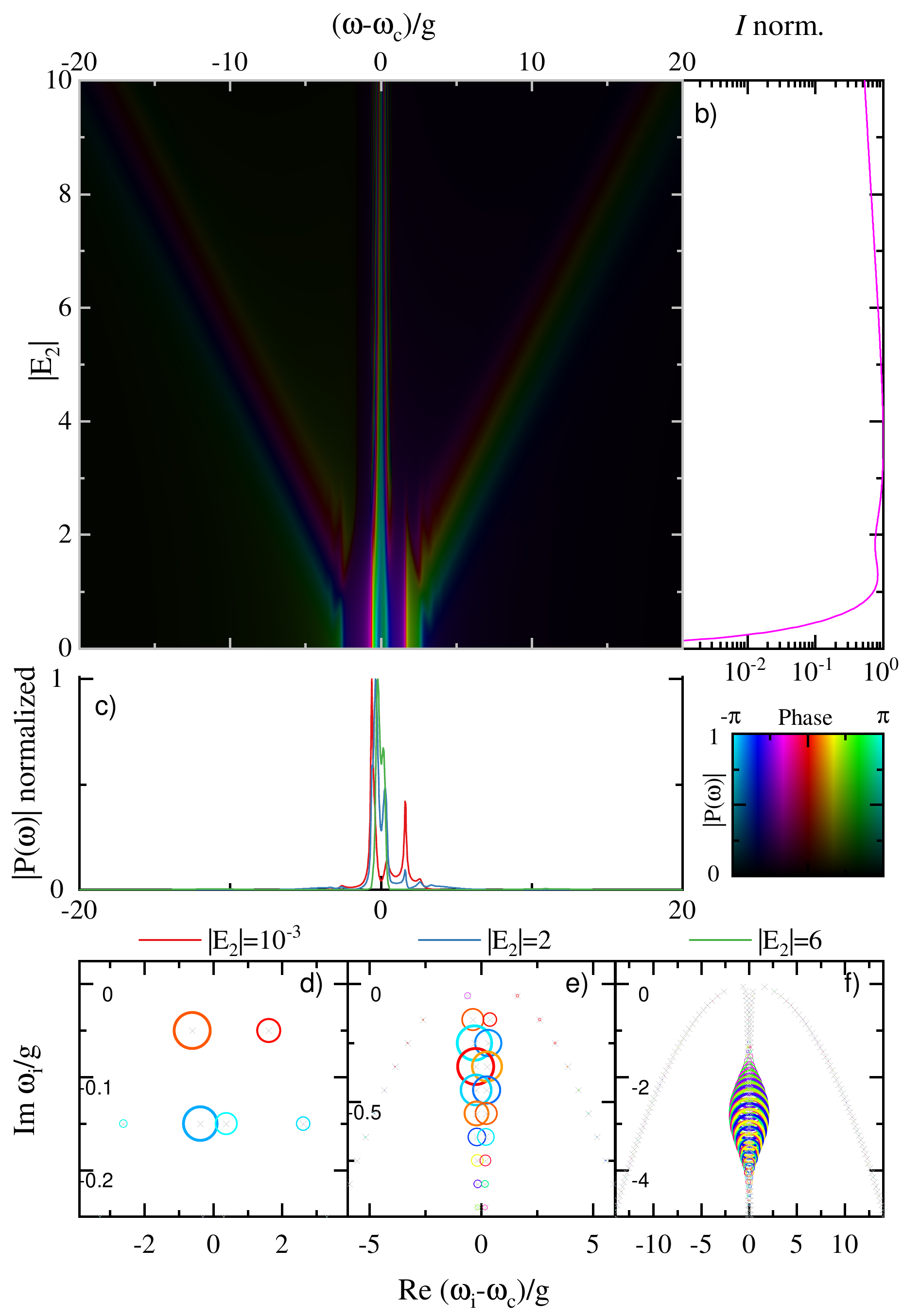}
	\caption{As \Fig{fig:e2d0g1} with alternate parameters $\delta=g$, $\gamma_C=g/20$.}
	\label{fig:e2d1g20}
\end{figure}

\begin{figure}[H]
	\centering
	\includegraphics[width=0.8\textwidth]{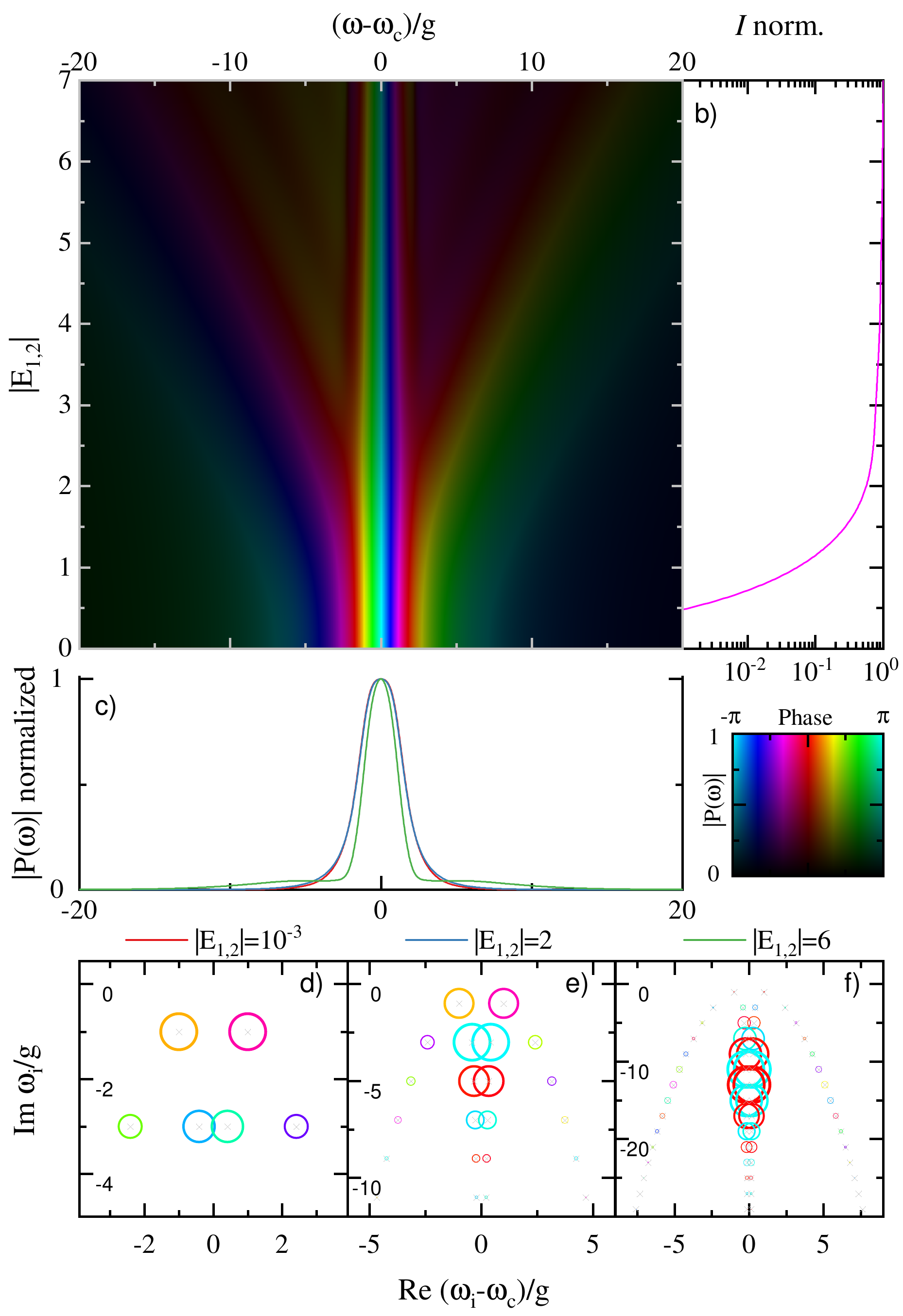}
	\caption{As Fig.\,1 with alternate parameters $\delta=0$, $\gamma_C=g$, and varying $|E_1|=|E_2|$.}
	\label{fig:e12d0g1}
\end{figure}

\begin{figure}[H]
	\centering
	\includegraphics[width=0.8\textwidth]{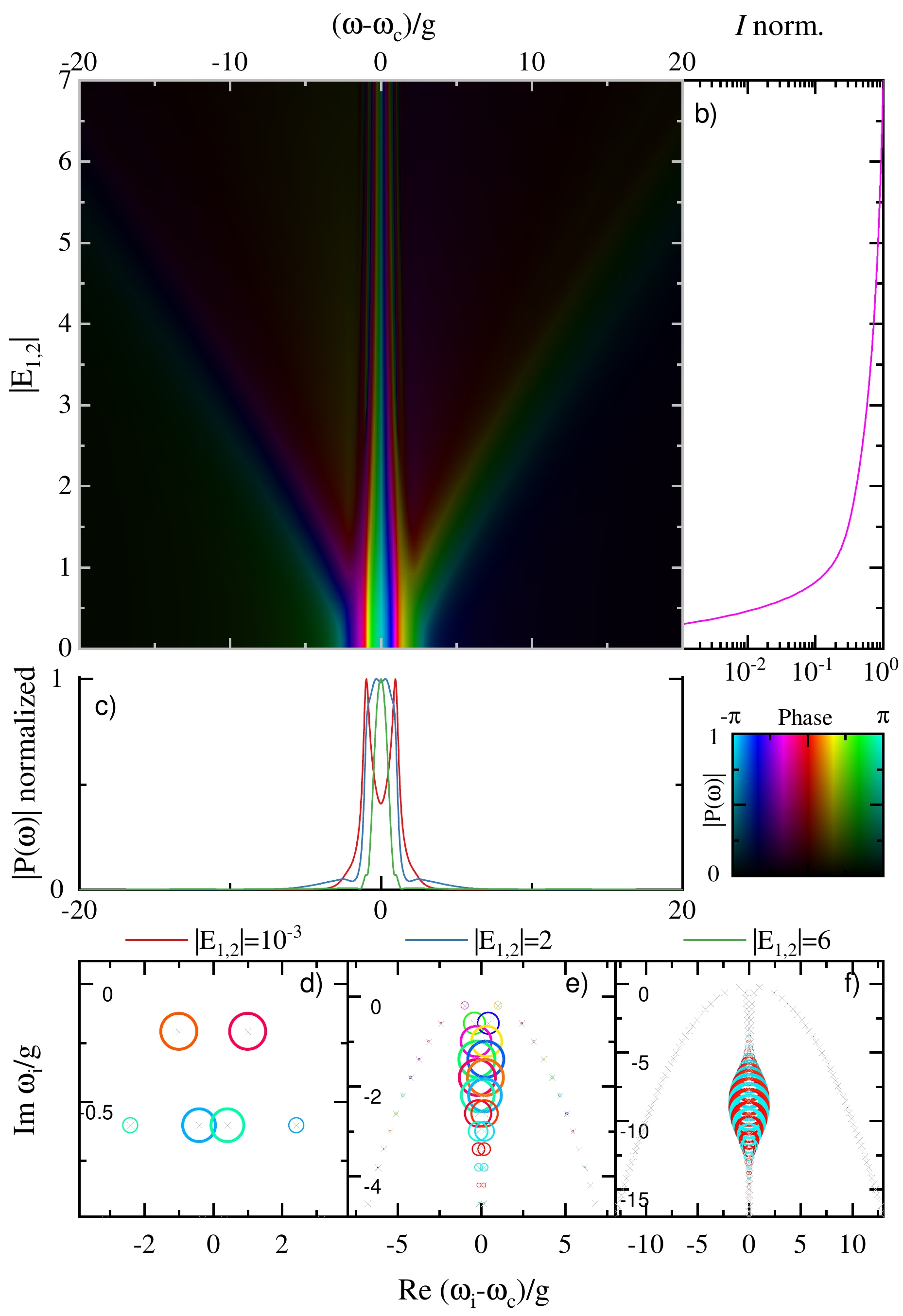}
	\caption{As \Fig{fig:e12d0g1} with alternate parameters $\delta=0$, $\gamma_C=g/5$.}
	\label{fig:e12d0g5}
\end{figure}

\begin{figure}[H]
	\centering
	\includegraphics[width=0.8\textwidth]{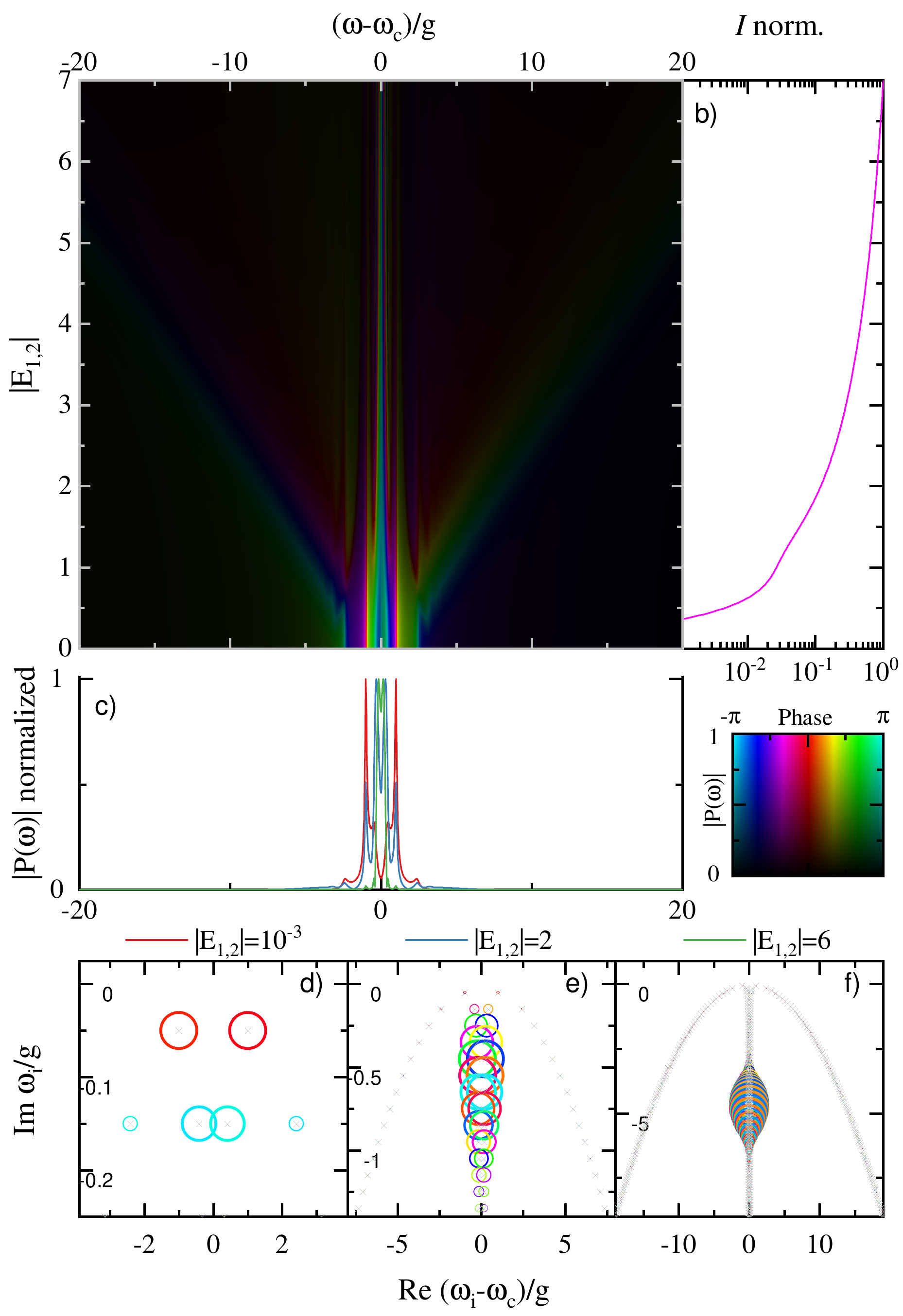}
	\caption{As \Fig{fig:e12d0g1} with alternate parameters $\delta=0$, $\gamma_C=g/20$.}
	\label{fig:e12d0g20}
\end{figure}

\begin{figure}[H]
	\centering
	\includegraphics[width=0.8\textwidth]{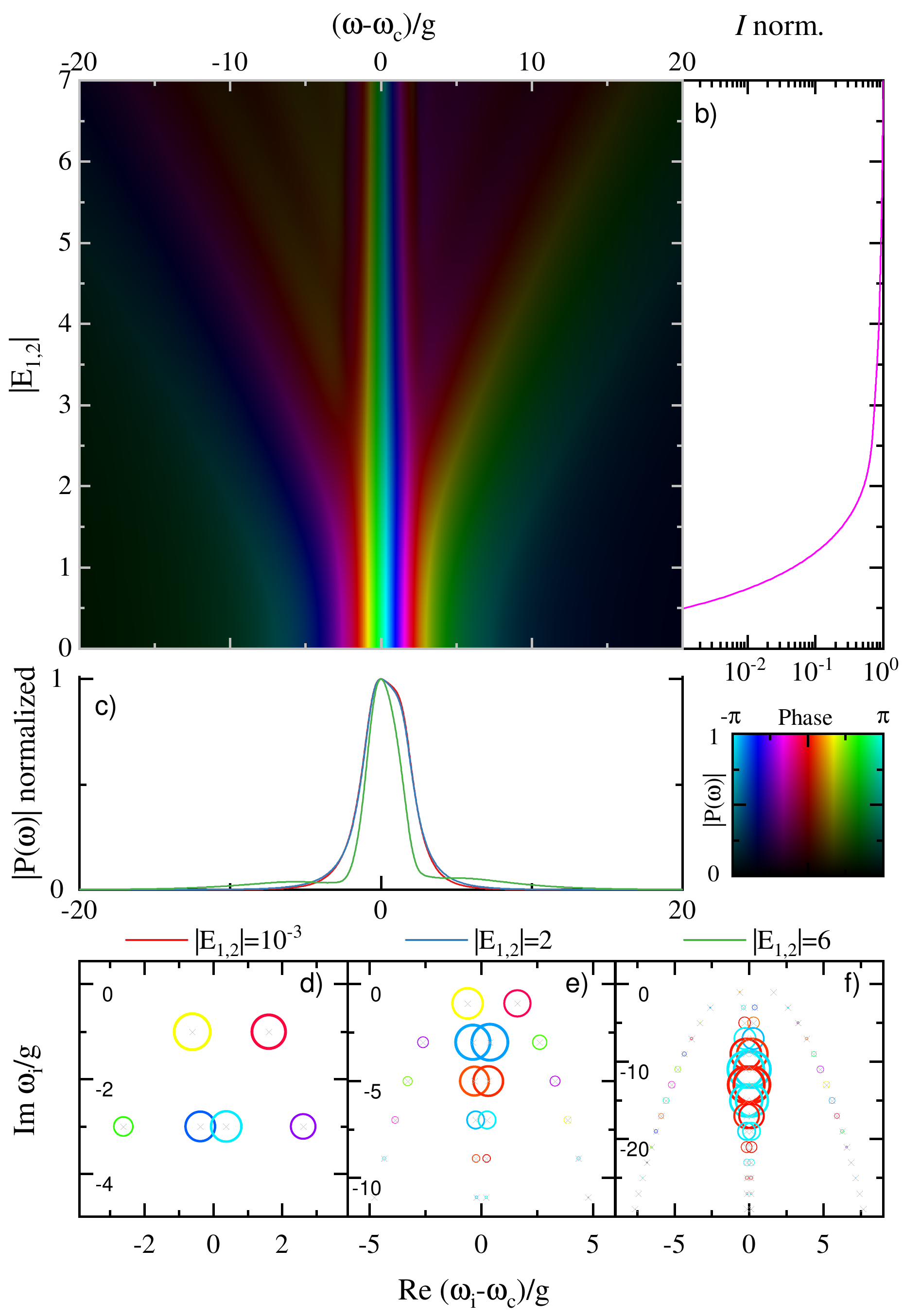}
	\caption{As \Fig{fig:e12d0g1} with alternate parameters $\delta=g$, $\gamma_C=g$.}
\label{fig:e12d1g1}
\end{figure}

\begin{figure}[H]
	\centering
	\includegraphics[width=0.8\textwidth]{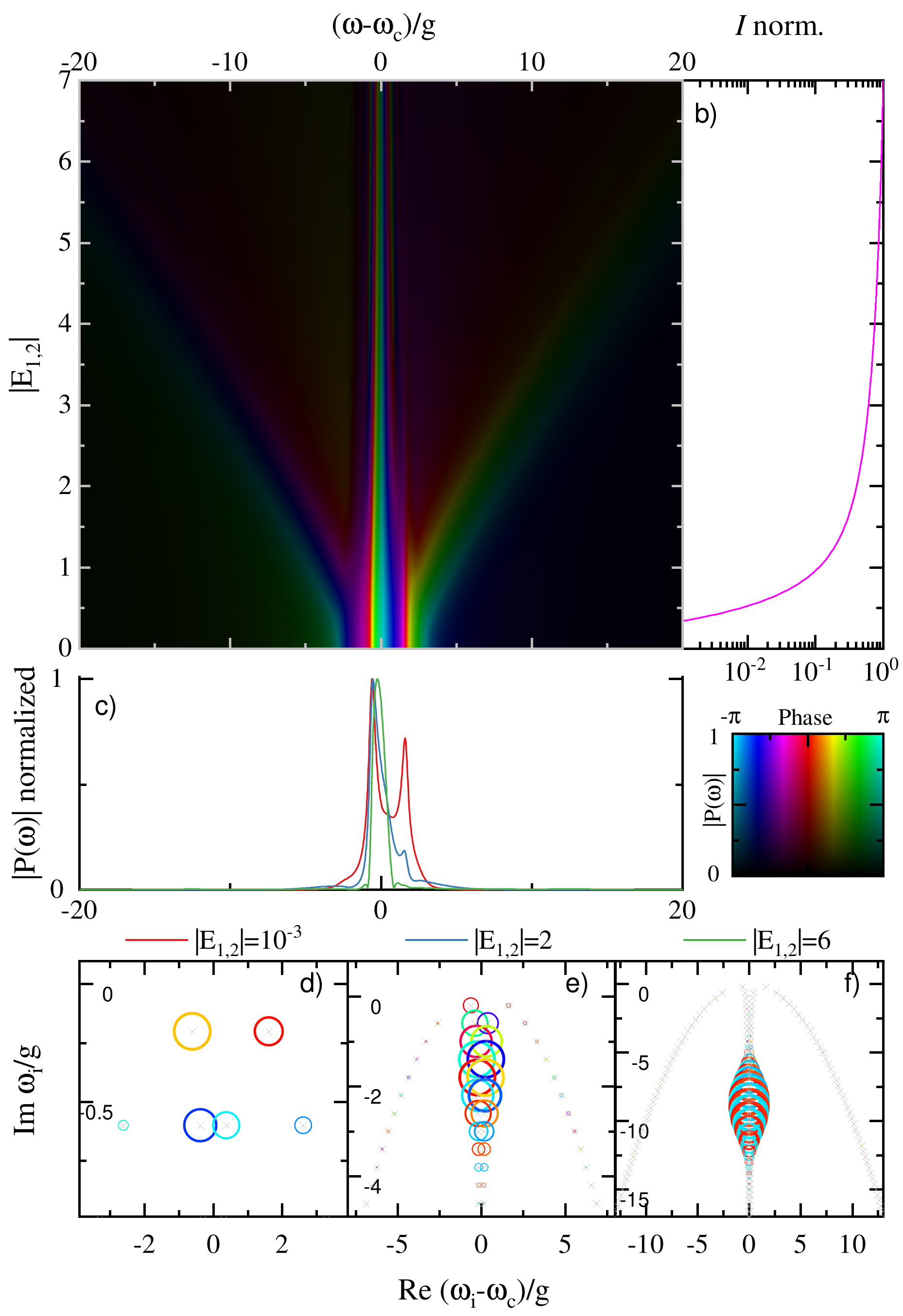}
	\caption{As \Fig{fig:e12d0g1} with alternate parameters $\delta=g$, $\gamma_C=g/5$.}
	\label{fig:e12d1g5}
\end{figure}

\begin{figure}[H]
	\centering
	\includegraphics[width=0.8\textwidth]{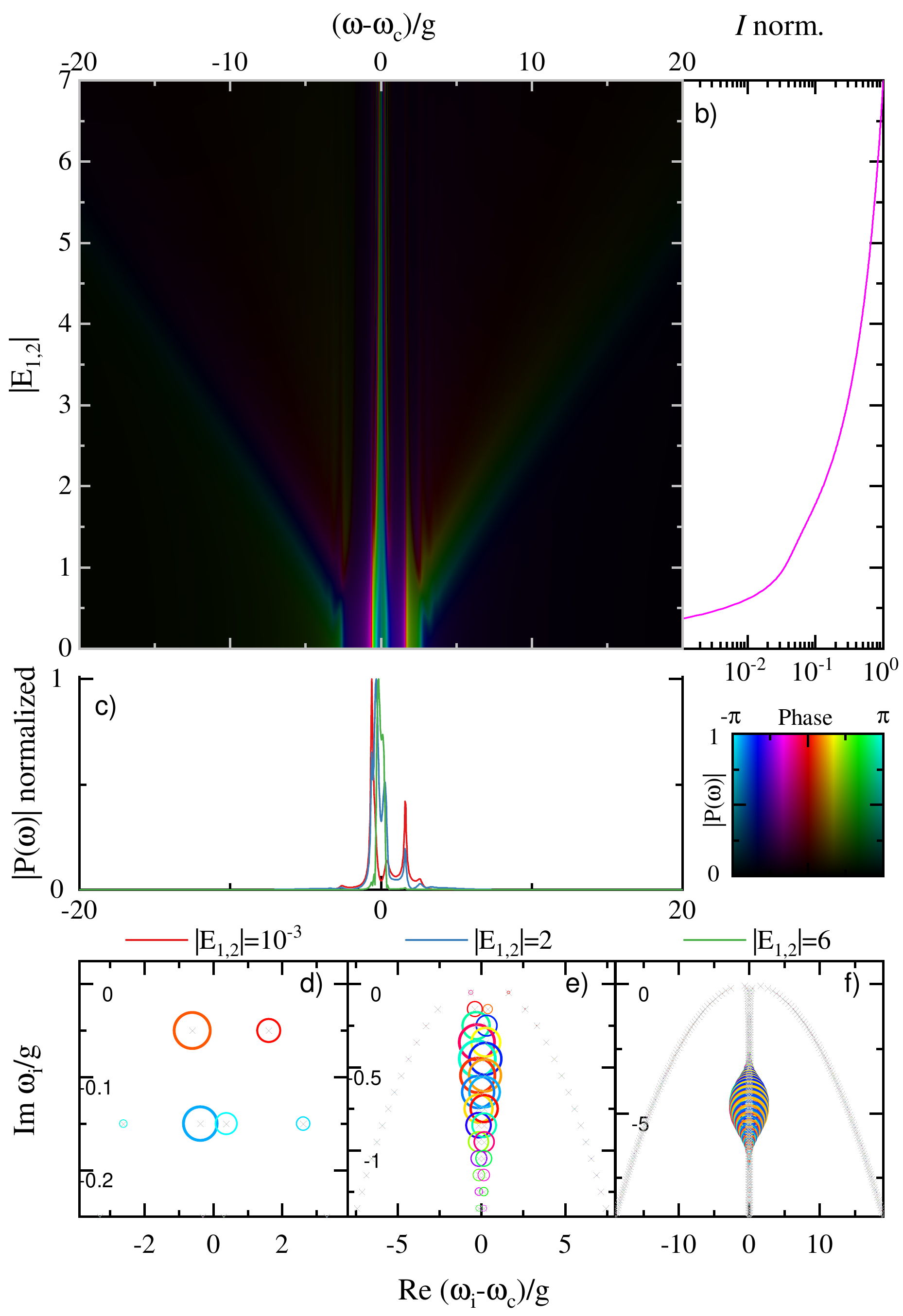}
	\caption{As \Fig{fig:e12d0g1} with alternate parameters $\delta=g$, $\gamma_C=g/20$.}
	\label{fig:e12d1g20}
\end{figure}

\clearpage
\section{Numerical Convergence and use of multiprecision}\label{sec:convergence}

In this section we discuss the numerical convergence and the need for multiprecision arithmetic when diagonalizing the Lindblad matrix for our system.

The condition number of a matrix $A$ is defined as ${\rm cond}_p(A)=\norm{A}_p\norm{A^{-1}}_p$ for any $p$-norm $\norm{A}_p$, and provides an estimate of lost precision when used in numerical calculations. Specifically, $\log_{2(10)}({\rm cond}(A))$ estimates the number of bits (digits) of numerical precision additionally required in the calculation with respect to the precision of the final result. We use the 1-norm here, which is found by taking the sum of the absolute values of the matrix elements of each column of a square matrix, then taking the maximum value of this sum.
We calculate the condition number of the eigenvectors of the Lindblad matrix, ${\rm cond}_1(U)=\norm{U}_1\norm{V}_1$ for varying rung truncation using the analytic form of $U$ and its inverse $V$ (see \Sec{Sec:Lindblad}). Our code uses 1000 bits (301 digits) of precision for all data presented in the main text and in this supplement.

\begin{figure}[H]
	\centering
	\includegraphics[width=11cm]{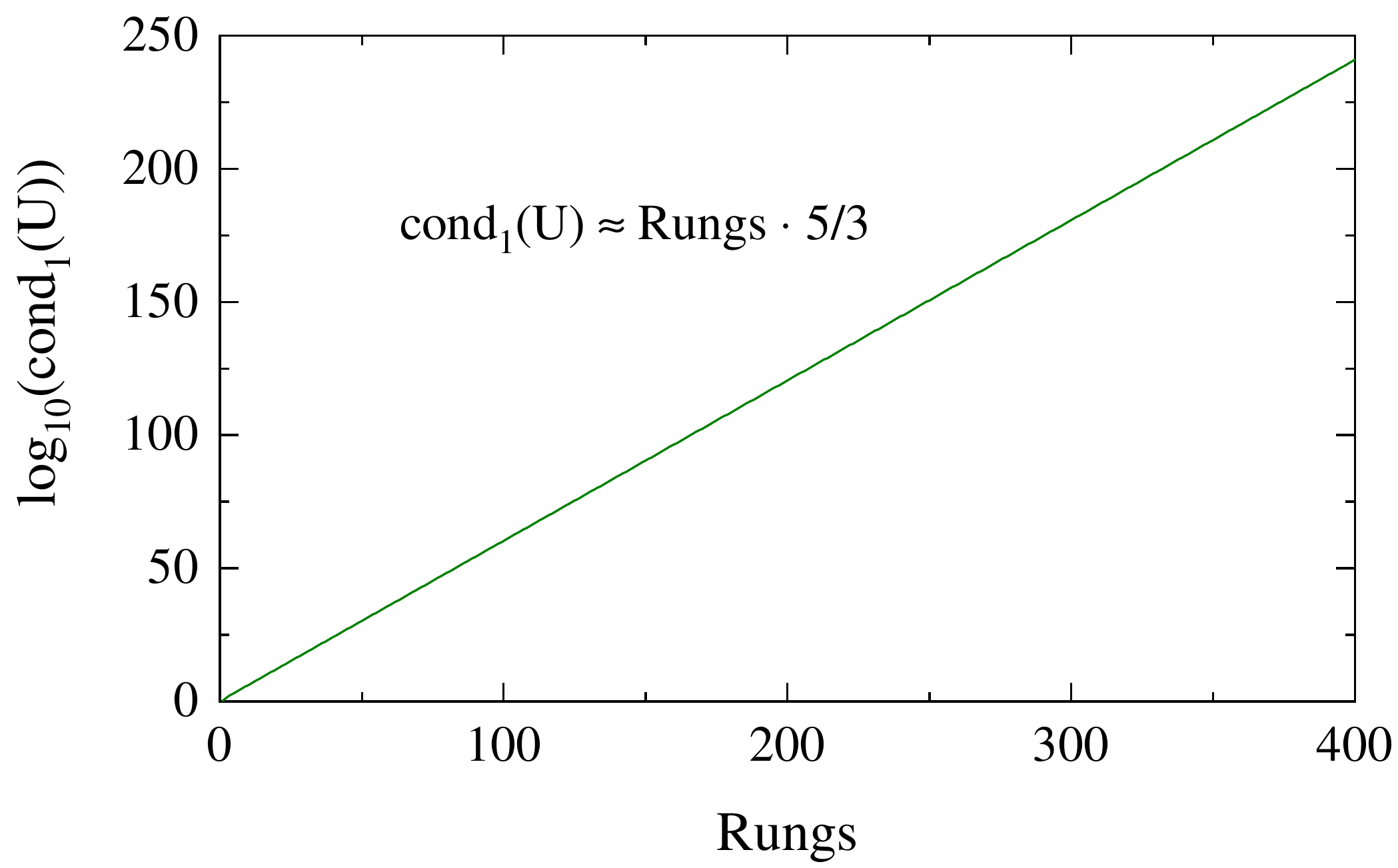}
	\caption{Upper bound for the number of additional digits required in the numerics, expressed in terms of the logarithm of the condition number, $\log_{10}({\rm cond}_1(U))$, as function of number of rungs considered. }
	\label{fig:Lcondition}
\end{figure}


The result shown in \Fig{fig:Lcondition} demonstrates that for each rung included, about 0.6 digits of extra precision are needed. Standard double precision arithmetic, which has some 15 digits of precision, is therefore expected to fail for more than 20 rungs included, and we have observed such behaviour. We thus did move to multi-precision calculations.

\begin{figure}[H]
    \centering
    \includegraphics[width=12cm]{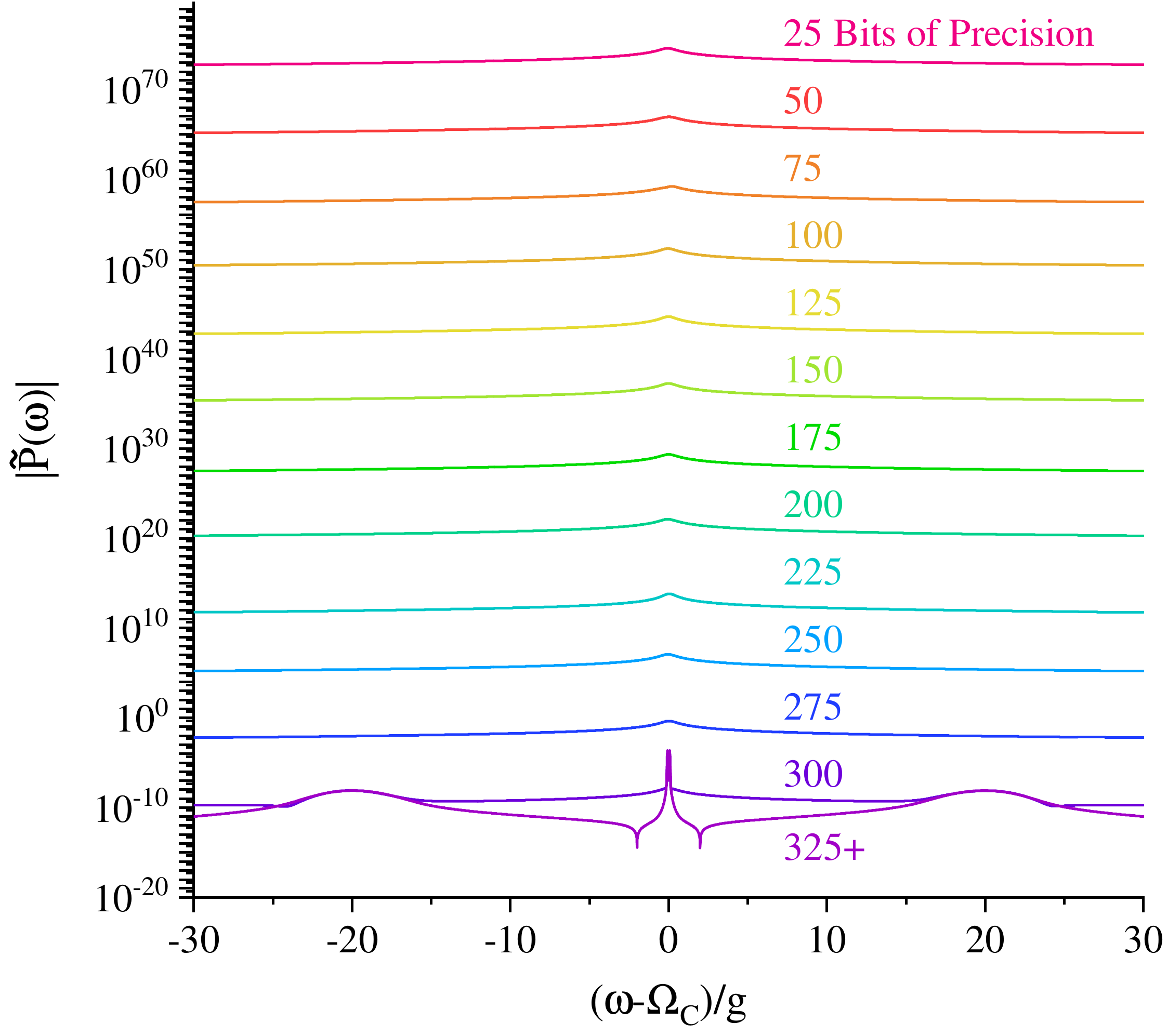}
    \caption{Calculated FWM spectrum $|\tilde{P}(\omega)|$ for $E_1=10$, $E_2=10^{-3}$, $\gamma_{C}=\gamma_{X}=10^{-3}$, $\delta=0$ and 500 rungs, as function of the number of bits precision as indicated.}
    \label{fig:bop_convergence}
\end{figure}

We show in \Fig{fig:bop_convergence} the calculated FWM polarization with $E_1=10$, $E_2=10^{-3}$, $\gamma_{C,X}=10^{-3}$, $\delta=0$ as function of the number of bits of precision. Convergence is found at $b=325$ bit precision. Notably, at lower precision the spectra are exponentially diverging proportional to $2^{-b}$ with nearly constant shape.

Assuming enough digits for numerical accuracy, the only additional limit to accuracy is the number of rungs considered. Clearly, all rungs which are significantly occupied will need to be taken into account, and this number is always more than the average number of photons, which in turn is roughly given by the square of the largest excitation pulse area.  Here we give some examples for the convergence of the \NWM\ polarization for different input parameters, and show the required number of rungs for a polarization curve to converge.

\begin{figure}
	\centering
	\includegraphics[width=16cm]{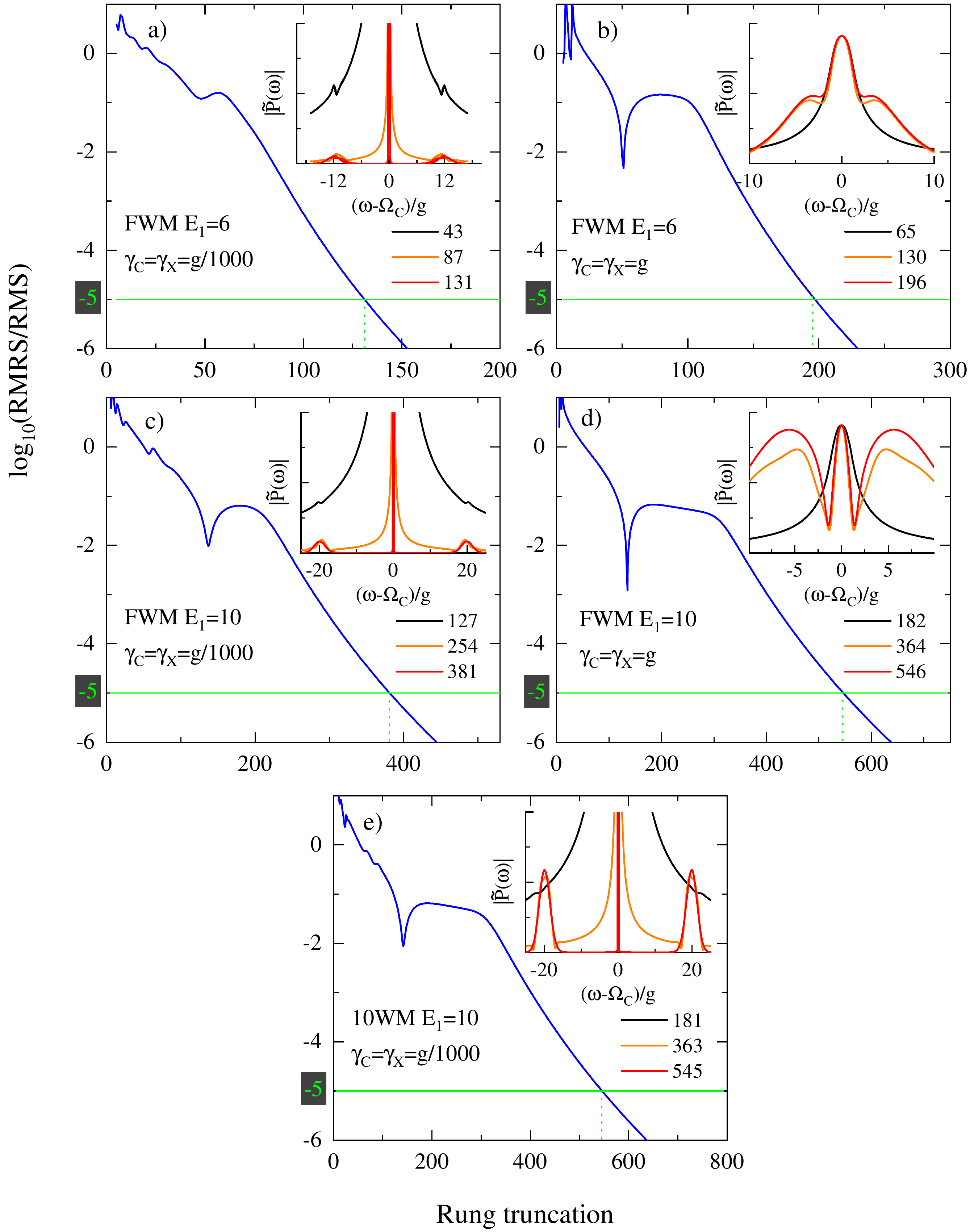}
	\caption{Error $\sigma_\eta$ versus number of rungs $\eta$. a) FWM, $|E_1|=6$,  $\gamma_C=\gamma_X=10^{-3}g$;
		b) FWM, $|E_1|=6$,  $\gamma_C=\gamma_X=g$; c) FWM, $|E_1|=10$,  $\gamma_C=\gamma_X=10^{-3}g$; d) FWM, $|E_1|=10$,  $\gamma_C=\gamma_X=g$; e) 10WM, $|E_1|=10$,  $\gamma_C=\gamma_X=10^{-3}g$. The insets show $|\tilde{P}(\omega)|$ normalized to a maximum of $1$ at three $\eta$ as given, corresponding to convergence (red), 2/3 of convergence (orange) and 1/3 of convergence (black).}
	\label{fig:convergence_main}
\end{figure}

To study the convergence we evaluate the change in the spectrum when increasing the number of rungs $\eta$ included in the calculation by one. We do this by introducing the relative root mean square error
\be \sigma_\eta=\sqrt{\frac{\int |\tilde{P}_{\eta+1}(\omega)-\tilde{P}_{\eta}(\omega)|^2 d\omega }{\int |\tilde{P}_{\eta+1}(\omega)|^2 d\omega }}\,.
\label{eqn:sigmaeta}\ee
We say the result is converged if $\sigma_\eta<10^{-5}$. Examples for the dependence of $\sigma_\eta$ on the number of rungs $\eta$ are shown in \Fig{fig:convergence_main} for FWM and 10WM, $|E_1|=6$ or 10, with $\gamma_C=\gamma_X=10^{-3}g$ or $g$, and zero detuning $\delta=0$.

We find that the error tends to a monotonous exponential decay with $\eta$, for $\eta$ above two to three times the average number of photons (given by $|E_1|^2$ here), reaching convergence at about four to five times $|E_1|^2$.
Increasing $\gamma_C$ and $\gamma_X$, and the order of the nonlinearity, the required number of rungs also increases.
The spectra shown in the insets for 1/3, 2/3 and 3/3 of the $\eta$ at convergence demonstrate a variety of behaviours depending on the parameters. We found that the inner doublet is the feature in the spectrum requiring the largest number of rungs to converge.

\begin{figure}
	\centering
	\includegraphics[width=10cm]{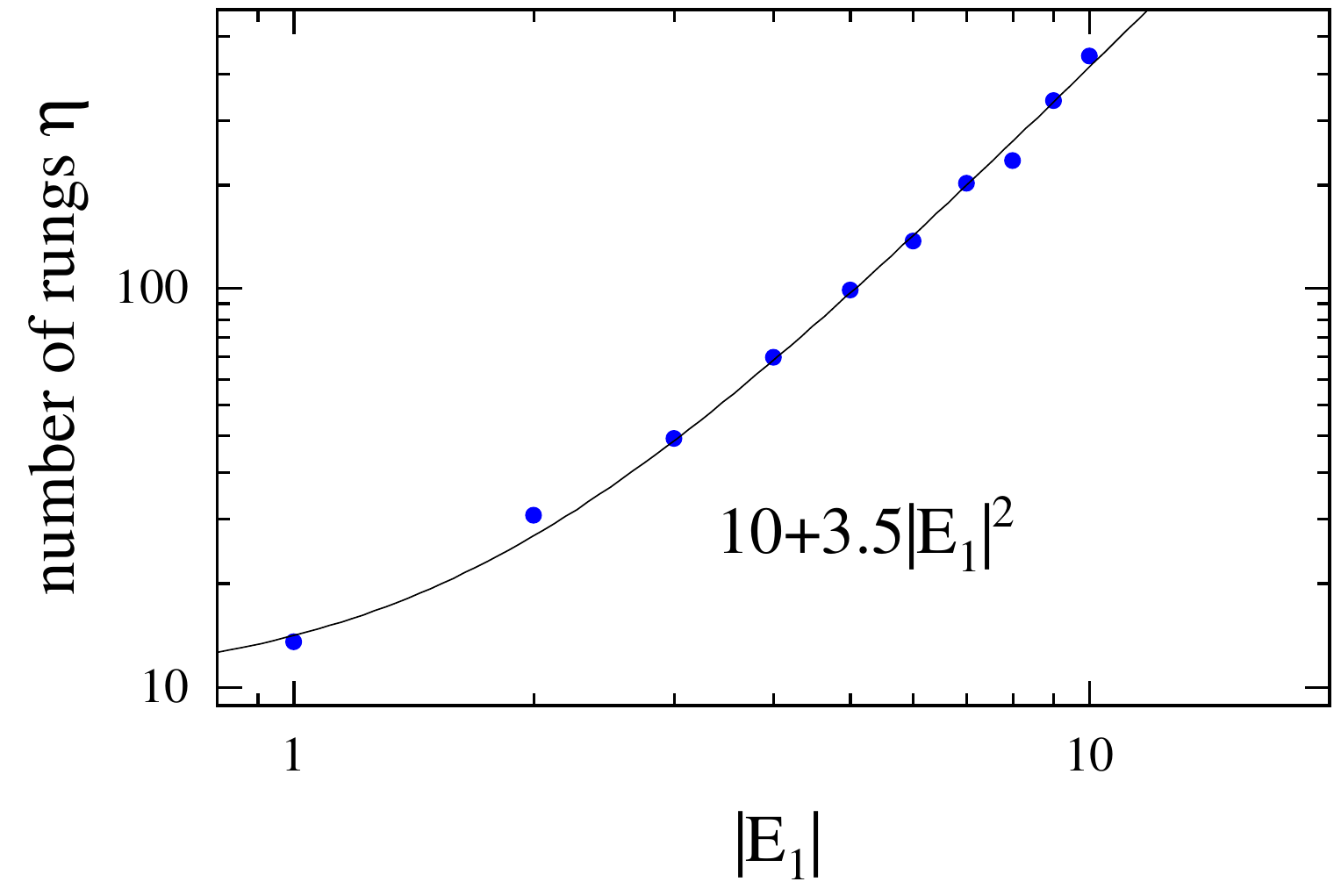}
	\caption{Number of rungs $\eta$ at which convergence for FWM is reached, as function of $|E_1|$, for $E_2=10^{-3}$, $\gamma_{C,X}=10^{-3}$, and $\delta=0$. The data (points) is described well by $10+3.5|E_1|^2$ shown as line.}
	\label{fig:convergence_lowg_rungreq}
\end{figure}

To optimize the numerical complexity of the simulations, it is important to choose a low rung number, while it has to be sufficiently high to provide convergence. For example, for Fig.\,1 we used a minimum number of rungs of 10 and maximum of 510. Interpolation over these two values for a total number of 501 curves (vertical resolution of a)) results in increasing rung truncation by 1 per curve. The absolute minimum number of rungs required for convergence in the low excitation regime $|E|\ll1$ is $1+{\cal N}/2$. \Fig{fig:convergence_lowg_rungreq} shows the required number of rungs to converge for arbitrary $|E_1|$ in the low damping regime.

%